\documentclass[opre]{informs4rad}

\newif\ifMS
\MStrue
\OneAndAHalfSpacedXI

\newif\iffigdraft
\figdrafttrue

\usepackage[suppress]{color-edits}
\addauthor{fe}{blue} 
\addauthor{yf}{purple}    

\usepackage{adjustbox}
\usepackage{amsmath}
 \usepackage{xfrac}
\usepackage{csquotes}
\makeatletter
\renewenvironment*{displayquote}
  {\begingroup\setlength{\leftmargini}{0.7cm}\csq@getcargs{\csq@bdquote{}{}}}
  {\csq@edquote\endgroup}
\makeatother
\usepackage{comment}
\usepackage{xcolor}
\usepackage[authoryear, round]{natbib}
\usepackage{fix-cm}  
\usepackage{bbm}
\usepackage{hhline}
\usepackage{mathtools}
\usepackage{ifthen}
\usepackage{etoolbox}

\usepackage{changepage}

\definecolor{cornellred}{rgb}{0.7, 0.11, 0.11}
\definecolor{maroon}{rgb}{0.52, 0, 0}
\definecolor{dgreen}{rgb}{0.0, 0.5, 0.0}
\definecolor{ballblue}{rgb}{0.13, 0.67, 0.8}
\definecolor{royalblue(web)}{rgb}{0.25, 0.41, 0.88}
\definecolor{bleudefrance}{rgb}{0.19, 0.55, 0.91}
\definecolor{royalazure}{rgb}{0.0, 0.22, 0.66}
\usepackage{hyperref}
\hypersetup{
	colorlinks = true,
	linkcolor=cornellred,
	citecolor=royalazure,
	urlcolor= maroon,
	linkbordercolor = {white}
}
\providecolor{revcolor}{RGB}{128,0,128}
\providecommand{\revcolor}[1]{{\color{revcolor}{#1}}}
\providecommand{\revcitep}[1]{\revcolor{\NoHyper\citep{#1}\endNoHyper}}

\usepackage{cleveref}
\crefname{algocf}{alg.}{algs.}
\Crefname{algocf}{Algorithm}{Algorithms}
\usepackage{enumitem}
\usepackage{multirow}

\usepackage{pgfplots}
\pgfplotsset{compat=1.18}
\usepackage{tikz}
\usetikzlibrary{patterns}
\usepgfplotslibrary{fillbetween}
\usetikzlibrary{intersections}
\usepackage[ruled,vlined,linesnumbered]{algorithm2e}

\SetCommentSty{mycommfont}
\definecolor{cornellred}{rgb}{0.7, 0.11, 0.11}
\definecolor{dgreen}{rgb}{0.0, 0.5, 0.0}
\definecolor{ballblue}{rgb}{0.13, 0.67, 0.8}
\definecolor{royalblue(web)}{rgb}{0.25, 0.41, 0.88}
\definecolor{bleudefrance}{rgb}{0.19, 0.55, 0.91}
\definecolor{royalazure}{rgb}{0.0, 0.22, 0.66}

\usepackage{relsize}

\tikzset{fontscale/.style = {font=\relsize{#1}}
    }

\allowdisplaybreaks
\newtheorem{theorem}{Theorem}[section]
\usepackage{thm-restate}
\newtheorem{lemma}[theorem]{Lemma}
\newtheorem{proposition}[theorem]{Proposition}
\newtheorem{corollary}[theorem]{Corollary}

\newtheorem{definition}{Definition}[section]

\newtheorem{remark}[definition]{Remark}
\newtheorem{example}[definition]{Example}
	\newtheorem{assumption}[definition]{Assumption}

	\usepackage{accents}

\newcommand{\consump}{b}

	\newcommand{\reals}{\mathbb{R}}
	
	\DeclareMathOperator{\OPT}{\textsc{OPT}}
	\DeclareMathOperator{\ALG}{\textsc{ALG}}

    \newcommand{\BALANCE}{{\sf BALANCE}}
    \newcommand{\estimated}{{projected}}

	\providecommand{\given}{}
	\DeclarePairedDelimiterX{\set}[1]\{\}{\renewcommand\given{\nonscript\:\delimsize\vert\nonscript\:\mathopen{}}#1}
	\let\Pr\relax
	\DeclarePairedDelimiterXPP{\Pr}[1]{\mathbb{P}}[]{}{\renewcommand\given{\nonscript\:\delimsize\vert\nonscript\:\mathopen{}}#1}
	\DeclarePairedDelimiterXPP{\Ex}[1]{\mathbb{E}}[]{}{\renewcommand\given{\nonscript\:\delimsize\vert\nonscript\:\mathopen{}}#1}

    \newcommand{\naturals}{\mathbb{N}}

\newcommand{\xhdr}[1]{\vspace{2mm} \noindent{\bf #1}}

\newcommand{\tpre}{\tau}
	
	\newcommand{\primed}{^\dagger}

	\newcommand{\inventory}{c}
	\newcommand{\mininventory}{{\inventory_{\min}}}

	\newcommand{\totaltime}{T}
	\newcommand{\reward}{r}

	\newcommand{\pen}{\Psi}

	\newcommand{\duration}{d}

	\newcommand{\inventorydual}{\theta}
	
	\newcommand{\probdual}{\lambda}

	\newcommand{\Rev}[2][]{\text{\bf Rev}\ifthenelse{\not\equal{}{#1}}{_{#1}}{}\!\left[{\def\givenn{\middle|}#2}\right]}

	\newcommand{\alloc}{x}

	\newcommand{\noaccents}[1]{#1}
	
	\newcommand{\newagentvar}[3][\noaccents]{%
		\expandafter\newcommand\expandafter{\csname #2\endcsname}{#1{#3}}%
		\expandafter\newcommand\expandafter{\csname #2s\endcsname}{#1{\boldsymbol{#3}}}%
		\expandafter\newcommand\expandafter{\csname #2smi\endcsname}[1][i]{#1{\boldsymbol{#3}}_{-##1}}%
		\expandafter\newcommand\expandafter{\csname #2i\endcsname}[1][i]{#1{#3}\agind[##1]}%
		\expandafter\newcommand\expandafter{\csname #2ith\endcsname}[1][i]{#1{#3}_{(##1)}}%
	}
	\newagentvar{typespace}{{\cal Z}}
	\newagentvar{typesubspace}{S}
	
	\newagentvar{type}{z}
	\newagentvar{othertype}{s}
	\newagentvar{val}{v}
	\newagentvar{hval}{\bar \val}
	\newagentvar{hbudget}{\bar \wealth}
	\newagentvar{budget}{B}
	\newagentvar{lbudget}{\underaccent{\bar}{ \wealth}}
	\newagentvar{lowestval}{0}
	\newagentvar{cumval}{V}
	\newagentvar{cumprice}{P}
	\newagentvar{welcurve}{W}
	\newagentvar{revcurve}{R}
	\newagentvar{outcome}{w}
	\newagentvar{outcomespace}{{\cal W}}

\newcommand*{\rom}[1]{\expandafter\romannumeral #1}
\newcommand{\Rom}[1]{\uppercase\expandafter{\romannumeral #1\relax}}

\newcommand{\curinventory}{\alpha}

\newcommand{\configspace}{\mathcal{S}}
\newcommand{\config}{S}

\newcommand{\allocconfig}{\alloc(i,  \config)}
\newcommand{\onlinedual}{\lambda}
\newcommand{\offlinedual}{\theta}

\newcommand{\timesetij}{\timeset_{ij}}

\newcommand{\FLB}{{\sf FLB}}
\newcommand{\OFF}{{\sf OPT}}
\newcommand{\GRD}{{\sf GREEDY}}
\newcommand{\FLBLong}{{\sf Forward-Looking BALANCE}}

\newcommand{\minduration}{\underline{\duration}}
\newcommand{\maxduration}{D}
\newcommand{\timeset}{\mathcal{T}}

\newcommand{\minreward}{\underline{\reward}}
\newcommand{\maxreward}{R}
\newcommand{\heteropair}{\maxreward\maxduration}
\newcommand{\heterosize}{\heteropair}

\newcommand{\neighbor}{N}
\newcommand{\inspecfreq}{\gamma}

\newcommand{\compratio}{\Gamma}

\newcommand{\penscalar}{\beta}
\newcommand{\penscalarTwo}{\eta}

\newcommand{\RHSlessOne}{\frac{\ln \left(1 - \frac{\maxreward}{\maxreward+\penscalarTwo}e^{\penscalarTwo\frac{d}{\maxreward+\penscalarTwo}}-\epsilon \right) 
}{\ln \left(\penscalar\right)}}
\newcommand{\RHSlessOnetau}{\frac{\ln \left(1-\frac{\maxreward}{\maxreward+\penscalarTwo}e^{\penscalarTwo\frac{\tau}{\maxreward+\penscalarTwo}}-\epsilon \right) 
}{\ln \left(\penscalar\right)}}

\newcommand{\RHSgreaterOne}{\frac{\ln \left(1 - \frac{\maxreward}{\maxreward+\penscalarTwo}e^{\penscalarTwo\frac{1}{\maxreward+\penscalarTwo}}-\epsilon\right)
-\frac{\maxreward}{\maxreward+\penscalarTwo}\ln\left(d \right) + \ln(1-\delta)
}{\ln \left(\penscalar\right)}}
\newcommand{\RHSgreaterOnetau}{\frac{\ln \left(1 - \frac{\maxreward}{\maxreward+\penscalarTwo}e^{\penscalarTwo\frac{1}{\maxreward+\penscalarTwo}}-\epsilon \right)
-\frac{\maxreward}{\maxreward+\penscalarTwo}\ln\left(\tau \right) + \ln(1-\delta)
}{\ln \left(\penscalar\right)}}

\newcommand{\estimateinventoryij}{\curinventory_{i, t_j \rightarrow \tpre}}

\newcommand{\estimateinventoryijt}{\curinventory_{i, t_j \rightarrow t}}

\newcommand{\estimateinventoryi}{\curinventory_{i, t \rightarrow \tpre}}

\newcommand{\event}{\mathcal{E}}

\newcommand{\logcons}{\mathcal{C}_\textsc{log-limit}}

\newcommand{\chargemapping}{\sigma}

\newcommand{\offlinedualdecomp}{\offlinedual_{(j,\tpre)}}
\newcommand{\offlinedualdecompi}{\offlinedualdecomp(i)}

\newcommand{\timesetiS}{\timeset_{i\config}}

\newcommand{\cseparation}{\textsc{separation}}
\newcommand{\cfeasibilitya}{\textsc{feasibility-i}}
\newcommand{\cfeasibilityb}{\textsc{feasibility-ii}}

\newcommand{\ked}{^{(k)}}

\newcommand{\FLBOPTInt}[1]{\text{\hyperref[eq:FLB parameter integer duration]{$\mathcal{P}_{\textsc{FLB-Int}}[#1]$}}}
\newcommand{\FLBOPTReal}[1]{\text{\hyperref[eq:FLB parameter real duration]{$\mathcal{P}_{\textsc{FLB-Real}}[#1]$}}}

\newcommand{\sumofreciprocals}{H}

\newcommand{\edges}{E}

\newcommand{\durationij}{\duration_{ij}}

\newcommand{\rewardij}{\reward_{ij}}

\renewcommand{\qed}{$\hfill\square$}

\newenvironment{proofof}[1]{%
  \Trivlist
  \item[\hskip\labelsep {\it #1.}]\ignorespaces
}{\hfill \qed
\endTrivlist
\addvspace{0pt}
}

\newcommand{\reducedreward}{\pi_{\pen}}
\newcommand{\jobconfig}{A}
\newcommand{\jobconfigspace}{\mathcal{A}}
\newcommand{\maxleg}{L}
\renewcommand{\revcolor}[1]{{\color{black}#1}}

\newcommand{\prob}[2][]{\text{Pr}\ifthenelse{\not\equal{}{#1}}{_{#1}}{}\!\left[{\def\givenn{\middle|}#2}\right]}
\newcommand{\expect}[2][]{\mathbb{E}\ifthenelse{\not\equal{}{#1}}{_{#1}}{}\!\left[{\def\givenn{\middle|}#2}\right]}

\newcommand{\tparen}{\big}
\newcommand{\tprob}[2][]{\text{Pr}\ifthenelse{\not\equal{}{#1}}{_{#1}}{}\tparen[{\def\given{\tparen|}#2}\tparen]}
\newcommand{\texpect}[2][]{\mathbb{E}\ifthenelse{\not\equal{}{#1}}{_{#1}}{}\tparen[{\def\given{\tparen|}#2}\tparen]}

\newcommand{\sprob}[2][]{\text{Pr}\ifthenelse{\not\equal{}{#1}}{_{#1}}{}[#2]}
\newcommand{\sexpect}[2][]{\mathbb{E}\ifthenelse{\not\equal{}{#1}}{_{#1}}{}[#2]}

\newcommand{\indicator}[1]{{\mathbbm{1}\left\{ #1 \right\}}}

\usetikzlibrary{arrows.meta, positioning}
\MANUSCRIPTNO{}
\begin{document}
\RUNTITLE{Online Job Assignment}
\TITLE{Online Job Assignment}
\RUNAUTHOR{Ekbatani, Feng, Kash, Niazadeh}

\ARTICLEAUTHORS{%
\AUTHOR{Farbod Ekbatani}
\AFF{University of Chicago Booth School of Business, Chicago, \EMAIL{fekbatan@chicagobooth.edu}} 

\AUTHOR{Yiding Feng}
\AFF{University of Chicago Booth School of Business, Chicago, \EMAIL{yiding.feng@chicagobooth.edu}}

\AUTHOR{Ian Kash}
\AFF{
 Computer Science, University of Illinois at Chicago, Chicago,
\EMAIL{iankash@uic.edu}
}
\AUTHOR{Rad Niazadeh}
\AFF{University of Chicago Booth School of Business, Chicago, IL, \EMAIL{rad.niazadeh@chicagobooth.edu}}

}

\ABSTRACT{
Motivated primarily by applications in cloud computing marketplaces, we study a simple, yet powerful, adversarial online allocation problem in which jobs of varying durations arrive over continuous time and must be assigned immediately and irrevocably to one of the available offline servers. Each server has a fixed initial capacity, with assigned jobs occupying one unit for their duration and releasing it upon completion. The algorithm earns a reward for each assignment upon completion. We consider a general \emph{heterogeneous} setting where both the reward and duration of a job assignment depend on the job-server pair. The objective of the online algorithm is to maximize the total collected reward while remaining competitive against an omniscient benchmark that knows all job arrivals in advance. Our main contribution is the design of a new family of online assignment algorithms, termed \emph{Forward-Looking BALANCE (FLB)}, and using primal-dual framework to establish its competitive ratio. After proper selection of parameters, this family is (asymptotically) optimal-competitive in various regimes.

In summary, this meta-algorithm has two important primitives: (i) keeping track of the capacity used for each server at each time and applying a penalty function to this quantity, and (ii) adjusting the reward of assigning an arriving job to a server by subtracting the total penalty of a particularly chosen subset of future times (referred to as \emph{inspection times}), in contrast to just looking at the current time. The FLB algorithm then assigns the arriving job to the server with the maximum adjusted reward. In the general setting, if $\maxreward$ and $\maxduration$ are the ratios of maximum over minimum rewards and durations, we show that there exists a choice of these primitives so that the FLB algorithm obtains a competitive ratio of $\ln\left(\maxreward\maxduration\right)+3\ln\ln\left(\maxreward\vee\maxduration\right)+\mathcal{O}(1)$ as the initial capacities converge to infinity. Furthermore, in the special case of fixed rewards ($\maxreward=1$), we show that the FLB algorithm with a different choice of primitives obtains a near-optimal asymptotic competitive ratio of $\ln\left(\maxduration\right)+\mathcal{O}(1)$. Our main analysis combines a dual-fitting technique, which leverages the configuration LP benchmark for this problem, and a novel inductive argument to establish the capacity feasibility of the algorithm, which might be of independent interest.

}
\maketitle

\newpage

\section{Introduction}
\label{sec:intro}

The real-time allocation of available supply to incoming demand is crucial for efficient service operations and revenue management of many modern online platforms. Traditional applications, such as ad allocation \citep{KVV-90,MSVV-05,BJN-07,AGKM-11} and assortment planning \citep{GNR-14,MS-20}, typically involve assignments where a demand request consumes a unit of supply for the entire decision-making horizon. For instance, in display advertising, once an impression is assigned to an ad, the advertiser's budget is immediately and irreversibly reduced by the bid amount. Similarly, in online retail, when a consumer purchases a product from the displayed assortment, that unit is permanently removed from the retailer's inventory. However, this paradigm shifts in modern e-commerce platforms and online marketplaces, such as cloud computing services like AWS and Microsoft Azure. In these environments, computing resources---such as virtual machines or memory---are allocated to jobs for their duration and are released upon completion, making them available for future tasks.


This novel feature of cloud platforms introduces emerging challenges in efficiently allocating computing resources in real time to users who compensate the platform for processing their jobs. Motivated primarily by this application, we study a simple, yet fundamental, sequential allocation problem, which we call \emph{online job assignment}. In this problem, a platform aims to manage the assignment of jobs to available servers. Jobs arrive online in continuous time and must be immediately assigned to a compatible server for processing or rejected (no advance booking is allowed). Once assigned, a server must process the job continuously and without interruption (no preemption) for a certain ``duration'' of time. Each server has a capacity (of at least $\mininventory\in\mathbb{N}$), which represents the maximum number of concurrent jobs that it can process at any given time. Upon assigning a job to a server, the platform receives a ``reward'' for the completion of each time unit of the job (which can represent monetary payments to the platform, throughput, or user satisfaction). The objective of the platform is to assign arriving jobs to servers in an online fashion, considering their per-unit rewards and durations, to maximize the total collected rewards while respecting the servers' capacity constraints.

We aim to design competitive online algorithms for our problem and compare them to the optimal offline benchmark, which has full knowledge of the entire set of arriving jobs (thus clearly providing an upper bound on the expected reward achievable by any online algorithm). Importantly, we consider the fully \emph{heterogeneous} setting, where both the duration required to complete a job and the platform's reward may vary depending on the specific job and the server to which it is assigned. This is primarily motivated by the fact that different jobs may have distinct computational requirements, and servers typically differ in specifications, resulting in varied compatibilities regarding these requirements. Consequently, both the completion time of a job on a server and the reward generated from the assignment (e.g., the price paid by the user) may differ significantly.

Specifically, after normalizing the minimum (non-zero) reward and duration to one, we specify the heterogeneity of a given problem instance by the parameters $\maxreward\geq 1$ and $\maxduration\geq 1$, representing the maximum reward and maximum duration among all jobs, respectively. We say an algorithm is \emph{$\compratio(\maxreward,\maxduration)$-competitive} for some $\compratio(\maxreward,\maxduration)\geq 1$ if its total reward for any instance with heterogeneity parameters $(\maxreward,\maxduration)$ is at least $\frac{1}{\compratio(\maxreward,\maxduration)}$ times the total reward obtained by the optimal offline benchmark. Focusing on the practically relevant and widely studied ``large capacity regime'' or ``asymptotic regime,'' where $\mininventory$ is large (see, e.g., the classic AdWords problem in \citealp{MSVV-05}), we address the following research question regarding the competitive ratio and heterogeneity parameters $\maxreward$ and $\maxduration$:

\vspace{1mm}
\begin{displayquote}
\emph{\textbf{Question~(i):} In the online job assignment problem, given heterogeneity parameters $\maxreward\geq 1$ (for rewards) and $\maxduration\geq 1$ (for durations), can we design an online algorithm—ideally simple and practical—with an asymptotically optimal competitive ratio $\compratio(\maxreward,\maxduration)$?}
\end{displayquote}
\vspace{1mm}

Based on the prior literature studying special cases of our model---in particular, (i) single or multiple non-reusable resources with heterogeneous rewards \citep{BQ-09,MS-20,FKMMP-09,EFN-23}, and (ii) single or multiple reusable resources with heterogeneous rewards and durations \citep{HC-22,RSTB-23}---and from the literature on general online packing and covering models---in particular, (i) online routing, load balancing, and interval scheduling \citep{BN-09,AAP-93,LT-94,AKPPW-97}, and (ii) offline combinatorial auctions \citep{BG-15}---we know that the asymptotic optimal competitive ratio should exhibit at least a logarithmic dependency on $R$ and $D$, or more specifically, $\compratio(\maxreward,\maxduration)=\Omega\left(\log(\maxreward\maxduration)\right)$.\footnote{Here, the standard \emph{Bachmann–Landau notations} (or \emph{asymptotic notations})---including Big $\mathcal{O}$, small ${o}$, Big $\Omega$, and small $\omega$---are used to describe the limiting behaviors of our upper and lower performance ratio bounds.} See \Cref{apx:further related work} for details on these results for the mentioned special cases of our problem. Given these lower bounds, for any online algorithm $\ALG$ in the online job assignment problem, we define its \emph{asymptotic logarithmic constant} $\logcons(\ALG)$ as the smallest constant $C>0$ satisfying:
\begin{align*}
\compratio(\maxreward,\maxduration)= C \cdot \ln(\maxreward \maxduration) + o\left(\log(\maxreward \maxduration)\right),
\end{align*}
when the minimum server capacity $\mininventory$ is sufficiently large. By convention, we set $\mathcal{C}_\textsc{log-limit}(\ALG)=+\infty$ if $\compratio(\maxreward,\maxduration)=\omega\left(\log(\maxreward\maxduration)\right)$. We can now rephrase our research question as follows:

\begin{displayquote}
\emph{\textbf{Question~(ii):} In the online job assignment problem, can we characterize the optimal asymptotic logarithmic constant $\logcons^*$, defined as the smallest asymptotic logarithmic constant $\logcons(\ALG)$ achievable by some online algorithm $\ALG$?}
\end{displayquote}

\subsection{Our Main Contributions}
In this paper, we provide compelling answers to both of these questions. In a nutshell, we establish that $\logcons^*=1$, and that it can be achieved by a simple, interpretable, and practical algorithm. More formally, for the online job assignment problem, we show the following main result:

\smallskip
\begin{displayquote}
\emph{\textbf{[Main Result]} We propose a simple, practically relevant, and polynomial-time online algorithm (\Cref{alg:FLB}) that attains the asymptotically optimal competitive ratio (up to lower-order terms) of $\ln(\maxreward\cdot\maxduration)$ for general choices of parameters $\maxreward$ and $\maxduration$ (and $\frac{e}{e-1}$ when $\maxreward=\maxduration=1$), when the minimum server capacity $\mininventory$ is sufficiently large. We further show that this bound is the best achievable by any online algorithm, thus establishing that $\logcons^*=1$. See \Cref{tab:results}.}
\end{displayquote}
\vspace{-1mm}


\begin{table}[ht]
\centering
\begin{adjustwidth}{1.2cm}{}
\smallskip
\small\addtolength{\tabcolsep}{-3pt}
\caption{Summary of the main results (optimal asymptotic logarithmic constant of $\logcons^*=1$).}
\label{tab:results}
\renewcommand\arraystretch{1.5}
\renewcommand{\thefootnote}{\fnsymbol{footnote}}
\small

\begin{tabular}{|c|c|c|c|}
  \hhline{~~--}
  \multicolumn{2}{c|}{}%
    & {\color{blue}\shortstack{\\\footnotesize{\textbf{Upper Bound}}}}%
    & {\color{maroon}\shortstack{\\\footnotesize{\textbf{Lower Bound}}}} \\
  \hhline{--==}
  
  \multirow{2}{*}{\shortstack{\\\footnotesize{General}\\ \footnotesize{($\maxreward>1$, $\maxduration>1$)}}}%
    & \footnotesize{\shortstack{\\Real\\Durations}}%
    & {\color{blue}\shortstack{\\\footnotesize{$\ln(\maxreward\maxduration)+3\ln\ln(\maxreward\vee\maxduration)+\mathcal{O}(1)$}\\
      \footnotesize{\textbf{(Theorem.~\ref{thm:FLB competitive ratio general})}}}}%
    & {\color{maroon}\shortstack{\footnotesize{~$\ln(\maxreward\maxduration)+\Omega(1)$~}\\
      \footnotesize{\textbf{(Proposition.~\ref{prop:optimal the competitive ratio divisible lower bound})}}}} \\
  \hhline{~---}
  
  & \footnotesize{\shortstack{\\Integer\\Durations}}%
    & {\color{blue}\shortstack{\\\footnotesize{~$\ln(\maxreward\maxduration)+\ln\ln(\maxreward\vee\maxduration)+\mathcal{O}(1)$~}\\
      \footnotesize{\textbf{(Theorem.~\ref{thm:competitive ratio integer duration})}}}}%
    & {\color{maroon}\footnotesize{$(\mathrm{same\;as\;above})$}} \\
  \hhline{|=|=|=|=|}
  
  \multirow{2}{*}{\shortstack{\\\footnotesize{Fixed-Reward}\\ \footnotesize{($R=1$, $D>1$)}}}%
    & \footnotesize{\shortstack{\\Real\\Durations}}%
    & {\color{blue}\shortstack{\footnotesize{$\ln(\maxduration)+4$}\\
      \footnotesize{\textbf{(Proposition.~\ref{prop:CR identical reward real duration})}}}}%
    & {\color{maroon}\shortstack{\footnotesize{$\ln(\maxduration)+2$}\\
      \footnotesize{\textbf{(Proposition.~\ref{prop:CR identical reward real duration})}}}} \\
  \hhline{~---}
  
  & \footnotesize{\shortstack{\\Integer\\Durations}}%
    & {\color{blue}\shortstack{\footnotesize{$H(\maxduration)+2$}\\
      \footnotesize{\textbf{(Proposition.~\ref{prop:CR identical reward integer duration})}}}}%
    & {\color{maroon}\shortstack{\footnotesize{$H(\maxduration)$}\\
      \footnotesize{\textbf{(Proposition.~\ref{prop:CR identical reward integer duration})}}}} \\
   \hhline{|=|=|=|=|}
  
  \multicolumn{2}{|c|}{\shortstack{\\\footnotesize{Fixed-Reward, Fixed-Duration}\\ \footnotesize{($\maxreward=1$, $\maxduration=1$)}}}%
    & {\color{blue}\shortstack{\\\footnotesize{$\frac{e}{e-1}$}\\
      \footnotesize{\textbf{(Remark.~\ref{rmk:HomogenousRewardsDurations})}}}}%
    & {\color{maroon}\shortstack{\footnotesize{$\frac{e}{e-1}$} \\ \footnotesize{\textbf{\cite{KVV-90,MSVV-05}}}}} \\
  \hhline{----}
\end{tabular}
\smallskip

\noindent
\begin{minipage}{0.85\textwidth}
  \footnotesize
  \emph{\underline{Note}: $H(k) \triangleq \sum_{i=1}^{k} i^{-1} \in [\ln(k),\,\ln(k)+1]$ denotes 
         the $k$-th harmonic number; operator $\vee$ is \emph{max} and all 
         $\ln(\cdot)$ are natural logarithms. The upper bound of $\frac{e}{e - 1}$ is known in \citet{FNS-19,GIU-20} for the setting of fixed reward, fixed duration and discrete-time arrival.}
\end{minipage}
\end{adjustwidth}
\end{table}

\vspace{-2mm}
Our paper extends and unifies the scope of several previous works in the literature on online resource allocation, each focusing on specific aspects of the problem. Within our comprehensive model, we achieve asymptotically near-optimal competitive ratios (with almost optimal dependencies on the parameters $\maxreward$ and $\maxduration$). Our technical developments yield a significant improvement in terms of the asymptotic logarithmic constant over prior work concerning special cases of our problem, using a simple and practical algorithm. Moreover, we recover the exact asymptotic optimal bounds in the special cases corresponding to $\maxreward=1$ and to $\maxreward=\maxduration=1$, studied previously in \cite{FNS-19,GGISUW-22,DFNSU-22,GIU-20}. We also note that although our primary analysis addresses the base setting in which the heterogeneity parameters are independent of the server choice, all of our results naturally extend to the server-dependent heterogeneous setting where each server $i$ has its own reward range $\maxreward_i$ and duration range $\maxduration_i$, with $\maxreward\triangleq\max \maxreward_i$ and $\maxduration\triangleq\max \maxduration_i$. See \Cref{apx:server-dependent}. \revcolor{In addition, we generalize our algorithm and its competitive ratio analysis to the network revenue management and multi-unit online matching models in \Cref{apx:NRM,apx:multi-unit}, respectively.}

Finally, in \Cref{sec:numerical}, we evaluate the numerical performance guarantee of our proposed algorithm both on the class of worst-case instances we identify in this paper, and on randomized instances beyond the worst-case. We compare the algorithm with other existing benchmarks in the literature in terms of their instance-wise competitive ratio.  We observe that not only our algorithm both theoretically and numerically has the best performance in the worst-case instance, but also in almost all of our simulations, our algorithm outperforms the other benchmarks in most randomized instances. 

\vspace{-1mm}
\subsection{Overview of the Techniques}
Before elaborating on the algorithmic development that leads to our main result, it is helpful to provide some context from the literature. As a special case of our model, \citet{KP-00} introduce the simple and elegant {\BALANCE} algorithm for the online bipartite $b$-matching \revcolor{(where at most $b$ jobs can be matched to any server)} problem involving non-reusable resources with \revcolor{unweighted} rewards. Subsequent works extend this idea to other variants of online resource allocation, such as those with non-reusable resources and essentially identical rewards over time, e.g.,~\cite{MSVV-05,AGKM-11} (corresponding to $\maxreward=1$ and infinite-duration jobs), as well as to reusable resources with identical rewards and durations over time, e.g.,~\cite{FNS-19,FNS-21} (corresponding to $\maxreward=\maxduration=1$). 

The central idea behind the {\BALANCE} algorithm is to evenly distribute the load over time by computing a reduced reward for each resource (in our setting, each server), based on the resource's current available capacity at each point in time. The algorithm then greedily assigns each arriving request (or job, in our setting) to the resource with the highest positive reduced reward. Notably, this reward adjustment can be interpreted through a primal-dual perspective, where the available capacity is used as a surrogate by the online algorithm to maintain an approximate dual solution, and then rewards are adjusted according to this dual solution following complementary slackness. 

The {\BALANCE} algorithm achieves the asymptotically optimal competitive ratio of $\frac{e}{e - 1}$ when $\maxreward=\maxduration=1$ in the large-capacity regime. However, it remains unclear whether this approach can be successfully applied or extended to the heterogeneous online job assignment setting (i.e., when $R>1$ or $D>1$). Indeed, we have both strong theoretical (see \Cref{example:FLB}) and empirical (see \Cref{sec:numerical}) \revcolor{evidence} suggesting that such an extension would fail to maintain the algorithm's performance guarantees, as it only considers the current server utilizations (equivalently, the available capacity), without accounting for how \emph{future utilizations} may evolve. Our proposed algorithm aims to \emph{precisely} address this limitation.

\xhdr{\FLBLong: ``forecasting'' future utilizations.} Our algorithm is a novel generalization of {\BALANCE}. We introduce a new parametric family of online algorithms, called {\FLBLong} ({\FLB}), that extends the original idea of achieving a ``balanced'' allocation based on adjusted rewards to the online job assignment problem with arbitrary heterogeneity in rewards and durations of arriving jobs. To handle the reusability of servers and heterogeneity across jobs, a balanced allocation must consider not only the current available capacities of servers but also the (projected) future trajectories of these available capacities---thus being ``forward-looking.'' Intuitively speaking, while the current number of jobs assigned to a server (reflected in its current available capacity) indicates how busy the server is, it provides an incomplete picture in settings with heterogeneous durations: currently assigned jobs will finish one by one at different future times, increasing the server's available capacity at each of those times. Accounting for this future capacity recovery can potentially improve current assignment decisions.

To illustrate this point, consider a simple scenario involving two servers with identical rewards and durations for a particular job, and imagine that the current available capacities of these servers are also identical. In this situation, the classic {\BALANCE} would calculate the same reduced reward for both servers and is thus indifferent to assigning the job to either of them. However, if server $1$ is expected to regain all currently unavailable capacity units within the next second, while the unavailable units of server $2$ become available much later or never (due to significantly longer-duration jobs assigned to this server in the past), it becomes intuitively preferable to prioritize server $1$, which will soon have higher available capacity, over server $2$. This forward-looking prioritization maintains balanced resource utilization over time. {\FLB} formalizes this intuitive idea by incorporating both the current available capacity as well as the anticipated future available capacities into the calculation of the reduced reward for each server. For additional details, see the discussion in \Cref{sec: general_alg}, particularly \Cref{example:FLB} and \Cref{fig:example}.



\xhdr{Customizing the penalty function and inspection times.} Given the inherent heterogeneity of the environment, particularly due to varying job durations, accurately capturing the trade-off between a server's immediate availability and its anticipated future availability through reward adjustments poses intricate challenges, necessitating novel algorithmic flexibility. As a first form of flexibility, we allow {\FLB} to use the parameters $\maxreward$ and $\maxduration$ to determine how it adjusts rewards. Specifically, we employ the following \emph{family of exponential penalty functions} $\pen:[0,1]\longrightarrow\reals_{+}$, parameterized by $(\penscalar,\penscalarTwo)$, to translate the (projected) normalized available capacity at each current and future inspection time into a dual-based penalty:
$$
    \pen(x) = 
    \penscalarTwo \left(
    \penscalar^{(1 - x)} - 1
    \right),
$$
where parameters $\penscalar\geq 1$ and $\penscalarTwo\geq 0$ may depend on $\maxreward$ and $\maxduration$. As a second form of flexibility, {\FLB} addresses duration heterogeneity by selecting only a (possibly infinite) subset of future times---referred to as \emph{inspection times}---to anticipate the trajectory of available capacity when making the current assignment decision. At any given continuous time $t$, these inspection times for an arriving job on each server are periodic points at a fixed frequency $\inspecfreq\geq 0$, chosen from the set of future times during which this job causes resource conflicts on that server (i.e., the interval $[t,t+\duration]$, where $\duration$ is the job's duration on that server); see \Cref{sec: general_alg} for a formal definition. Similar to $(\penscalar,\penscalarTwo)$, the frequency parameter $\inspecfreq$ may depend on the heterogeneity parameters $\maxreward$ and $\maxduration$. Ultimately, the reward adjustment at time $t$ reduces the original reward by the sum of all dual-based penalties corresponding to current and future inspection times. For technical reasons elaborated further in \Cref{sec:competitive ratio analysis}, the choices of the penalty function and frequency of future inspections both play critical roles in our primal-dual analysis, which we briefly describe next.

\xhdr{Primal-dual analysis with configuration LP.}
We rely on a novel primal-dual analysis to establish the competitive ratios of {\FLB}. Similar to prior works on variants and special cases of the online job assignment problem, we consider a linear programming (LP) relaxation \ref{eq:opt lp} of the optimal offline policy and its corresponding dual program. However, unlike the standard LP relaxations used for primal-dual analysis in this literature, where each primal variable represents the probability of matching a job-server pair \citep{FNS-19,HC-22}, we introduce a \emph{configuration linear program} (see \ref{eq:opt lp}) to help with a refined analysis of {\FLB} and establish the (almost) optimal dependencies of the competitive ratio on both $\maxreward$ and $\maxduration$.


In this configuration LP, each server is assigned to a collection of feasible configurations, where each feasible configuration is a subset of non-overlapping jobs that can all be processed using the same unit of capacity on that server. Each job is used in exactly one configuration, and each unit of server capacity is assigned precisely to one feasible configuration (the units are identical, so no server has a collection of feasible configurations larger than its capacity). Such a configuration linear program can encode any feasible assignment of jobs to servers given their capacities, regardless of duration heterogeneity. This granular, ``higher-dimensional representation'' of our assignment is one of the key technical insights that allows us to construct proper dual solutions, certifying the approximate optimality of the primal solution generated by {\FLB}.


More specifically, in the dual program of our configuration LP (see \ref{eq:opt lp}), there is a non-negative dual variable associated with each job and server. The dual objective function is simply the sum of all these dual variables, while the dual constraints impose limitations on the sum of dual variables corresponding to each server and each subset of jobs that can simultaneously be assigned to the same unit of capacity on that server. The primal-dual method aims to construct a dual assignment satisfying two key properties: (i) its objective value is at most $\compratio(\maxreward,\maxduration)$ times the total reward collected by {\FLB}; and (ii) all dual constraints are satisfied. To achieve this, we explicitly construct a dual assignment guided by the execution of {\FLB}. We establish property~(i) using the specific closed-form expression the algorithm employs to compute the reduced reward and demonstrate property~(ii) via a novel constructive charging argument.

As will become clear later in \Cref{sec:competitive ratio analysis}, proving property~(ii)---and thus obtaining our refined bounds with near-optimal dependencies on $\maxreward$ and $\maxduration$---is quite subtle. The key insight behind our improved analysis is to \emph{not} force feasibility of the primal solution generated by {\FLB} through an overly conservative choice of the penalty function (which sets the reduced reward to zero whenever the current capacity reaches zero). Instead, we adopt an alternative penalty function and show that feasibility naturally arises as an \emph{indirect} consequence of running our algorithm. This reasoning relies on maintaining a specific invariant property throughout the algorithm's execution, representing another novel aspect of our approach, which we briefly discuss next.

\xhdr{Primal capacity feasibility using an invariant.}
In both the classic {\BALANCE} and our proposed {\FLB}, the penalty function must be carefully designed to ensure capacity feasibility. Analyzing the capacity feasibility of {\BALANCE} for non-reusable resources is relatively straightforward, as it involves considering a single scenario where the entire capacity is exhausted. However, in our setting, verifying the capacity feasibility of {\FLB} involves infinitely many scenarios---capacity can become temporarily exhausted, and future capacity levels vary across these scenarios. To guarantee the feasibility in all such scenarios, the algorithm designer might be forced to adopt a pessimistic penalty function, which, as previously discussed, would lead to poor competitive performance. To improve competitive performance, we introduce an \emph{invariant property} that essentially characterizes the possible future capacity levels of {\FLB} given different choices of the penalty function. This invariant allows us to exclude pessimistic scenarios where capacity feasibility could fail, as these scenarios will never occur in any execution path of {\FLB}. To prove this invariant property, we develop a novel inductive argument that also clarifies the selection of specific parameters $(\inspecfreq,\penscalarTwo,\penscalar)$ for {\FLB}, and reveals how they are influenced by the parameters $\maxreward$ and $\maxduration$.


Our work is connected to various lines of literature in operations research and computer science. A comprehensive discussion of further related work can be found in \Cref{apx:further related work}.


\section{Problem Formulation}

\label{sec:prelim}
We study the \emph{online job assignment} problem in a cloud computing platform, where jobs arrive online and servers are available offline. In this setting, there are $n$ servers, indexed by $[n] \triangleq \{1, 2, \dots, n\}$. Each server $i$ has a capacity denoted by $\inventory_i \in \naturals$, representing the maximum number of jobs the server can simultaneously process at any given time. We consider a finite continuous-time horizon $[0,T]$, during which $m$ jobs arrive at times $0 \leq t_1 \leq \dots \leq t_m \leq \totaltime$. The set $\edges \subseteq [n] \times [m]$ encodes the compatibility between servers and jobs. Each edge $(i,j)\in \edges$ is associated with a per-period reward $\rewardij \in \reals_{+}$, representing the reward the platform receives per unit time if job $j$ is assigned to server~$i$. Additionally, each edge $(i,j)\in \edges$ has an associated duration $\durationij \in \reals_{+}$, representing the time required for job $j$ to complete on server $i$ (which may differ across servers). 

Upon the arrival of job $j$, the set of compatible servers $\neighbor(j) \triangleq \{i\in [n] : (i, j)\in \edges\}$ and their corresponding rewards and durations $\{\rewardij, \durationij\}_{i\in \neighbor(j)}$ are revealed to an online algorithm. The algorithm then immediately and irrevocably assigns job $j$ to one of the compatible servers with available capacity, if any. We refer to the tuple $(\neighbor(j), \{\rewardij, \durationij\}_{i\in \neighbor(j)})$ as the type of job $j$. The platform receives a total reward of $\rewardij\durationij$ for assigning (and completing) job $j$ on server $i$. The completion time $t_j + \durationij$ may exceed the time horizon $T$, in which case the server's capacity remains occupied until time $T$. The goal of the online algorithm is to maximize the total reward collected during the time interval $[0,T]$. We highlight that this model is similar to the online matching with reusable resources~\citep{Udw-21,FNS-19}, but we diverge in that the rewards and durations are allowed to be \revcolor{\emph{arbitrarily heterogeneous (over time and resources)}, meaning that the reward of allocating a unit of an offline resource---i.e., a unit of server capacity---could depend on both the resource and the time that it has been allocated, or equivalently, the reward could depend on the pair of the resource and the job that is assigned to that resource. Additionally, our framework accommodates job arrivals in continuous time.}

We make the following regularity assumption to bound the heterogeneity of rewards and durations, without which meaningful performance guarantees are not possible~\citep{AKLMNY-15}.
\begin{assumption}[Bounded Heterogeneity]
\label{assumption:bounded-het}
    For every job $j\in[m]$ and every compatible server $i\in \neighbor(j)$, the associated reward $\rewardij$ and duration $\durationij$ satisfy $\rewardij \in [1, \maxreward]$ and $\durationij\in[1, \maxduration]$, where $\maxreward$ and $\maxduration$ denote the maximum reward and maximum duration across all job-server pairs, respectively.
\end{assumption}



\begin{remark}[Known Parameters $\boldsymbol{R}$ and $\boldsymbol{D}$]
We assume that the online algorithm knows the ranges of rewards $R$ and durations $D$. This assumption is well-motivated in practical applications and is also theoretically necessary. Specifically, \Cref{apx:agnostic}, together with \Cref{thm:FLB competitive ratio general}, shows a strict separation in performance between algorithms with and without knowledge of $(\maxreward,\maxduration)$. Similar assumptions have also been previously made in the literature for special cases of our problem~\citep[e.g.,][]{BN-09,CMT-19,MS-20}.\end{remark}

\begin{remark}[Uncertain Rewards and Durations]\label{rmk:uncertainRewardAndDuration}
\revcolor{While our baseline model assumes perfect information regarding job rewards and durations, both the theoretical framework and our algorithm extend naturally to environments featuring interval uncertainty, where the decision-maker observes only the upper and lower bounds of these parameters. As established in \Cref{apx:noisy}, accommodating these intervals scales the competitive ratio by a multiplicative factor of $\kappa_d^2\kappa_r$, where $\kappa_d$ and $\kappa_r$ denote the maximum-to-minimum ratios of the duration and reward intervals, respectively (measuring the amount of uncertainty in these quantities). Notably, this extension also accommodates broader settings where the total reward does not scale linearly with job duration; our framework captures such non-linear dynamics provided that the true reward function can be bounded from above and below by appropriate linear functions of the duration $d$.}
\end{remark}

Given parameters $\maxreward \geq 1$ and $\maxduration \geq 1$, we evaluate the performance of an online algorithm by its \emph{competitive ratio} against the optimal offline benchmark with full knowledge of the sequence of job types in advance. Specifically, the competitive ratio is the worst-case ratio of the total reward obtained by the online algorithm to that of the optimal offline benchmark within instances with rewards at most $\maxreward$ and durations at most $\maxduration$. Throughout this work, we focus mainly on the \emph{large capacity regime} when analyzing competitive ratios, where $\mininventory \triangleq \min_{i\in[n]}\inventory_i$ tends to infinity.\footnote{This regime is practically relevant in cloud marketplaces and standard in theoretical analyses within the online matching literature, e.g., \cite{KP-00,MSVV-05,GNR-14,FNS-19,MS-20}.} 

More formally, letting $\mathcal{I}(R,D,\mininventory)$ denote the subset of problem instances with $1\leq \rewardij\leq R$ and $1\leq \durationij\leq D$ for $j\in[m],i\in \neighbor(j)$,\footnote{In \Cref{assumption:bounded-het}, we normalize both the minimum reward and the duration to 1 for compatible server-job pairs. This is without loss of generality (particularly for durations, since we consider a continuous-time horizon), and our results naturally extend to a more general setting, where each server $i\in[n]$ is associated with parameters $(\minreward^{(i)}, \minduration^{(i)})$, and the reward (resp. duration) range between server $i$ and its compatible jobs $j\in N^{-1}(i)$ is $[\minreward^{(i)}, \maxreward\cdot\minreward^{(i)}]$ (resp. $[\minduration^{(i)}, \maxduration\cdot\minduration^{(i)}]$). See \Cref{apx:server-dependent} for a detailed discussion.}  and minimum capacity $\mininventory$, we have the following definition.
\begin{definition}[Asymptotic Competitive Ratio]
	\label{def:competitive ratio}
	Given any $\maxreward,\maxduration\geq 1$,
        a (possibly randomized) online algorithm $\ALG$ is said to be ``(asymptotically) $\compratio(\maxreward,\maxduration)$-competitive'' if the following holds:
    \begin{align*}
          \underset{\mininventory\to\infty}{\limsup}\left(\sup_{I\in\mathcal{I}(R,D)}
		\frac{\OPT(I)}{\mathbb{E}\left[\ALG(I)\right]}\right)
		\leq \compratio({\maxreward,\maxduration})
	\end{align*}
	where 
    $\ALG(I)$ and $\OPT(I)$, respectively, are the total rewards of the online algorithm $\ALG$ and the optimal offline benchmark $\OPT$ under the problem instance $I$.  
\end{definition}

\section{The Forward-Looking BALANCE Algorithm}
\label{sec: general_alg}
Our main result is a new online algorithm, known as {\FLBLong} ({\FLB}). In this section, we describe this algorithm and all its ingredients. The competitive ratio analysis is deferred to \Cref{sec:Heterogeneous weights,apx:FLB analysis real duration} (upper bound) and \Cref{sec:lower bound construction} (lower bound).

The main new idea behind {\FLB} is simple and natural: When a job $j$ arrives, the assignment decision for this job is made not only based on the type of job $j$ and the current available capacities of compatible servers, but also based on the \emph{{\estimated} available capacities} 
of each compatible server $i$ at the current time $t_j$ for each future time $\tpre \geq t_j$ (or a subset of these times). 
\begin{definition}[Projected Available Capacity]
\label{def:projected available capacity}
    For each server $i\in[n]$ and time points $t, \tpre \in [0,T]$ with $t \leq \tpre$, the (normalized) {\estimated} available capacity $\estimateinventoryi\in[0, 1]$ is defined as
    \begin{align*}
        \estimateinventoryi \triangleq 
        1 - \frac{1}{\inventory_i}\sum\nolimits_{j\in[m]: t_j < t}
        \indicator{\event(i, j)}
        \cdot 
        \indicator{t_j + \duration_{ij} > \tpre},
    \end{align*}
    where $\event(i, j)$ is the event that a capacity unit of server $i$ is assigned to job $j$.
\end{definition}
In other words, the {\estimated} available capacity $\estimateinventoryi$ represents the proportion of available units for server $i$ at time $\tpre$, taking into account the job assignments made before time $t$. For any assignment decisions, the resulting $\estimateinventoryi$ is weakly increasing in $\tpre$ and weakly decreasing in $t$. Furthermore, $\curinventory_{i, t\rightarrow t}$ denotes the proportion of available units for server $i$ at time $t$. 


To use the {\estimated} available capacities in its assignments,  {\FLB} is equipped with two important technical ingredients:~\emph{(i) inspection-time subset} and \emph{(ii) projected-utilization-based reduced reward}.

\xhdr{(i)~Inspection-time subset.}
As its name suggests, {\FLB} makes job assignment decisions by taking a forward-looking perspective and examining the available capacity in the future. Specifically, at the arrival time $t_j$ of each job $j$, {\FLB} computes the {\estimated} available capacity $\estimateinventoryij$ for every inspection time $\tpre \in \timesetij(\inspecfreq)$ and every server $i\in[n]$. Here, the inspection-time subset $\timesetij(\inspecfreq)\subseteq[t_j, t_j + \durationij)$, parameterized by an \emph{inspection-frequency scalar $\inspecfreq\in[1,\infty)$}, is a subset of current and future time points defined as
\begin{align*}
    \timesetij(\inspecfreq) \triangleq
    \left\{\tpre 
    \in [t_j, t_j + \durationij):
    \exists \ell\in\naturals
    \text{ s.t. }
    \tpre = 
    t_j + \tfrac{\ell}{\inspecfreq}
    \right\}.
\end{align*}
Recall that when a unit of server $i$ is assigned to job $j$, it remains occupied from time $t_j$ and becomes available again at the beginning of time $t_j + \durationij$. Therefore, it is natural for the algorithm to examine the {\estimated} available capacity $\estimateinventoryij$ at inspection times $\tpre \in \timesetij(\inspecfreq) \subseteq [t_j, t_j + \durationij)$.

\xhdr{(ii)~Projected-utilization-based reduced reward.} 
Due to future uncertainty and job heterogeneity, it may be preferable to assign jobs less frequently to servers with high utilization (or equivalently, low available capacity) at the current time, or those projected to have high utilization at a future time, unless the immediate reward is substantial. To capture this inherent trade-off, we introduce the (projected-utilization-based) reduced reward for each job $j$ and server~$i\in N(j)$, defined as:
\begin{align*}
    \rewardij \durationij -
    \sum\nolimits_{\tpre \in \timesetij(\inspecfreq)}
    \pen\left(\estimateinventoryij\right),
\end{align*}
where $\pen:[0,1]\rightarrow\reals_+$ is a weakly decreasing and convex \emph{penalty function} satisfying $\pen(1) = 0$. As a sanity check, when all units of server $i$ are available for processing job $j$ throughout the entire time horizon, the reduced reward equals the original reward. Furthermore, the reduced reward decreases as fewer units become available. Notably, we make \emph{no} explicit assumptions on $\pen(0)$, and therefore we do not enforce the reduced reward of a server with no available units to be negative. We also focus on a specific form of exponential penalty function $\pen$ with a varying base, parameterized by \emph{penalty parameters} $(\penscalarTwo,\penscalar)$, where $\penscalar \geq 1$ and $\penscalarTwo \geq 0$:
\begin{align*}
    \pen(x) = 
    \penscalarTwo \left(
    \penscalar^{(1 - x)} -
    1
    \right).
\end{align*}

Similar to the classic {\BALANCE} algorithm~\citep{KP-00,MSVV-05}, {\FLB} computes a score for all compatible servers upon the arrival of each job~$j$. It then makes a greedy-style decision, assigning job $j$ to the server $i^*_j$ with the highest positive score, if such a server exists (otherwise, the job remains unassigned). The key difference is that {\FLB} uses the projected-utilization-based reduced rewards as its scores, while the scores in {\BALANCE} are based solely on current utilization. See \Cref{alg:FLB} for a formal description of the algorithm.
\vspace{-2mm}
\begin{algorithm}
 	\caption{\FLBLong\ (\FLB)}
 	\label{alg:FLB}
 	\KwIn{inspection-frequency scalar $\inspecfreq$,
  penalty parameters $(\penscalarTwo,\penscalar)$
  }
  \For{each job $j=1$ to $m$}{ 
 	\tcc{job $j$ with type $(N(j), \{\rewardij, \durationij\}_{i\in N(j)})$ arrives
 	} 
        \If{$\max_{i\in N(j)}\rewardij \durationij -
    \sum\nolimits_{\tpre \in \timesetij(\inspecfreq)}
    \pen\left(\estimateinventoryij\right)
    > 0$ }{
    let $i^*_j \gets \argmax_{i\in N(j)}
    ~
    \rewardij \durationij -
    \sum\nolimits_{\tpre \in \timesetij(\inspecfreq)}
    \pen\left(\estimateinventoryij\right)
    $
    
    assign job $j$ to server $i_j^*$}
 	 }
\end{algorithm}
\vspace{-5mm}
Importantly, we do not enforce feasibility constraints (due to servers having limited capacity) \emph{explicitly} within {\FLB} by controlling $\pen(0)$. Instead, as highlighted in the next remark, its feasibility is achieved \emph{implicitly} through the structure of the algorithm and the appropriate parameter choices. This approach is one of the key insights underlying our design, enabling {\FLB} to be a polynomial-time algorithm with correct dependence of the competitive ratio on $\maxreward$ and $\maxduration$.
\begin{remark}[Capacity Feasibility and Running Time]
\label{remark:feasibility and runtime}
In later sections, we identify conditions (\Cref{lem:FLB feasibility real duration,prop:FLB feasibility integer duration}) on parameters $(\inspecfreq,\penscalarTwo,\penscalar)$ that ensure the ``capacity feasibility'' of {\FLB}, meaning it never assigns a job to a server without available capacity by following lines (2-4) in \Cref{alg:FLB}. Moreover, {\FLB} can be implemented in polynomial time, even when its inspection-time subset $\timesetij(\inspecfreq)$ is large or continuous, because there are only a polynomial number of time points at which the value of the {\estimated} available capacity changes.
\end{remark}

\xhdr{Comparison with {\BALANCE}.}
If we set the inspection-frequency scalar $\inspecfreq = 1$ for instances with maximum duration $\maxduration=1$, the inspection-time subset $\timesetij(1)$ reduces to a singleton $\{t_j\}$. This implies that the reduced reward is computed solely based on the available capacity $\curinventory_{i, t_j\rightarrow t_j}$ at the current time $t_j$, causing {\FLB} to coincide with {\BALANCE}. However, for general instances with $\maxduration>1$, {\FLB} may behave differently from {\BALANCE}, as it also takes projected utilization levels at future times into account. The following example illustrates that this distinction plays a key role in enabling {\FLB} to potentially outperform {\BALANCE}.
\begin{example}
\label{example:FLB}
Consider an example with $n=2$ servers $\{1,2\}$, each having capacity $\inventory_i= 4$. At time $t = 4$, new job $j$ arrives with durations $\duration_{1j}=\duration_{2j} = 2$ and rewards $\reward_{1j} = 1$, $\reward_{2j} = 1+\epsilon$. The state of the servers before assigning job $j$ is illustrated in \Cref{fig:example}: server 1 (on the left) is processing two jobs ending at $\{4.8,5\}$, while server 2 (on the right) is handling two jobs completing at $\{5.2,6.3\}$. Since both servers have identical current available capacity, {\BALANCE} prioritizes the higher immediate reward and assigns job~$j$ to server~2. However, {\FLB} penalizes server~2 for its lower {\estimated} available capacity in the near future, and therefore assigns job $j$ to server~1 instead. This choice better ``hedges'' against the risk of future (adversarial) job arrivals by maintaining more balanced future utilization. For instance, in the scenario where three jobs represented by the dashed rectangles in \Cref{fig:example} arrive in the future, both servers would retain sufficient available capacity to accept all three under {\FLB}. This is indeed needed if these jobs are compatible with only one of the servers and generate high rewards.
    
\end{example}
\begin{figure}
    \centering
\scalebox{0.8}{

\begin{tikzpicture}[
    job/.style={draw, fill=blue!30, rectangle, minimum height=0.6cm, minimum width=1cm} 
]
\begin{scope}
\draw[->, thick] (0, 0) -- (7, 0) node[below] {$t$};
\foreach \x in {0, 2, 4, 6} {
    \draw[thick] (\x, -0.2) -- (\x, 0);
    \node[below] at (\x, -0.2) {\x};
}
\draw[thick] (-0.2, 2.4) -- (0, 2.4);
\node[left] at (-0.2, 2.4) {1};
\draw[->, thick] (0, 0) -- (0, 3) node[above] {$\alpha_{1,4\rightarrow t}$};
\node[job, minimum width=5cm,minimum height=.6cm, fill=red!50] (job2) at (2.5, 0.3) {};
\node[job, minimum width=3cm,minimum height=.6cm, fill=blue!50] (job2) at (3.3, 0.9) {};

\draw[dashed, ultra thin] (0, 2.4) -- (7, 2.4) node[right]{};
\draw[white, thick] (5.2,2.4) -- (6.2,2.4);

\draw[dashed, thick] (5.5,0.6) rectangle (6.65,1.2);
\draw[white, thick] (5.5,1.2) -- (6.65,1.2);
\draw[dashed, thick] (5.35,1.2) rectangle (7,1.8);
\draw[white, thick] (5.35,1.8) -- (6.2,1.8);
\draw[dashed, thick] (5.2,1.8) rectangle (6.2,2.4);

\node[job, minimum width=2cm,minimum height=.6cm, fill=dgreen!50] (job2) at (5, 3) {$\textrm{job}~j$};
\draw[dashed, ultra thin,color=dgreen] (4, 4) -- (4, 0) node[right]{};
\draw[dashed, ultra thin,color=dgreen] (6, 4) -- (6, 0) node[right]{};
\end{scope}

\begin{scope}[shift={(10, 0)}]

\draw[->, thick] (0, 0) -- (7, 0) node[below] {$t$};
\foreach \x in {0, 2, 4, 6} {
    \draw[thick] (\x, -0.2) -- (\x, 0);
    \node[below] at (\x, -0.2) {\x};
}
\draw[thick] (-0.2, 2.4) -- (0, 2.4);
\node[left] at (-0.2, 2.4) {1};
\draw[->, thick] (0, 0) -- (0, 3) node[above] {$\alpha_{2,4\rightarrow t}$};

\draw[dashed, ultra thin] (0, 2.4) -- (7, 2.4) node[right]{};
\draw[white, thick] (5.2,2.4) -- (6.2,2.4);
\draw[dashed, thick] (5.5,0.6) rectangle (6.65,1.2);
\draw[white, thick] (5.5,1.2) -- (6.65,1.2);
\draw[dashed, thick] (5.35,1.2) rectangle (7,1.8);
\draw[white, thick] (5.35,1.8) -- (6.2,1.8);
\draw[dashed, thick] (5.2,1.8) rectangle (6.2,2.4);

\node[job, minimum width=4cm,minimum height=0.6cm] (job1) at (3.3, 0.3) {};
\node[job, minimum width=3.1cm,minimum height=0.6cm, fill=yellow!30] (job2) at (3.75, 0.9) {};
\node[job, minimum width=0.9cm,minimum height=0.6cm, fill=yellow!30] (job2) at (5.75, 0.3) {};

\node[job, minimum width=2cm,minimum height=.6cm, fill=dgreen!50] (job2) at (5, 3) {$\textrm{job}~j$};
\draw[dashed, ultra thin,color=dgreen] (4, 4) -- (4, 0) node[right]{};
\draw[dashed, ultra thin,color=dgreen] (6, 4) -- (6, 0) node[right]{};
\end{scope}
\end{tikzpicture}
}
\caption{Graphical illustration of \Cref{example:FLB}: Projected available capacity of the two servers at time $t = 4$ before new job $j$ is assigned.}
\label{fig:example}
\vspace{-8mm}
\end{figure}

\revcolor{Having illustrated the potential benefits of incorporating estimated future utilization into an online policy in \Cref{example:FLB}, we now present our primary theoretical result. Specifically, while the classic \BALANCE\ algorithm fails to differentiate between high-reward, short-duration jobs and low-reward, long-duration jobs---thereby suffering a performance gap\footnote{\revcolor{Consider a single-server instance ($n=1$) with capacity $\inventory$, where $\maxduration=\maxreward\in\naturals$. Initially, $\inventory$ low-value jobs arrive in the time window $[0,1]$, each with long duration $\maxduration$ and reward $1$. Subsequently, for each $k\in[\maxduration]$, there are $\inventory$ high-value jobs arriving in the time window $[k,k+1]$, each with short duration $1$ and reward $\maxduration$. From the perspective of {\BALANCE}, all jobs have the same total reward, so it will accept at most $\inventory$ of the low-value jobs arriving in $[0,1]$. Once these jobs are assigned, the reduced rewards of all remaining jobs become negative, and the algorithm rejects them. Hence, the total reward of {\BALANCE} is at most $\inventory\cdot\maxduration$. In contrast, the optimal offline benchmark rejects the initial low-value jobs and instead accepts all $\maxduration\cdot \inventory$ high-value jobs, obtaining a total reward of $\inventory\cdot\maxduration\cdot\maxreward$. This yields a competitive ratio of $\Omega(\maxduration)$.}} of order $\maxduration$---we demonstrate that, under an appropriate parameterization, \FLB\ completely mitigates this vulnerability and achieves an asymptotically optimal competitive ratio.}


\begin{restatable}[Competitive Ratio]{theorem}{crgeneral}
\label{thm:FLB competitive ratio general}
There exists a choice of parameters $(\inspecfreq^*,\penscalarTwo^*,\penscalar^*)$ such that the asymptotic competitive ratio of {\FLB}, with inspection-frequency scalar $\inspecfreq^*$ and penalty parameters $(\penscalarTwo^*,\penscalar^*)$, is at most
\begin{align*}
    \ln(\maxreward\maxduration) + 3\ln\ln(\maxreward\vee\maxduration) + O(1).
\end{align*}
Moreover, there exist instances involving a single server with $\maxreward,\maxduration \geq 1$, for which the asymptotic competitive ratio of any online algorithm (possibly fractional or randomized) against the optimal offline benchmark is at least $\ln(\maxreward\maxduration) + \Omega(1)$. As a corollary, $\logcons(\FLB)=\logcons^*=1$.
\end{restatable}
We remark that there is an explicit construction for $(\inspecfreq^*,\penscalarTwo^*,\penscalar^*)$ in the above theorem (and a closed form for these parameters when $\mininventory$ approaches infinity in \Cref{apx:FLB parameter optimize}), which we detail later.

The proof of the upper bound result in the above theorem is deferred to \Cref{apx:FLB analysis real duration}, and the lower bound result is proved in \Cref{sec:lower bound construction}.  We also note that, although the theorem above is stated under the large-capacity assumption, our analysis additionally characterizes the non-asymptotic competitive ratio for any given capacity (see \Cref{prop:FLB competitive ratio integer duration,lem:FLB competitive ratio real duration}).

In the next section, we focus on proving a similar upper bound on the competitive ratio in a practical special case of our problem with integer-valued durations, which is slightly stronger than the upper bound in \Cref{thm:FLB competitive ratio general}. Although this analysis relies on similar techniques, it simplifies the exposition of the main ideas used in the proof of \Cref{thm:FLB competitive ratio general}. Later in \Cref{subsec:proof sketch real duration,apx:FLB analysis real duration}, we explain additional ideas needed in the upper bound analysis of {\FLB} in the general case with real-valued durations. 






\section{Competitive Ratio Analysis under Integer-valued Durations}
\label{sec:Heterogeneous weights}
\label{sec:competitive ratio analysis}
Consider a special case of our problem where, for every job $j\in[m]$ and compatible server $i\in N(j)$, the associated duration takes an integer value $\durationij\in\naturals$. We refer to this setting as the \emph{integer-duration environment}. This setting is practically motivated by scenarios in which the processing time for each job is a multiple of a minimum allowed processing interval,\footnote{This scenario is common in cloud marketplaces, where usage is typically sold or charged in fixed time increments.} and hence, after normalizing the minimum duration to one (without loss of generality), we can assume durations to be integers.\footnote{While durations $\durationij$ are assumed to be integer-valued, we still allow each job's arrival time $t_j$ and reward $\reward_{ij}$ to be real-valued.}

Besides its practical relevance, this special case simplifies exposition of the core technical ideas required for the general case. Specifically, in this integer-duration environment, {\FLB} can simply choose the inspection-frequency scalar $\inspecfreq = 1$ and still achieve the asymptotically optimal competitive ratio, which simplifies the analysis. This contrasts with the general setting of real-valued durations, in which the algorithm must carefully select the inspection-frequency scalar $\inspecfreq$.





\begin{theorem}[Competitive Ratio under Integer-valued Durations]
\label{thm:competitive ratio integer duration}
    In the integer-duration environment, there exists a choice of parameters $(\penscalarTwo^*,\penscalar^*)$ such that the asymptotic competitive ratio of {\FLB}, with inspection-frequency scalar $\inspecfreq^* = 1$ and penalty parameters $(\penscalarTwo^*,\penscalar^*)$\footnote{\label{footnote:choice-of-parameter}\revcolor{As we will show in \Cref{subsec:CR optimization} the choices are $\penscalarTwo = \frac{1}{\ln(\maxreward\vee\maxduration)}$ and $\penscalar = (\prod\nolimits_{k \in[\maxduration]}(1-\frac{\maxreward\ln(\maxreward\vee\maxduration)}{k(\maxreward\ln(\maxreward\vee\maxduration)+1
        )})  
        )^{-1}$}.}, is at most
    \begin{align*}
        \ln(\maxreward\maxduration) + \ln\ln(\maxreward\vee\maxduration) + O(1)
    \end{align*}
\end{theorem}
Similar to \Cref{thm:FLB competitive ratio general}, the improved competitive ratio presented in \Cref{thm:competitive ratio integer duration} is asymptotically optimal, as the competitive ratio lower bound established in \Cref{thm:FLB competitive ratio general} remains valid for the integer-duration environment. An explicit construction (along with asymptotic closed-form parameter choices) for $(\penscalarTwo^*,\penscalar^*)$ to achieve the upper bound stated above is detailed in \Cref{subsec:CR optimization}.


\xhdr{Overview of the analysis.} In the remainder of this section, we present the formal proof of \Cref{thm:competitive ratio integer duration}. At a high level, our analysis follows a three-step approach. Recall that {\FLB} is parameterized by penalty parameters $(\penscalar,\penscalarTwo)$ and the inspection-frequency scalar $\inspecfreq$. In the first step (\Cref{subsection: primal feasibility formal proof}), we characterize the set of parameter choices that guarantee the capacity feasibility of {\FLB}, i.e., ensuring that the algorithm never assigns a job to a server lacking available capacity.

\begin{restatable}[Capacity Feasibility of FLB under Integer-valued Durations]{proposition}{PrimalFeasibility}
\label{prop:FLB feasibility integer duration}
    In the integer-duration environment, {\FLB} with inspection-frequency scalar $\inspecfreq = 1$ is capacity feasible if penalty parameters $(\penscalarTwo,\penscalar)$ satisfy:\footnote{In fact, as we show in the proof of \Cref{prop:FLB feasibility integer duration}, this sufficient condition for the capacity feasibility becomes necessary as well when $\mininventory$ approaches infinity.}
    \begin{align*}
        \ln(\penscalar) \geq - \ln\left(\prod\nolimits_{k \in[\maxduration]}\left(1-\frac{\maxreward}{k(\maxreward+\penscalarTwo
        )}\right)  
        - 
        \frac{(\maxreward+\penscalarTwo)\ln(\penscalar)}{\maxreward\mininventory}
        \right)
    \end{align*}
\end{restatable}
In the second step (\Cref{subsec:configuration LP analysis}), we introduce a configuration LP, which upper bounds the optimum offline benchmark. We then use a primal-dual analysis based on this configuration LP to express the competitive ratio of the capacity feasible {\FLB} as a function of its parameters.
\begin{restatable}[Competitive Ratio of Capacity Feasible FLB under Integer-valued Durations]{proposition}{FLBCompetitiveRatio}
\label{prop:FLB competitive ratio integer duration}
    In the integer-duration environment, for every $\penscalarTwo > 0$ and $\penscalar \geq e$, the competitive ratio of a capacity feasible {\FLB} with inspection-frequency scalar $\inspecfreq = 1$ is at most 
    \begin{align*}
        \ln(\penscalar)\cdot \left(1 + \penscalarTwo\left(1 + \penscalar\left(\penscalar^{\frac{1}{\mininventory}}-1\right)\right)\right)
    \end{align*}
\end{restatable}
Combining the two previous steps, the third step of our analysis (\Cref{subsec:CR optimization}) formulates the problem of determining the best (asymptotic) competitive ratio of {\FLB} as a constrained optimization problem over parameters $(\penscalarTwo,\penscalar)$, while fixing $\inspecfreq = 1$. In \Cref{subsec:CR optimization}, we formally present this optimization problem and analytically solve it in the large capacity regime (i.e., $\mininventory\to\infty$). The solution for finite minimum capacity, $\mininventory < \infty$, is provided in \Cref{apx:finiteCmin}. \revcolor{Importantly, we demonstrate that our analysis extends to any finite capacity satisfying $\mininventory = \Omega(\maxduration\ln(\maxduration))$.}
These analytical solutions, along with a simple continuity argument, complete the proof of \Cref{thm:competitive ratio integer duration}. 

We highlight that the proof of \Cref{thm:FLB competitive ratio general} for general instances (with real-valued durations) follows the same structure as the proof of \Cref{thm:competitive ratio integer duration}. However, both the derivation of sufficient conditions for capacity feasibility and the primal-dual analysis used to establish the competitive ratio become more involved. \revcolor{Moreover, as is evident from comparing the statements of \Cref{thm:competitive ratio integer duration} and \Cref{thm:FLB competitive ratio general}, there is a performance gap between the two regimes. The performance gap between the integer-duration and real-duration regimes stems from a fundamental discontinuity in the adversary's power to block future arrivals. With integer-valued durations, job completions align exactly with discrete time boundaries; with real-valued durations, an assignment may spill over a temporal boundary by an arbitrarily small amount, enabling the adversary to block one additional wave of arrivals. To counter this fractional spillover, \FLB\ must monitor capacity at a finer granularity via an inspection frequency $\inspecfreq>1$ . While this improves tracking precision, it introduces an approximation error into the primal-dual charging argument. Raising the frequency above 1  increases the lower-order term $\ln\ln(\maxreward\vee \maxduration)$  in the competitive ratio; however, as $\maxduration$ grows large, the relative impact of this boundary fractionality diminishes, and the competitive gap between the two regimes vanishes.} See \Cref{apx:FLB analysis real duration} for details.

\subsection{{Sufficient Condition for the Capacity Feasibility}}
\label{subsection: primal feasibility formal proof}

The ultimate goal of this section is to prove \Cref{prop:FLB feasibility integer duration}. To this end, we show that the current (normalized) available capacity $\curinventory_{i,t\rightarrow t}$ remains non-negative throughout the execution of {\FLB} if the parameters $(\penscalarTwo,\penscalar)$ satisfy the condition in the statement of \Cref{prop:FLB feasibility integer duration}. 

As we illustrated in \Cref{example:FLB}, a good online algorithm in our problem (such as {\FLB}) should aim to evenly distribute resource utilization over time; in this way, it can avoid overcommitting to short jobs when the current utilization is high, thereby preserving the capacity for longer jobs that generate higher rewards. An algorithm can materialize this idea by maintaining a smooth trajectory for the {\estimated} available capacity $\curinventory_{i,t\rightarrow \tau}$ as a function of $\tau$. \revcolor{Specifically, we aim to regulate the temporal variations in the estimated available capacity, represented by the difference} $\curinventory_{i,t\rightarrow \tau+d}-\curinventory_{i,t\rightarrow \tau}$ for any $d\in\mathbb{N}$. We will formalize this property in the following lemma---which is maintained by {\FLB} as an \emph{invariant} throughout its execution. This invariant property is crucial for characterizing the admissible parameter space $(\penscalar,\penscalarTwo)$ that guarantees the capacity feasibility of \FLB. \revcolor{Throughout this section, we restrict our choice of pairs $(\penscalar,\penscalarTwo)$ to those satisfying
\begin{equation*}
    \prod\nolimits_{k \in[\maxduration]}\left(1-\frac{\maxreward}{k(\maxreward+\penscalarTwo)}\right) \geq \frac{(\maxreward+\penscalarTwo)\ln(\penscalar)}{\maxreward\mininventory},
\end{equation*}
a condition that can always be satisfied given a sufficiently large minimum inventory $\mininventory>0$.}

\begin{lemma}[Invariant for Projected Available Capacities]
\label{lem:inventory invariant integer duration}
Consider a hypothetical scenario in which {\FLB} can assign a job to a server that has no available capacity (which could result in a negative {\estimated} capacity level). In the integer-duration environment with inspection-frequency scalar $\inspecfreq = 1$, the {\estimated} available capacity under {\FLB} satisfies the following property: for every server $i\in[n]$, time points $t,\tpre\in\reals_+$ such that $\tpre - t \in \naturals_+$, and \revcolor{any} integer duration $\duration\in\naturals$,
    \begin{align*}
        \left(\curinventory_{i,t\rightarrow \tau + \duration} - 
        \curinventory_{i,t\rightarrow \tau}\right)
        \cdot \ln(\penscalar)
        \leq 
        - \ln\left(\prod\nolimits_{k \in[\duration]}\left(1-\frac{\maxreward}{k(\maxreward+\penscalarTwo
        )}\right)  
        - 
        \frac{(\maxreward+\penscalarTwo)\ln(\penscalar)}{\maxreward\mininventory}
        \right).
    \end{align*}
\end{lemma}
\begin{proofof}{Proof of \Cref{lem:inventory invariant integer duration}}
    We prove the lemma by an induction over the sequence of currently arrived jobs $[0:j]$. \revcolor{The high-level intuition is to bound the number of active jobs that overlap with time $\tau$ but do not extend to time $\tau+d$.}

\smallskip
    \noindent\emph{Base case ($j = 0$):} Before the first job arrives, the {\estimated} available capacity $\curinventory_{i,0\rightarrow \tpre}$ is $1$ for all servers $i\in[n]$ and all future time points $\tpre \geq 0$ by definition. Therefore, the inequality in lemma statement holds trivially as the left-hand side is zero and the right-hand side is non-negative (for large enough $\mininventory$). 
    
\smallskip
    \noindent\emph{Inductive step ($j \geq 1$):} Fix an arbitrary server $i\in[n]$ and suppose the inequality in the lemma statement holds for the arrival of first $j-1$ jobs. Now consider the arrival of job $j$ at time $t_j$ with duration $\durationij$ for server $i$.  We show the inequality holds after {\FLB} determines the assignment for job $j$. The only non-trivial case happens when $t = t_j \leq \tau < t + \durationij \leq t + d$ and {\FLB} assigns job $j$ to server $i$. In this case, the reduced reward of server $i$ is strictly positive,
   that is, $\rewardij\durationij - \sum_{\tpre'\in\timesetij(1)}\pen(\alpha_{i,t_j\rightarrow\tau'}) > 0$. Combining the facts that $\maxreward\geq \rewardij$ and $\timesetij(1) = \{t_j + k:k\in[0:\durationij - 1]\}$ in the integer-duration environment, we obtain
    \begin{align*}
        \frac{1}{\penscalarTwo}\maxreward \durationij
         >{}  &
         \frac{1}{\penscalarTwo}\sum\nolimits_{k' \in [0:\durationij   - 1]}
        \pen\left(\curinventory_{i,t_j\rightarrow t_j + k'}\right)
        \\
        = {}& \frac{1}{\penscalarTwo}\sum\nolimits_{k' \in [0:\tau -t_j-1]}
         \pen\left(\curinventory_{i,t_j\rightarrow t_j + k'}\right) +
         \frac{1}{\penscalarTwo}\sum\nolimits_{k' \in [0:\durationij   +   t_j - \tau- 1]}
         \pen\left(\curinventory_{i,t_j\rightarrow \tau + k'}\right)\\
         \geq {} &
         \left(\tau -t_j \right) \left(\penscalar^{(1-\curinventory_{i,t_j\rightarrow \tau})} -1 \right) +
         \sum\nolimits_{k' \in [0:\durationij   +   t_j - \tau- 1]}
         \left(\penscalar^{\left(1-\curinventory_{i,t_j\rightarrow \tau + k'}\right)}-1\right)\\
         \overset{(a)}{\geq} {} & -   \durationij + \left(\tau - t_j \right) \penscalar^{\left(1-\curinventory_{i,t_j\rightarrow \tau}\right)}
         + \sum\nolimits_{k' \in [0:\durationij + t_j - \tau- 1]}
         \penscalar^{1-\curinventory_{i,t_j\rightarrow \tau}+ \frac{\ln\left(\prod\nolimits_{k \in[k']}\left(1-\frac{\maxreward}{k(\maxreward+\penscalarTwo
        )}\right)  
        - \frac{(\maxreward+\penscalarTwo)\ln(\penscalar)}{\maxreward\mininventory}
        \right)}{\ln (\penscalar)}}\\
         \revcolor{\overset{(b)}{\geq} {}} &-   \durationij + \penscalar^{(1-\curinventory_{i,t_j\rightarrow \tau})} \sum\nolimits_{k' \in [0:\durationij   - 1]}
         \left(\prod\nolimits_{k \in[k']}\left(1-\frac{\maxreward}{k(\maxreward+\penscalarTwo
        )}\right)  - 
        \frac{(\maxreward+\penscalarTwo)\ln(\penscalar)}{\maxreward\mininventory}
        \right) \\
         \overset{(c)}{=}{} &-   \durationij + \penscalar^{(1-\curinventory_{i,t_j\rightarrow \tau})} 
         \left(\frac{\durationij}{\frac{\penscalarTwo}{\maxreward+ \penscalarTwo}}\prod\nolimits_{k \in[\durationij]}\left(1-\frac{\maxreward}{k(\maxreward+\penscalarTwo)}\right)  
         - \frac{\durationij(\maxreward+\penscalarTwo)\ln(\penscalar)}{\maxreward\mininventory}\right)
    \end{align*}
    where inequality~(a) holds due to the induction hypothesis and the fact that $\tpre - \revcolor{t_j} \leq \durationij$, and \revcolor{inequality~\revcolor{(b)} follows by lower-bounding the scalar $\tau - t_j$ via $\tau - t_j$ individual terms in the summation, each of which is bounded above by one. Finally,} equality~\revcolor{(c)} holds due to the following identity 
    (with proof provided in \Cref{apx: rho lemma}) \revcolor{with $d = \durationij, \ell = k'$ and $z = \frac{\maxreward}{\maxreward + \penscalarTwo}$}.
    \begin{restatable}[Identity]{lemma}{rholemma} \label{lem: rho lemma}
        For any $d\in\naturals$ and $z\geq 0$,
        $\sum_{\ell \in[0:d-1]} \prod_{k\in [\ell]} (1-\frac{z}{k}) = \frac{d}{1-z} \prod_{k\in [d]} (1-\frac{z}{k})$.
    \end{restatable}
Now, after moving $\durationij$ to the left-hand side and canceling $\left(\frac{\maxreward}{\penscalarTwo} + 1\right) \durationij$ from both sides, and then taking a logarithm from both sides and rearranging the terms, we have:
\begin{align*}
        -\ln \left(\prod\nolimits_{k \in[\durationij]}\left(1-\frac{\maxreward}{k(\maxreward+\penscalarTwo)}\right)  
         - \frac{\penscalarTwo\ln(\penscalar)}{\maxreward\mininventory}\right)\geq \left(1-\curinventory_{i,t_j\rightarrow \tau}\right) \cdot \ln (\penscalar)~.
    \end{align*}
and consequently,
    \begin{align*}
        &- \ln\left(\prod\nolimits_{k \in[\duration]}\left(1-\frac{\maxreward}{k(\maxreward+\penscalarTwo
        )}\right)  -\frac{(\maxreward+\penscalarTwo) \ln(\penscalar)}{\maxreward\mininventory}
        \right) 
        \overset{(a)}{\geq}
        -\ln \left(\prod\nolimits_{k \in[\durationij]}\left(1-\frac{\maxreward}{k(\maxreward+\penscalarTwo)}\right)  
         - \frac{\penscalarTwo\ln(\penscalar)}{\maxreward\mininventory}\right) + \frac{\ln(\penscalar)}{\mininventory}\\
         \overset{(b)}{\geq}& \left(1-\left(\curinventory_{i,t_j\rightarrow \tau}-\frac{1}{\mininventory}\right)\right) \cdot \ln (\penscalar)
         \overset{(c)}{\geq}{} 
         \left(\curinventory_{i,t\rightarrow \tau + \duration}-\left(\curinventory_{i,t\rightarrow \tau}-\frac{1}{\inventory_i}\right)\right) \cdot \ln (\penscalar)~.
    \end{align*}
    where inequality~(a) holds due to the concavity of the logarithm function and the case assumption that $\duration \geq \durationij$, inequality~(b) holds as we argued above, and inequality~(c) holds since $\curinventory_{i,t\rightarrow \tau + \duration} \leq 1$ and $\inventory_i \geq \mininventory$.
    Note that terms $\curinventory_{i,t\rightarrow \tau + \duration}$ and $\curinventory_{i,t\rightarrow \tau}-\frac{1}{\inventory_i}$ above are the updated {\estimated} available capacities after the assignment of job $j$ to server $i$ in {\FLB}. Therefore, we \revcolor{have shown} the induction hypothesis after {\FLB} determines the assignment for job $j$, which finishes the proof of the induction.
\end{proofof}

The above key lemma provides an upper bound on the difference $\curinventory_{i,t\rightarrow t+d}-\curinventory_{i,t\rightarrow t}$ in terms of the parameters $(\penscalar,\penscalarTwo)$ of {\FLB}, which is maintained throughout the execution of the algorithm. To understand how this upper bound relates to the capacity feasibility, consider the time $t_j$ at which job~$j$ arrives, and let $\curinventory_{i,t_j\rightarrow \tpre}$ denote the {\estimated} available capacity after {\FLB} finishes processing job $j$. Recall that $\curinventory_{i,t_j\rightarrow \tpre}$ is weakly increasing in $\tpre$ and reaches $1$ no later than $\tau=t_j+D$, since all current jobs will certainly be finished by time $t_j+D$. Combining this observation with \Cref{lem:inventory invariant integer duration}, we obtain the following corollary.
\vspace{-3mm}
\begin{corollary}[Capacity Feasibility Condition]
    \label{cor:inventory lower bound integer duration}
   Consider a hypothetical scenario in which {\FLB} can assign a job to a server that has no available capacity (which could result in a negative {\estimated} capacity level). In the integer-duration environment with inspection-frequency scalar $\inspecfreq = 1$, the current (normalized) available capacity under {\FLB} satisfies the following property: 
    \begin{align*}
        \forall i\in[n], j\in[m]:~~~\curinventory_{i,t_j\rightarrow t_j} \geq 1 + 
        \frac{\ln\left(\prod\nolimits_{k \in[\maxduration]}\left(1-\frac{\maxreward}{k(\maxreward+\penscalarTwo
        )}\right)  
        - 
        \frac{(\maxreward+\penscalarTwo)\ln(\penscalar)}{\maxreward\mininventory}
        \right)}{\ln(\penscalar)}
    \end{align*}
\end{corollary}
\yfedit{\begin{proofof}{Proof of \Cref{prop:FLB feasibility integer duration}}
The capacity feasibility is ensured if $\curinventory_{i,t_j\rightarrow t_j}$ remains non-negative for all jobs. Thus, a sufficient condition is that the lower bound on this quantity given in \Cref{cor:inventory lower bound integer duration} is non-negative, which is equivalent to the condition in \Cref{prop:FLB feasibility integer duration}.
\end{proofof}}
\vspace{-2mm}

\subsection{Primal-Dual Analysis of the Competitive Ratio}

\label{subsec:configuration LP analysis}
In this section, we aim to prove \Cref{prop:FLB competitive ratio integer duration}. To this end, we first introduce a new linear programming relaxation of the optimal offline benchmark, which we refer to as the \emph{configuration LP} (\ref{eq:opt lp}). This LP provides a more refined relaxation than the standard LP, as it (fractionally) assigns each unit of server capacity to a feasible schedule of jobs that can run on this unit. We then conduct a primal-dual analysis based on this configuration LP, resulting in a closed-form characterization of an upper bound on the competitive ratio of {\FLB} as a function of its parameters $(\penscalarTwo,\penscalar)$.

\xhdr{Configuration LP.} 
Any feasible assignment of jobs to servers can be represented by explicitly specifying a feasible \emph{configuration} for each capacity unit of each server $i\in[n]$. A feasible configuration for server $i$ is defined as a subset of non-overlapping jobs that can be assigned to run on a single capacity unit of this server. Formally, a subset $\config\subseteq[m]$ is a feasible configuration for server $i$ if $(i,j)\in \edges$ for every job $j\in \config$, and for every pair of distinct jobs $j,j' \in \config$, it holds that
\begin{align*}
    [t_{j'}, t_{j'} + \duration_{ij'}] \cap 
    [t_{j}, t_{j} + \durationij] = \emptyset.
\end{align*}
We denote by $\configspace_i\subseteq 2^{[m]}$ the set of all feasible configurations for server $i\in[n]$. We now introduce the configuration LP and its dual program as follows:
\begin{align}
\tag{${\mathcal{P}_{\textsc{OPT}}}$}
\label{eq:opt lp}
    \arraycolsep=5.4pt\def\arraystretch{1}
    &\begin{array}{lllllll}
    \underline{\textit{Primal}}:\qquad\qquad&\max\limits_{\boldsymbol\alloc\geq \mathbf 0} &
    \displaystyle\sum\nolimits_{i\in[n]}
    \displaystyle\sum\nolimits_{\config\in \configspace_i}
     \displaystyle\sum\nolimits_{j\in\config}
     \rewardij \durationij
     \cdot 
    \allocconfig  
    & \text{s.t.} &
    &
    & 
    \\
    & &
    \displaystyle\sum\nolimits_{i\in[n]}
    \displaystyle\sum\nolimits_{\config\in\configspace_i:j\in \config}
     \allocconfig \leq 1
     \quad
     & 
     j \in [m]
     &
     &
     &
     \\
     &
     &\displaystyle\sum\nolimits_{
     \config \in \configspace_i} 
     \allocconfig \leq \inventory_i
     \quad
     & 
     i\in[n]
     &
     &
     &
     \\
        \\
        \underline{\textit{Dual}}:\qquad \qquad& \min\limits_{\boldsymbol{\probdual,\inventorydual}\geq \mathbf 0} &
    \displaystyle\sum\nolimits_{j\in[m]}
    \onlinedual(j)
    +
    \displaystyle\sum\nolimits_{i\in[n]}
    \inventory_i
    \offlinedual(i)
    & \text{s.t.} 
    &
    &
    &\\
     &&
     \displaystyle\sum\nolimits_{j\in\config}
     \onlinedual(j)
     +
     \offlinedual(i)
     \geq 
     \displaystyle\sum\nolimits_{j\in\config}
     \rewardij \durationij
     &
     i\in[n],\config\in\configspace_i
    & 
     & 
     &
    \end{array}
\end{align}
Here, the decision variable $\allocconfig$ can be interpreted as the probability of assigning a (feasible) configuration $\config$ to a single unit of server $i$. The first set of constraints addresses feasibility from the job perspective: for each job $j\in[m]$, the total assignment probability across all feasible configurations containing job $j$, summed over all units of all servers, must not exceed 1. The second set of constraints addresses feasibility from the server perspective: for each server $i\in[n]$, the sum of assignment probabilities across all feasible configurations assigned to its units must not exceed its total capacity $\inventory_i$. The following lemma establishes that this linear program provides a relaxation for the optimal offline benchmark. We defer its formal proof to \Cref{apx:relaxation}.

\begin{restatable}{lemma}{lemrelax}
\label{lem:relaxation}
For any sequence of jobs with types $\{\rewardij,\durationij\}_{(i, j) \in \edges}$, the total reward of the optimal offline benchmark is upper bounded by the optimal objective value of the linear program \ref{eq:opt lp}.
\end{restatable}
We emphasize that prior related works~\citep[e.g.,][]{BN-09,GNR-14,FNS-19,HC-22} have employed a standard LP formulation for primal-dual analysis, in which decision variables represent the probability of assigning each individual job (rather than configurations) to each server. While this standard LP suffices for obtaining optimal competitive ratios in certain special cases (e.g., $\maxreward = \maxduration = 1$), it becomes technically challenging---and possibly even theoretically infeasible---to achieve optimal competitive ratios with precise logarithmic dependency on $\maxreward\maxduration$ (particularly, obtaining the exact constant before $\ln(\maxreward\maxduration)$) in the general setting~\citep[cf.,][]{HC-22}. To circumvent this difficulty, we propose and analyze the configuration LP \ref{eq:opt lp}, which allows us to obtain tight, optimal bounds on the competitive ratio.

\smallskip
\begin{proofof}{Proof of \Cref{prop:FLB competitive ratio integer duration}}
%
We upper bound the competitive ratio of {\FLB} (with inspection-frequency scalar $\inspecfreq = 1$) by constructing a dual solution based on its job assignments as follows. 

Initially, set $\onlinedual(j) \gets 0$ and $\offlinedual(i) \gets 0$ for each job $j\in[m]$ and server $i\in [n]$. Now consider all assignments made by {\FLB}. For each job $j\in[m]$, if {\FLB} assigns job $j$ to server $i$, update the dual variables as follows:
\begin{align*}
    \onlinedual(j) &\gets
    \rewardij\durationij
    -
    \sum\nolimits_{\tpre \in \timesetij}
 	\pen\left(
 	\estimateinventoryij
 	\right),\quad
    \offlinedual(i) \gets
    \offlinedual(i) 
    +
    \sum\nolimits_{\tpre \in \timesetij}
 	\left(
    \pen\left(
 	\estimateinventoryij - \frac{1}{\inventory_i}
 	\right)
  -
    \pen\left(
 	\estimateinventoryij
 	\right)
    \right),
\end{align*}
where $\timesetij$ is shorthand for the inspection-time subset $\timesetij(\inspecfreq)$ with $\inspecfreq = 1$.
Note that the above dual solution is well-defined under the assumption that {\FLB} is capacity feasible, meaning that the input argument to the penalty function $\pen$ is non-negative. This holds since when {\FLB} assigns job $j$ to server $i$, there is at least one available unit of server~$i$. Therefore, we have $\curinventory_{i, t_j \rightarrow t_j} \geq \frac{1}{\inventory_i}$. Additionally, for every $\tpre \in \timesetij$, we have $\estimateinventoryij \geq \curinventory_{i, t_j\rightarrow t_j} \geq \frac{1}{\inventory_i}$, as $\estimateinventoryij$ is increasing in $\tpre$ and $\tpre \geq t_j$.

The rest of the proof is done in two steps:

\smallskip
\noindent[\emph{Step i}]
\emph{Comparing objective values in primal and dual.}
Here, we show that the total reward of {\FLB} is a $\Gamma$-approximation to the objective value of the constructed dual solution, where $\Gamma = \ln(\penscalar)\cdot \left(1 + \penscalarTwo\left(1 + \penscalar\left(\penscalar^{\frac{1}{\mininventory}}-1\right)\right)\right)$, as stated in \Cref{prop:FLB competitive ratio integer duration}.
To establish this, we analyze the increments of the reward in {\FLB} and the objective value of the constructed dual solution resulting from each job assignment decision separately.

Suppose {\FLB} assigns job $j$, arriving at time $t_j$, to server $i$. The increment of the total reward in {\FLB} due to this assignment is $\Delta(\textsc{primal}) = \rewardij \durationij$, while the increment of the objective value of the constructed dual solution can be upper bounded as:
\begin{align*}
    \Delta(\textsc{Dual}) 
    ={} &
    \rewardij \durationij
    -
    \sum\nolimits_{\tpre \in \timesetij}
 	\pen\left(
 	{\estimateinventoryij}
 	\right)
  +
  \sum\nolimits_{\tpre \in \timesetij}
  \inventory_i
  \left(
    \pen\left(
    \estimateinventoryij - 
 	\frac{1}{\inventory_i}
 	\right)
  -
    \pen\left(
 	{\estimateinventoryij}
  \right)
  \right)
  \\
  \overset{(a)}{\leq}{}&
  \ln\left(\penscalar\right)
  \left(
  \rewardij \durationij
    -
    \sum\nolimits_{\tpre \in \timesetij}
 	\pen\left(
 	{\estimateinventoryij}
 	\right)
  \right)
  -
  \sum\nolimits_{\tpre \in \timesetij}
  \inventory_i
  \left(
    \pen\left(
 	{\estimateinventoryij}
  \right)
    -
    \pen\left(
    \estimateinventoryij - 
 	\frac{1}{\inventory_i}
 	\right)
  \right)
  \\
  ={} &
  \rewardij \durationij
  \ln\left(\penscalar\right)
  -
  \sum\nolimits_{\tpre \in \timesetij}
  \left(
  \ln\left(\penscalar\right)
  \pen\left(
 	{\estimateinventoryij}
 	\right)
  +
  \inventory_i
  \left(
    \pen\left(
 	{\estimateinventoryij}
  \right)
    -
    \pen\left(
    \estimateinventoryij - 
 	\frac{1}{\inventory_i}
 	\right)
  \right)
  \right)
  \\
  \overset{(b)}{\leq}{} &
  \rewardij \durationij
  \ln\left(\penscalar\right)
  -
  \sum\nolimits_{\tpre \in \timesetij}
  \left(
  \ln\left(\penscalar\right)
  \pen\left(
 	{\estimateinventoryij}
 	\right)
  +
    \pen'\left(
    \estimateinventoryij - 
 	\frac{1}{\inventory_i}
 	\right)
  \right)
  \\
  \overset{(c)}{\leq}{} &
  \rewardij \durationij
  \ln\left(\penscalar\right)
  +
  \sum\nolimits_{\tpre \in \timesetij}
 	\penscalarTwo 
      \ln\left(\penscalar\right) 
  \left(
  1 + 
  \penscalar
    \left(
    \penscalar^{\frac{1}{\mininventory}
    }
    - 1
    \right)
  \right)
  \\
  \overset{}{=}{} &
  \rewardij \durationij
  \ln\left(\penscalar\right)
  +
  |\timesetij|
 	\penscalarTwo 
      \ln\left(\penscalar\right) 
  \left(
  1 + 
  \penscalar
    \left(
    \penscalar^{\frac{1}{\mininventory}
    }
    - 1
    \right)
  \right)
  \\
  \overset{(d)}{\leq}{} &
  \rewardij \durationij
  \ln\left(\penscalar\right)
  +
  \rewardij\durationij
 	\penscalarTwo 
      \ln\left(\penscalar\right) 
  \left(
  1 + 
  \penscalar
    \left(
    \penscalar^{\frac{1}{\mininventory}
    }
    - 1
    \right)
  \right)
  \overset{}{=}
   \Gamma\cdot\rewardij\durationij
  =  \Gamma\cdot\Delta(\textsc{primal})~.
\end{align*}
In the above derivation, inequality~(a) holds since $\ln(\penscalar) \geq 1$ for all $\penscalar \geq e$, and the reduced reward of job $j$ for server~$i$, 
$\rewardij \durationij - \sum\nolimits_{\tpre \in \timesetij}\pen\left({\estimateinventoryij}\right),$ is positive by the assumption that the {\FLB} algorithm assigns job $j$ to server $i$. Moreover, inequality~(b) holds since the penalty function $\pen$ is convex, and thus $\inventory_i \left(
\pen\left({\estimateinventoryij}\right) -
\pen\left(\estimateinventoryij - \frac{1}{\inventory_i}\right)\right)
\geq \pen'\left(\estimateinventoryij - \frac{1}{\inventory_i}\right).$ Also, inequality~(c) follows from algebraic simplifications, as detailed below:
\begin{align*}
    &
    \ln\left(\penscalar\right) 
 	\pen\left(
 	{\estimateinventoryij}
 	\right)
    +
    \pen'\left(
    \estimateinventoryij -
    \frac{ 1}{\inventory_i}
    \right)
    \\
    =
    {} &
    \left(\ln\left(\penscalar\right) 
 	\pen\left(
 	{\estimateinventoryij}
 	\right)
    +
    \pen'\left(
    {\estimateinventoryij}{}
    \right)
    \right)
    +
    \left(
    \pen'\left(
    \estimateinventoryij -
    \frac{ 1}{\inventory_i}
    \right)
    -
    \pen'\left(
    {\estimateinventoryij}{}
    \right)
    \right)
    \\
    =
    {}&
 	-\penscalarTwo
      \ln\left(\penscalar\right)
  -
 	\penscalarTwo
  \ln\left(\penscalar\right) 
  \penscalar^{
  \left(1 - \frac{\estimateinventoryij}{\inventory_i}\right)}
    \left(
    \penscalar^{\frac{1}{\inventory_i}
    }
    - 1
    \right)
    \geq 
 	-\penscalarTwo
      \ln\left(\penscalar\right) 
  \left(
  1 + 
  \penscalar
    \left(
    \penscalar^{\frac{1}{\mininventory}
    }
    - 1
    \right)
  \right)
\end{align*}
Finally, inequality~(d) holds since 
$|\timesetij| = \durationij$ and $\rewardij \geq 1$.

\smallskip
\noindent[\emph{Step ii}]
\emph{Checking the feasibility of dual.}
Now we show that the constructed dual solution is feasible. First, note that clearly $\forall j\in[m]:\onlinedual(j)\geq 0$ (as the reduced reward of job $j$ on server $i$ is always positive if {\FLB} assigns job $j$ to server $i$) and $\forall i\in[n]:~\offlinedual_i\geq 0$. Now fix an arbitrary server $i\in[n]$ and a feasible configuration $\config\in\config_i$. By the assignment rule of {\FLB} and the construction of our dual solution, it is guaranteed that for every job $j\in\config$, the dual variable $\onlinedual(j)$ is at least as large as the reduced reward of job $j$ for server $i$, that is
$
\onlinedual(j) \geq \rewardij \durationij -
    \sum\nolimits_{\tpre \in \timesetij}
    \pen\left(\estimateinventoryij\right).
$
Therefore, the dual constraint associated with the primal variable $\allocconfig$ is satisfied if the following inequality holds:
\begin{align}
\label{eq:feasible}
    \offlinedual(i) 
    -
    \sum\nolimits_{j\in{\config}}
    \sum\nolimits_{\tpre\in\timesetij}
 	\pen\left(
 	{\estimateinventoryij}{}
 	\right)
    \geq 0~.
\end{align}
\revcolor{Note that the primary analytical complexity of the problem stems from the fact that arrivals occur in continuous time. If arrivals were restricted to discrete integer time steps, the offline dual objective $\offlinedual(i)$ could be straightforwardly expressed as the direct summation of marginal penalties across those discrete points. However, real-valued arrival times introduce temporal overlap, which complicates this accounting and necessitates a more sophisticated argument.} We prove \revcolor{\eqref{eq:feasible}} with a ``charging argument.'' At a high level, our goal is to identify values $\{\offlinedualdecompi\}_{j\in\config,\tpre \in \timesetij}$
such that 
\begin{align}
\label{eq:decomposed offline dual divisible}
  \forall j\in\config, \tpre\in\timesetij:~~ \offlinedualdecompi - \pen(\estimateinventoryij)  \geq 0
    \qquad 
    \text{and}
    \qquad
\sum\nolimits_{j\in\config}\sum\nolimits_{\tpre\in\timesetij}\offlinedualdecompi \leq \offlinedual(i)~,
\end{align}
which then implies inequality~\eqref{eq:feasible}. In the following, we provide details of this charging argument.

Let us introduce an auxiliary notation
$\timesetiS \triangleq
\{(j,\tpre): j\in \config, \tpre \in \timesetij\}$
for each server $i\in[n]$ and configuration $\config\in \configspace_i$.
Recall that for every configuration $\config\in\configspace_i$,
the processing-time intervals of any two distinct jobs $j_1, j_2\in\config$
do not overlap, and thus
$\timeset_{ij_1} \cap \timeset_{ij_2} = \emptyset$.
Consequently,
for every pair $(j_1, \tpre_1), (j_2,\tpre_2) \in \timesetiS$,
if $\tpre_1 \not = \tpre_2$, then $j_1 \not = j_2$.
We now identify a one-to-many  correspondence $\chargemapping: \timesetiS\to[m]\times [T]$ that satisfies the following three properties.
\begin{itemize}
    \item[-] {\cseparation:}
    For every pair of distinct elements $(j_1, \tpre_1), (j_2,\tpre_2)\in \timesetiS$,
    it holds that $\chargemapping(j_1,\tpre_1) \cap \chargemapping(j_2,\tpre_2) = \emptyset$.
    
    \item[-] {{\cfeasibilitya:}}
    For every $(j, \tpre)\in\timesetiS$, 
    we have $|\chargemapping(j, \tpre)| = \inventory_i - \inventory_i\estimateinventoryij$.
    Moreover, for any two distinct elements $(j_1,\tpre_1), (j_2, \tpre_2) \in \chargemapping(j, \tpre)$,
    it holds that $j_1 \neq j_2$.
    
    \item[-] {{\cfeasibilityb:}}
    For every $(j, \tpre)\in\timesetiS$ 
    and every $(j', \tpre') \in \chargemapping(j, \tpre)$, 
    we have $j' < j$,
    $\tpre' \leq \tpre$,
    $\tpre'\in\timeset_{ij'}$,
    and a unit of server $i$ is assigned to 
    job $j'$ with duration $\duration_{ij'} > \tpre - t_{j'}$.
\end{itemize}
\revcolor{Intuitively, the mapping $\sigma$ serves to synchronize the duration of different jobs to a common reference time frame.} Before demonstrating the existence of such a correspondence $\chargemapping$, we first illustrate how to construct 
values $\{\offlinedualdecompi\}_{j\in\config,\tpre \in \timesetij}$
that satisfy condition~\eqref{eq:decomposed offline dual divisible} given the correspondence $\chargemapping$. Specifically, for every $j\in\config$ and $\tpre\in\timesetij$, define
\begin{align*}
    \offlinedualdecompi \triangleq 
    \sum\nolimits_{(j', \tpre')\in\chargemapping(j, \tpre)} 
    \left(
    \pen\left(
    \curinventory_{i, t_{j'}\rightarrow\tpre'}
    - \frac{1}{\inventory_i}
    \right)
    -
    \pen\left(
    \curinventory_{i, t_{j'}\rightarrow\tpre'}
    \right)
    \right)
\end{align*}
Note that property~\cseparation\ and property~\cfeasibilityb\ of the correspondence $\chargemapping$ guarantee the second half of condition~\eqref{eq:decomposed offline dual divisible}, i.e., 
$\sum_{j\in\config}\sum_{\tpre\in\timesetij}\offlinedualdecompi \leq \offlinedual(i)$.
To establish the first half of condition~\eqref{eq:decomposed offline dual divisible}, i.e., 
$\offlinedualdecompi -\pen(\estimateinventoryij) \geq 0$, observe that
\begin{align*}
    \offlinedualdecompi - \pen(\estimateinventoryij)
    &=
    \sum\nolimits_{(j', \tpre')\in\chargemapping(j, \tpre)} 
    \left(
    \pen\left(
    \curinventory_{i, t_{j'}\rightarrow\tpre'}
    - \frac{1}{\inventory_i}
    \right)
    -
    \pen\left(
    \curinventory_{i, t_{j'}\rightarrow\tpre'}
    \right)
    \right)
    -
    \pen(\estimateinventoryij)
    \\
    &\overset{(a)}{\geq}
    \sum\nolimits_{(j', \tpre')\in\chargemapping(j, \tpre)} 
    \left(
    \pen\left(
    \curinventory_{i, t_{j'}\rightarrow\tpre}
    - \frac{1}{\inventory_i}
    \right)
    -
    \pen\left(
    \curinventory_{i, t_{j'}\rightarrow\tpre}
    \right)
    \right)
    -
    \pen(\estimateinventoryij)
    \\
    &\overset{(b)}{\geq}
    \sum\nolimits_{\ell \in [\inventory_i - \inventory_i\estimateinventoryij]}
    \left(
    \pen\left(
    \frac{\inventory_i - \ell}{\inventory_i}
    \right)
    -
    \pen\left(
    \frac{\inventory_i - \ell + 1}{\inventory_i}
    \right)
    \right)
    -
    \pen\left(\estimateinventoryij\right)
    \overset{(c)}{=}
    -\pen(1)
    \overset{(d)}{=}
    0.
\end{align*}
In the above derivation, inequality~(a) holds since the {\estimated} available capacity $\curinventory_{i,t_{j'}\rightarrow\tpre'}$ is weakly increasing in $\tpre'$, property~\cfeasibilityb\ of the correspondence $\chargemapping$ ensures $\tpre' \leq \tpre$, and the penalty function $\pen$ is convex. For inequality~(b), properties~\cfeasibilitya\ and~\cfeasibilityb\ of the correspondence $\chargemapping$ imply that there exist $s \triangleq \inventory_i - \inventory_i\estimateinventoryij$ jobs $j_1 < j_2 < \dots < j_s < j$, each of which is assigned to a different unit of server $i$ and occupies that unit at time $\tpre$. This further implies that, for each $\ell\in[s]$, $\curinventory_{i,t_{j_\ell}\rightarrow\tpre} \leq \frac{\inventory_i - \ell + 1}{\inventory_i}$. Applying the convexity of the penalty function $\pen$ completes the argument for inequality~(b). Finally, equality~(c) holds by simplifying algebra (noticing the telescopic summation), and equality~(d) follows since $\pen(1) = 0$.

Now we show the existence of the correspondence $\chargemapping$ satisfying properties~\cseparation, \cfeasibilitya, and \cfeasibilityb\ through the following explicit construction. Fix an arbitrary $(j,\tpre)\in\timesetiS$. As mentioned earlier, by definition of the {\estimated} available capacity $\estimateinventoryij$, there exist $s \triangleq \inventory_i - \inventory_i\estimateinventoryij$ jobs $j_1 < j_2 <  \dots <  j_s < j$, each of which is assigned to a distinct unit of server $i$ and occupies this unit at time $\tpre$. In other words, for every $\ell\in[s]$, we have $t_{j_\ell} + \duration_{ij_\ell} \geq \tpre$. Moreover, there exists a unique $\tpre_\ell\in\timeset_{ij_\ell}$ such that $\tpre_\ell \leq \tpre < \tpre_\ell + 1$. By defining 
\[
\chargemapping(j, \tpre) \triangleq 
\{(j_\ell, \tpre_\ell)\}_{\ell\in[s]},
\] 
properties~\cfeasibilitya\ and~\cfeasibilityb\ are straightforwardly satisfied.

To show property~\cseparation, fix two distinct elements $(j', \tpre'), (j'', \tpre'') \in \timesetiS$ such that $j' \leq j''$. It suffices to show that $|\tpre'' - \tpre'| \geq 1$. We consider two cases separately. If $j' = j''$, then for $\tpre' \neq \tpre''$, the construction of inspection-time subset $\timeset_{ij'}$ directly implies that $|\tpre'' - \tpre'| \geq 1$, as desired. If $j' < j''$, then since both $j', j''\in \config$, the inspection-time subsets $\timeset_{i, j'}$ and $\timeset_{i, j''}$ do not overlap, meaning that $t_{j'} + \duration_{ij'} \leq t_{j''}$. Thus, we have
$
\tpre' \leq \max\timeset_{i j'} \leq t_{j'} + \duration_{ij'} \leq t_{j''} = \min\timeset_{i j''} \leq \tpre''.
$
Moreover, since all durations are integer-valued, we have
$\max \timeset_{ij'} = t_{j'} + \duration_{ij'} - 1$. Hence, 
$
|\tpre'' - \tpre'| = \tpre'' - \tpre' \geq t_{j''} - \max\timeset_{ij'} \geq 1,
$
as desired.

Finally, since the total reward of {\FLB} is a $\Gamma$-approximation of the objective value of the constructed dual solution (Step i), and
the constructed dual solution is feasible (Step ii),
invoking the weak duality of the linear program concludes the proof.
\end{proofof}

\subsection{Putting All the Pieces Together: Proof of Theorem~\ref{thm:competitive ratio integer duration}}
\label{subsec:CR optimization}
\Cref{prop:FLB feasibility integer duration} and \Cref{prop:FLB competitive ratio integer duration} allow us to formulate the best competitive ratio upper bound obtained from the primal-dual analysis of {\FLB} (with inspection-frequency scalar $\inspecfreq = 1$) as the following optimization problem over the penalty parameters $(\penscalarTwo,\penscalar)$, denoted by \ref{eq:FLB parameter integer duration}:
\begin{align}
\tag{$\FLBOPTInt{\maxreward,\maxduration,\mininventory}$}
\label{eq:FLB parameter integer duration}
    &\begin{array}{lll}
    \min\limits_{\penscalarTwo,\penscalar}\quad\quad &
    \ln(\penscalar)\cdot \left(1 + \penscalarTwo\left(1 + \penscalar\left(\penscalar^{\frac{1}{\mininventory}}-1\right)\right)\right)
    &
    \text{s.t.}
    \vspace{5pt}
    \\
    &
    \ln(\penscalar) \geq - \ln\left(\prod\nolimits_{k \in[\maxduration]}\left(1-\frac{\maxreward}{k(\maxreward+\penscalarTwo
    )}\right)  
    - 
    \frac{(\maxreward+\penscalarTwo)\ln(\penscalar)}{\maxreward\mininventory}
    \right)
    &
    \vspace{5pt}
    \\
    &
    \penscalarTwo > 0,\penscalar\geq e~
    &
    \end{array}
\end{align}
Here, the first constraint follows from the capacity feasibility condition stated in \Cref{prop:FLB feasibility integer duration}, while the objective and the second constraint follow from \Cref{prop:FLB competitive ratio integer duration}.

In practice, the decision-maker (e.g., a cloud computing platform) can select the parameters of {\FLB} based on the platform's maximum reward $\maxreward$, maximum duration $\maxduration$, and minimum capacity~$\mininventory$, by numerically solving~\ref{eq:FLB parameter integer duration} in the integer-duration environment. 
To prove the asymptotically optimal competitive ratio stated in \Cref{thm:competitive ratio integer duration}, we provide the following simple analytical analysis of \ref{eq:FLB parameter integer duration} when $\mininventory$ goes to $\infty$. The analysis for $\mininventory < \infty$ can be found in \Cref{apx:finiteCmin}.

\begin{proofof}{Proof of \Cref{thm:competitive ratio integer duration}}
We focus on the large capacity regime case where  $\mininventory\to\infty$ in \ref{eq:FLB parameter integer duration}.\footnote{Due to continuity, examining the case where $\mininventory = \infty$ yields the asymptotic competitive ratio as $\mininventory \rightarrow \infty$. For further details, refer to \cref{apx:finiteCmin}.} We consider two cases based on the magnitude of $\maxreward \vee\maxduration$.

\noindent[\emph{Case i}]
Suppose $\ln(\maxreward\vee\maxduration) \geq (e - 1)$.
In this case, we assign $\penscalarTwo = \frac{1}{\ln(\maxreward\vee\maxduration)}$. We set $\penscalar$ such that the first constraint in program~$\FLBOPTInt{\maxreward,\maxduration,\infty}$ binds, i.e., 
\begin{align*}
        \ln(\penscalar) &= - \ln\left(\prod\nolimits_{k \in[\maxduration]}\left(1-\frac{\maxreward\ln(\maxreward\vee\maxduration)}{k(\maxreward\ln(\maxreward\vee\maxduration)+1
        )}\right)  
        \right)
\end{align*}
It can be verified that constraint $\penscalar\geq e$~from ~$\FLBOPTInt{\maxreward,\maxduration,\infty}$ is satisfied.
Therefore, the competitive ratio of {\FLB} (with the aforementioned parameter values) is at most 
\begin{align*}
        ~~~&-\left(1 + \frac{1}{\ln(\maxreward\vee\maxduration)}\right)\ln\left(\prod\nolimits_{k \in[\maxduration]}\left(1-\frac{\maxreward\ln(\maxreward\vee\maxduration)}{k(\maxreward\ln(\maxreward\vee\maxduration)+1
        )}\right)  
        \right)
        \\
        =&- \left(1 + \frac{1}{\ln(\maxreward\vee\maxduration)}\right)\sum\nolimits_{k \in[\maxduration]}\ln\left(1-\frac{\maxreward\ln(\maxreward\vee\maxduration)}{k(\maxreward\ln(\maxreward\vee\maxduration)+1
        )}\right)  
        %
        = 
        \ln(\maxreward\maxduration) + \ln\ln(\maxreward\vee\maxduration) + O(1)
    \end{align*}
    where the last equality holds since
    \begin{align*}
        - \ln\left(1-\frac{\maxreward\ln(\maxreward\vee\maxduration)}{\maxreward\ln(\maxreward\vee\maxduration)+1
        }\right) = \ln(\maxreward) + \ln\ln(\maxreward\vee\maxduration) + O(1)
        \;\;
        \intertext{and}
        \;\;
        - \sum\nolimits_{k\in[2:\maxduration]}\ln\left(1-\frac{\maxreward\ln(\maxreward\vee\maxduration)}{k\maxreward\ln(\maxreward\vee\maxduration)+1
        }\right)
        \leq
        \sum\nolimits_{k\in[2:\maxduration]}\ln\left(\frac{k}{k-1}\right)
        =
        \ln(\maxduration)
    \end{align*}

\noindent[\emph{Case ii}]
    Suppose $\ln(\maxreward\vee\maxduration) \leq e - 1$. In this case, we assign $\penscalarTwo = \frac{\maxreward}{e - 1}$ and $\penscalar = e$. It can be verified that the first constraint in $\FLBOPTInt{\maxreward,\maxduration,\infty}$ is satisfied. Therefore the competitive ratio of {\FLB} (with the aforementioned parameter values) is at most $1+\frac{\maxreward}{e - 1} = \Theta(1)$.
\end{proofof}

\xhdr{Improved competitive ratios for special cases.} 
\yfedit{We can also derive a better  bound on the competitive ratio for instances with heterogeneous durations but homogeneous rewards (i.e., $\maxduration \geq 1$ but $\maxreward = 1$). This special case is motivated by practical considerations and, from a theoretical standpoint, it provides insights into the impact of duration heterogeneity in our model.\footnote{The impact of the reward heterogeneity for online resource allocation has been studied in the literature \citep[e.g.,][]{BQ-09,FKMMP-09,MS-20,EFN-23} for non-reusable resources.}
A similar result for (real-valued) durations is also given in \Cref{prop:CR identical reward real duration}.
}



\begin{proposition}[Competitive Ratio for Integer-valued Durations and Homogeneous Rewards]
\label{prop:CR identical reward integer duration}
\yfedit{For instances with homogeneous rewards (i.e., $\maxreward = 1$) in the integer-duration environment, 
there exists a choice of parameters $(\penscalarTwo^*,\penscalar^*)$ such that the asymptotic competitive ratio of {\FLB}, with inspection-frequency scalar $\inspecfreq^* = 1$ and penalty parameters $(\penscalarTwo^*,\penscalar^*)$, is at most
    $H({\maxduration}) + 2$, where $H(\maxduration) \triangleq \sum_{i \in [D]}\frac{1}{i}$ is the $\maxduration$-th harmonic number.
    Moreover, there exist instances involving a single server with $\maxreward=1,\maxduration \geq 1$, for which the asymptotic competitive ratio of any online algorithm (possibly fractional or randomized) against the optimal offline benchmark is at least $\sumofreciprocals({\maxduration})$. 
    }
    \yfdelete{
under large capacity, there exists $(\inspecfreq,\penscalarTwo,\penscalar)$ such that the competitive ratio of the {\FLB} with $(\inspecfreq,\penscalarTwo,\penscalar)$ is at most 
    $H({\maxduration}) + 2$, where $H(\maxduration) \triangleq \sum_{i \in [D]}\frac{1}{i}$ is the $\maxduration$-th harmonic number.
Moreover, the competitive ratio of any online algorithm (possibly fractional or randomized) is at least $\sumofreciprocals({\maxduration})$.}
\end{proposition}

\begin{proofof}{Proof of \Cref{prop:CR identical reward integer duration}} We first show the competitive ratio upper bound. Consider instances where all compatible job server pairs $(i, j)$ have the same per-period reward, normalized to $\rewardij = 1$. In the large capacity regime, program~$\FLBOPTInt{1,\maxduration,\infty}$ is simplified as 
\yfdelete{\begin{align*}
\begin{array}{lll}
    \min\nolimits_{\penscalarTwo,\penscalar}~ &
    \ln(\penscalar)\cdot \left(1 + \penscalarTwo\right)
    &
    \text{s.t.}
    \vspace{5pt}
    \\
    &
    \ln(\penscalar) \geq - \ln\left(\prod\nolimits_{k \in[\maxduration]}\left(1-\frac{1}{k(1+\penscalarTwo
    )}\right)
    \right)
    &
    \vspace{5pt}
    \\
    &
    \penscalarTwo > 0,\penscalar\geq e
    &
    \end{array}
\end{align*}}
\yfedit{
\begin{align*}
    \min\nolimits_{\penscalarTwo\geq 0,\penscalar\geq e}~
    \ln(\penscalar)\cdot \left(1 + \penscalarTwo\right)
    \quad
    \text{s.t.}
    \quad 
    \ln(\penscalar) \geq - \ln\left(\prod\nolimits_{k \in[\maxduration]}\left(1-\frac{1}{k(1+\penscalarTwo
    )}\right)
    \right)
\end{align*}}
Is is evident that the first inequality binds in the optimal solution, and therefore, the competitive ratio of \FLB\ can be expressed as 
    $- \ln\left(\prod\nolimits_{k \in[\maxduration]}\left(1-\frac{1}{k(1+\penscalarTwo
    )}\right)
    \right)
    \cdot 
    (1 + \penscalarTwo)$ as a function of  $\penscalarTwo$, which is decreasing in $\penscalarTwo$. Combining with the constraint that $\penscalar\geq e$, we obtain an upper bound for the competitive ratio of the {\FLB} is as
    $1 + \penscalarTwo^{(\maxduration)}$
where $\penscalarTwo^{(\maxduration)}$ is the solution of 
\begin{align*}
    \prod\nolimits_{k \in[\maxduration]}\left(1-\frac{1}{k(1+\penscalarTwo
    )}\right) = \frac{1}{e}
\end{align*}
In the special case of $\maxduration = 1$, we obtain $\penscalarTwo^{(1)} = \frac{1}{e - 1}$, $\penscalar = e$ and {\FLB} recovers the optimal competitive ratio of $\frac{e}{e - 1}$ \citep{FNS-19,GIU-20}. For $\maxduration \geq 2$, we can upper bound $\penscalarTwo^{(\maxduration)}$ with $H({\maxduration})$ + 2 and thus obtain a competitive ratio of $H(\maxduration) + 2$ with \Cref{lem: bounding theta} (see its proof in \Cref{apx: bounding theta}).

Next, we construct an adversarial instance involving a single server with capacity $c$, for which no online algorithm achieves a competitive ratio smaller than $H(\maxduration)$. Let $M > c$. Consider an arrival pattern where jobs arrive in $\maxduration$ batches, each of size $M$, at times $0, 0+\epsilon, 0+2\epsilon,\ldots$, for an infinitesimally small $\epsilon$. All jobs in batch $d\in[\maxduration]$ have duration $d$. We assume these batches arrive sequentially in the order $1,2,3,\ldots$ (thus, all jobs in batch $i>j$ arrive after jobs in batch $j$), and the index of the last batch is chosen adversarially. For any given online algorithm $\mathcal{A}$, let $x_{i}$ denote the (fractional) number of jobs from batch $i$ accepted by $\mathcal{A}$. Note that if the last batch is $d$, the optimal offline benchmark assigns all $c$ units exclusively to jobs in batch $d$ (and none to jobs from batches $i\in[d-1]$). Therefore, if $\mathcal{A}$ is $(H(\maxduration)-\delta)$-competitive for some $\delta > 0$, it follows that for all durations $\duration \in [\maxduration]$,
\begin{align*}
    \sum\nolimits_{i\in[\duration]} ix_i > \frac{c \cdot \duration}{H(\maxduration)}.
\end{align*}
Multiplying this inequality by $(\frac{1}{\duration}-\frac{1}{\duration+1})$ for $\duration < \maxduration$, and by $\frac{1}{\maxduration}$ for $\duration=\maxduration$, and summing all terms, we obtain $\sum\nolimits_{i\in[\maxduration]} x_i > c$. This inequality contradicts the capacity feasibility of $\mathcal{A}$ when all batches arrive.
\end{proofof}


\vspace{-5pt}

\begin{restatable}{lemma}{boundingtheta}
\label{lem: bounding theta}
Let $\lambda_{\maxduration}$ be the solution to the equation $e^{-1} = \prod_{k \in [\maxduration]} \left(1-\frac{\lambda_{\maxduration}}{k}\right).$ Then $\frac{1}{\lambda_{\maxduration}} \leq H(\maxduration) + 2.$
\end{restatable}

\begin{remark}[Competitive Ratio for Homogeneous Rewards and Durations]
\label{rmk:HomogenousRewardsDurations}
In the special case of $\maxreward=\maxduration = 1$, the competitive ratio of the {\FLB} with $\inspecfreq=1,\penscalarTwo=\frac{1}{e-1},\penscalar = e$ is $\frac{e}{e-1}$. Moreover, this ratio is  optimal among all (possibly fractional and randomized) online algorithms \citep{FNS-19,GIU-20}.
\end{remark}

\subsection{Proof Sketch of Theorem~\ref{thm:FLB competitive ratio general} for Real-valued Durations}
\label{subsec:proof sketch real duration}
In this section, we explain how to extend our previous analysis---which assumed integer-valued durations---to our main setting with real-valued durations. The formal analysis of \Cref{thm:FLB competitive ratio general} can be found in \Cref{apx:FLB analysis real duration}. The high-level approach resembles the one used for integer-valued durations: we first identify a sufficient condition for the capacity feasibility in \Cref{lem:FLB feasibility real duration}. Next we characterize its competitive ratio in \Cref{lem:FLB competitive ratio real duration}. Finally, we formulate the problem of identifying the asymptotically optimal competitive ratio as an optimization program~\ref{eq:FLB parameter real duration} induced by the propositions in the first two steps. One key difference is that the inspection frequency scalar $\inspecfreq$ is no longer fixed at one. In fact, the asymptotically optimal competitive ratio is achieved when $\inspecfreq \geq 2$.

\begin{restatable}[Capacity Feasibility of {\FLB}]{proposition}{PrimalFeasibilityReal}
\label{lem:FLB feasibility real duration}
    {\FLB} is capacity feasible if integer-valued inspection-frequency scalar $\inspecfreq\in\naturals$ and penalty parameters $(\penscalarTwo,\penscalar)\in\reals_+^2$ satisfy 
    \begin{align*}
        \ln(\penscalar) \geq &- \ln\left(\prod\nolimits_{k \in[\lceil\inspecfreq\maxduration\rceil]}\left(1-\frac{\maxreward}{k(\maxreward+\inspecfreq\penscalarTwo
        )}\right)  
        \right)
        -
        \ln\left(
        1+\left(\inspecfreq+\frac{\maxreward}{\penscalarTwo} \right)\left(1 - 
        \frac{(\penscalarTwo \inspecfreq + \maxreward)\ln \left(\penscalar \right)}{\maxreward\mininventory}
        \right) - \frac{\maxreward}{\penscalarTwo} \left(1+\frac{\penscalarTwo}{\maxreward+\inspecfreq\penscalarTwo}\right)^\inspecfreq
        \right)
        \\
        &~~~~+
        \ln\left(\frac{(\inspecfreq+1)(\maxreward+\inspecfreq\penscalarTwo)}{\inspecfreq\penscalarTwo}\right)
        +
        \ln\left(\prod\nolimits_{k \in[\inspecfreq+1]}\left(1-\frac{\maxreward}{k(\maxreward+\inspecfreq\penscalarTwo
        )}\right)  
        \right)
    \end{align*}
\end{restatable}
The proof of \Cref{lem:FLB feasibility real duration} extends the induction argument used in \Cref{prop:FLB feasibility integer duration}. We maintain a similar invariant for projected available capacities, as formalized in \Cref{lem:inventory invariant real duration}.
The key distinction arises from considering general $\inspecfreq \in \naturals$ in \FLB. This requires analyzing the projected available capacities difference $\curinventory_{i,t\rightarrow \tau+d} - \curinventory_{i,t\rightarrow \tau}$ for $\duration \in \reals_+$, which yields different closed-form expressions for $\duration \leq 1$ and $\duration \geq 1$. In contrast, the integer-valued duration case with $\inspecfreq = 1$ only needs to consider $\duration \in \naturals$, eliminating the need for $\duration \leq 1$ case.

\begin{restatable}[Competitive Ratio of Capacity Feasible {\FLB}]{proposition}{FLBCompetitiveRatioReal}
\label{lem:FLB competitive ratio real duration}
\yfedit{For every $\inspecfreq\geq 2$, $\penscalarTwo > 0$ and $\penscalar \geq e$, the competitive ratio of a capacity feasible {\FLB} is at most }
\yfdelete{    For every $\inspecfreq\geq 2$,  $\penscalarTwo > 0$ and $\penscalar \geq e$, the competitive ratio of \FLB is at most }
    \begin{align*}
        \frac{\inspecfreq}{\inspecfreq - 1}\cdot \ln(\penscalar)\cdot \left(1 + \inspecfreq\penscalarTwo\left(1 + \penscalar\left(\penscalar^{\frac{1}{\mininventory}}-1\right)\right)\right)
    \end{align*}
\end{restatable}
The proof of \Cref{lem:FLB competitive ratio real duration} follows a similar primal-dual analysis with the configuration LP~\ref{eq:opt lp}. The dual solution construction mirrors the one in the proof of \Cref{prop:FLB feasibility integer duration}. The argument which compares the objective values in the primal and dual is also similar. The key difference lies in the feasibility of the dual constraint. Unlike for the integer-valued durations, the same dual solution construction only guarantees \emph{approximate} dual feasibility, with a multiplicative factor of $\inspecfreq/(\inspecfreq - 1)$.\footnote{This factor is unavoidable in our dual construction, and thus we cannot use {\FLB} with inspection frequency scalar $\inspecfreq = 1$ for real-valued durations.} To establish this approximate feasibility, we develop a more involved charging argument through an extended construction.

\section{Tightness of the Competitive Ratio}
\label{sec:lower bound construction}
In this section, we demonstrate that the competitive ratio of {\FLB} is asymptotically optimal among all online algorithms, including those that allow randomization or fractional assignments. We construct an adversarial problem instance (illustrated in \Cref{fig:badExample}) with a single server. Jobs arrive sequentially with the shortest jobs with the lowest rewards arriving first. As time progresses, both job duration and reward steadily increase until arrivals abruptly stop at an adversarial time. Intuitively speaking, since the algorithm does not know when the arrivals will end, it must hedge against such an abrupt end of arrival by accepting a minimum number of arriving jobs to maintain the desirable competitive ratio. This necessity limits its ability to accept the (ex-post) most valuable jobs.

We now formalize this argument by restating and proving the lower bound result in \Cref{thm:FLB competitive ratio general}.

\begin{figure}
    \centering
      \centering
\scalebox{0.8}{
    \begin{tikzpicture}
    \draw[->, thick] (0, 0) -- (7, 0) node[below] {$t$};
    \draw[->, thick] (0, 0) -- (0, 5) node[above] {$r$};
    
    \draw[thick] (0, -0.2) -- (0, 0);
    \node[below] at (0, -0.2) {0};
    
    \draw[thick] (0.5, -0.2) -- (0.5, 0);
    \node[below] at (0.5, -0.2) {1};

    \draw[thick] (5, -0.2) -- (5, 0);
    \node[below] at (5, -0.2) {$D$};

    \foreach \i in {0,1,2,3, 6,7,8} {
    \pgfmathsetmacro{\scale}{pow(1.34, \i)}
    \pgfmathsetmacro{\xoffset}{0.05 * \i}
    \pgfmathsetmacro{\ybottom}{0.5 * \i}
    \pgfmathsetmacro{\ytop}{0.5 * (\i + 1)}
    \pgfmathsetmacro{\width}{0.5 * \scale}
    
   \pgfmathsetmacro{\shade}{250- \i * 15} 

    \definecolor{mygreen\i}{RGB}{0,\shade,0} 

    \fill[mygreen\i] (\xoffset, \ybottom) rectangle (\xoffset + \width, \ytop);
    \draw[thick] (\xoffset, \ybottom) rectangle (\xoffset + \width, \ytop);
    \ifnum\i=8
        \node at (\xoffset + \width/2, \ybottom + 0.25) {\large \( r = R \)};
    \fi
    }
    \draw[dashed] (0.5, -0.2) -- (0.5, 5);
    
    
    \end{tikzpicture}
    }
    \caption{The graphical illustration of the worst-case instance in \Cref{prop:optimal the competitive ratio divisible lower bound}. The darker green means a job with higher reward.}
    \label{fig:badExample}
\vspace{-8mm}
\end{figure}
\begin{restatable}[Restating Negative Result in \Cref{thm:FLB competitive ratio general}]{proposition}{CRdivisibleLB}
\label{prop:optimal the competitive ratio divisible lower bound}
    For instances involving a single server with $\maxreward \geq 1$ and $\maxduration \geq 1$, the competitive ratio of any online algorithm (possibly fractional or randomized) against the optimal offline benchmark is at least $\ln(\maxreward\maxduration) + \Omega(1)$.
\end{restatable}
\begin{proofof}{Proof of \Cref{prop:optimal the competitive ratio divisible lower bound}}
By Yao's lemma \citep{yao-77}, it suffices to construct a distribution over problem instances and argue no deterministic online algorithm achieves a competitive ratio better than $\ln(\heterosize)$.

Let $M\in \naturals$ be a sufficiently large integer. We construct a distribution over $M$ instances $\{I_k\}_{k\in[M]}$ as follows. In all $M$ instances, there are $m \triangleq M$ jobs and a single server with initial capacity $\inventory \triangleq 1$.\footnote{By duplicating jobs, our proof can also be extended to prove the same negative result for large capacities.} Jobs of instance $I_k$ are constructed as follows: every job $j\in [k]$ has reward $\reward_j\ked = \maxreward^{j/M}$ and duration $\duration_j\ked = \lfloor \maxduration^{j/M}\rfloor$, arriving at time $j/M$. Each job $j\in [k + 1: M]$ arrives at time $j/M$ but is incompatible with the server.

We set the probability $p_M$ of instance $I_M$ as $p_M \triangleq \frac{\reward_1^{(1)}\duration_1^{(1)}}{\reward_M^{(M)}\duration_M^{(M)}}$,
and the probability $p_k$ of instance $I_k$ as $p_k \triangleq 
\frac{\reward_1^{(1)}\duration_1^{(1)}}{\reward_k^{(k)}\duration_k^{(k)}}
-
\frac{\reward_1^{(1)}\duration_1^{(1)}}{\reward_{k+1}^{(k+1)}\duration_{k+1}^{(k+1)}}$ 
for every $k\in [M - 1]$. Under our construction, the expected total reward of both the optimal offline benchmark and the configuration LP benchmark~\ref{eq:opt lp} is $\sum_{k\in[M]} p_k\reward_k\ked\duration_k\ked
= M\left(1-\frac{1}{(\heterosize)^{1/M}}\right) + \Omega(1),$
where the $\Omega(1)$ term arises due to rounding in the duration construction. On the other hand, for an arbitrary deterministic online algorithm $\ALG$, it cannot distinguish among instances $\{I_k\}_{k=j}^M$ upon the arrival of job~$j$. Thus, the expected total reward of $\ALG$ can be upper bounded by
\begin{align*}
    \max_{\mathbf y \geq \mathbf 0}~
    \sum\nolimits_{k \in [M]}
    \left(\sum\nolimits_{\tpre \in [k: M]}
    p_\tpre
    \right)
    \reward_k\ked \duration_k\ked
    y(k)
    \quad
    \text{s.t.}
    \quad 
    \sum\nolimits_{k\in[M]}y(k) \leq 1,
\end{align*}
where variable $y(k)$ denotes the fraction of the server assigned to job $k$ by $\ALG$. Solving this simple LP, the total reward of $\ALG$ is at most 
$\max_{k\in[M]}  \left(\sum\nolimits_{\tpre \in [k: M]} p_\tpre \right)\reward_k\ked\duration_k\ked = 1.$ Therefore, the optimal competitive ratio is at least $M\left(1-\frac{1}{(\heterosize)^{1/M}}\right) + \Omega(1).$ By letting $M$ approach infinity, we obtain $\lim_{M \rightarrow \infty} M\left(1-\frac{1}{(\heterosize)^{1/M}}\right) + \Omega(1) = \ln(\heterosize) + \Omega(1),$ as desired.
\end{proofof}


\section{Numerical Experiments}

\label{sec:numerical}

In this section, we consider two types of instances for our numerical experiments. First, in \Cref{sec:worstCase}, we evaluate the performance of our algorithm on the worst-case instance presented in \Cref{sec:lower bound construction} \revcolor{along with {\GRD} and {\BALANCE}  benchmarks.} There, we numerically \revcolor{demonstrate} that (i) the benchmarks perform arbitrarily poorly on this instance, and (ii) {\FLB} achieves the worst-case theoretical guarantee outlined earlier in the paper. The results highlight how {\FLB} mitigates worst-case performance by strategically reserving sufficient capacity for potential high-duration, high-reward jobs in the future. \revcolor{Second, in \Cref{sec:RandomInstance}, we evaluate performance on more practical, beyond worst-case instances, and investigate how ``job congestion'' affects the results. There in addition to the simple benchmarks we compare {\FLB} with the algorithms designed in \cite{GIU-20} and \cite{HC-22}.} Our findings indicate that, although {\FLB} offers limited benefits in fully utilized or under-utilized scenarios, it has clear advantages in intermediate utilization regimes.



\subsection{Against the Worst-Case Instance} \label{sec:worstCase}
\xhdr{Setting.} Inspired by the worst-case instance introduced in \Cref{sec:lower bound construction}, we consider a family of $M=1000$ instances, each with a single server ($n=1$), capacity $c = 200$, and parameters $\maxreward = \maxduration = 10$. In instance $m\in[1:M]$, there are $m$ jobs arriving within the interval $[0,1]$, with durations and rewards increasing exponentially over time according to a geometric progression: each job $j\in [1:m]$ arrives at time $t_j = \frac{j-1}{1000}$ and has reward $\reward_j = \maxreward^{t_j}$ and duration $\duration_j = \lfloor \maxduration^{t_j}\rfloor$. Importantly, we require our algorithm and benchmarks to provide robust performance guarantees against all instances in this family, as though these $M=1000$ jobs could continue arriving indefinitely, yet the adversary may stop arrivals at any point (see \Cref{fig:badExample}).

\xhdr{Benchmarks.}
We compare the numerical performance of {\FLB} against several benchmarks:
\begin{itemize}
    \item \underline{{\OFF}:} \revcolor{An upper bound on the optimal offline benchmark for the given instance, computed by solving the standard Linear Programming (LP) relaxation of the offline assignment problem.}
    \item \underline{{\GRD}:} A greedy algorithm that assigns each arriving job $j$ to the compatible server with available capacity that offers the maximum total reward $r_{ij}d_{ij}$.
    \item \underline{{\BALANCE}:} The {\BALANCE} algorithm, which assigns each arriving job $j$ to the server with the highest positive reduced reward $\rewardij \durationij - \pen_{\text{B}}\left(\alpha_{i,t_j\rightarrow t_j}\right)$ (if any), where $\pen_{\text{B}}(x) = \frac{\maxreward \maxduration}{e-1}\left(e^{1-x}-1\right)$. Note that to maintain capacity feasibility, it must satisfy $\pen_{\text{B}}(0) \geq \maxreward \maxduration$.\footnote{\cite{GNR-14} define $\pen_{\text{B}}(x) = \frac{e^{1 - x}-1}{e-1}$, which matches our definition when $\maxreward = \maxduration = 1$.}
\end{itemize}

\begin{figure}
    \centering
    \includegraphics[width = 0.45\linewidth]{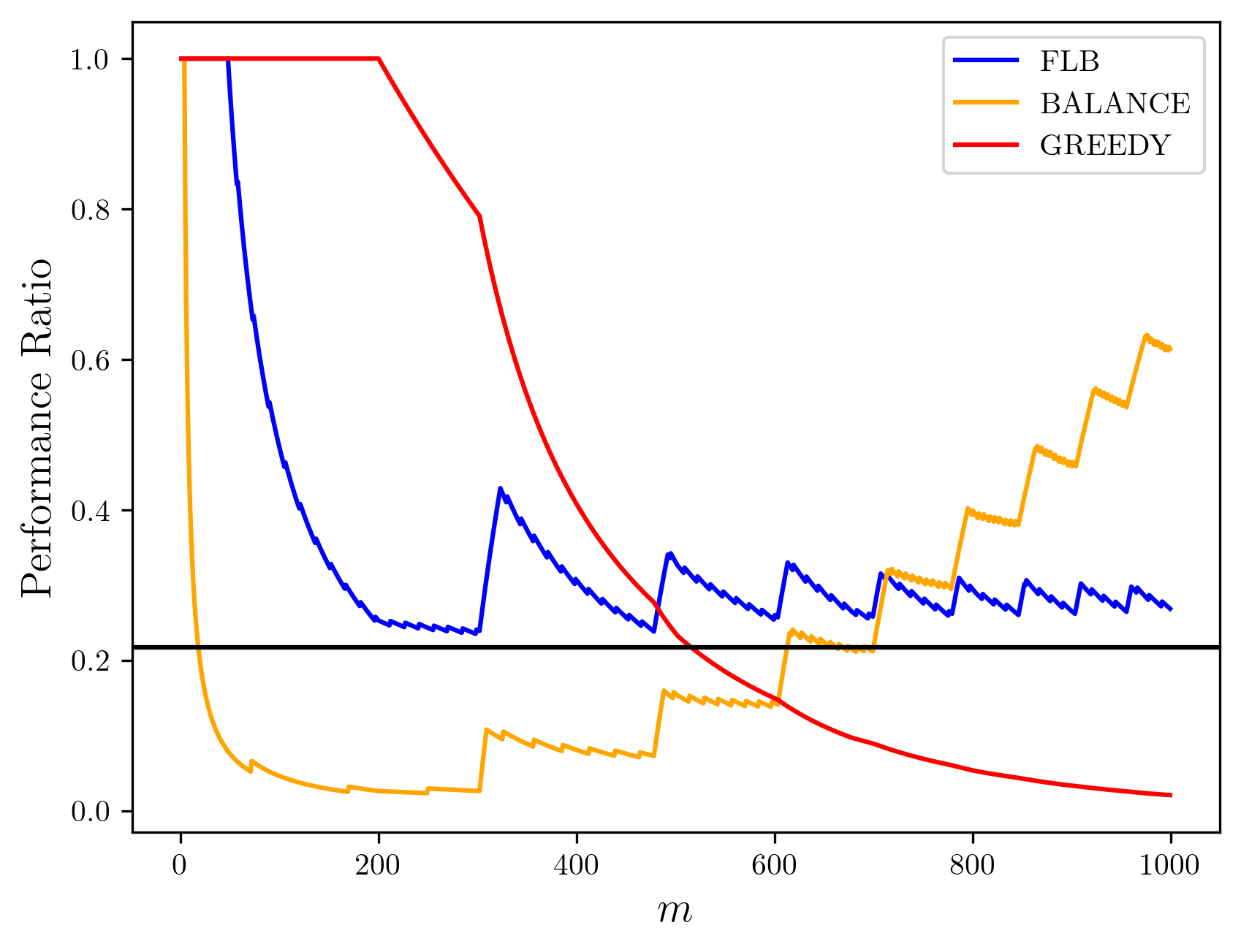}
    \includegraphics[width = 0.45\linewidth]{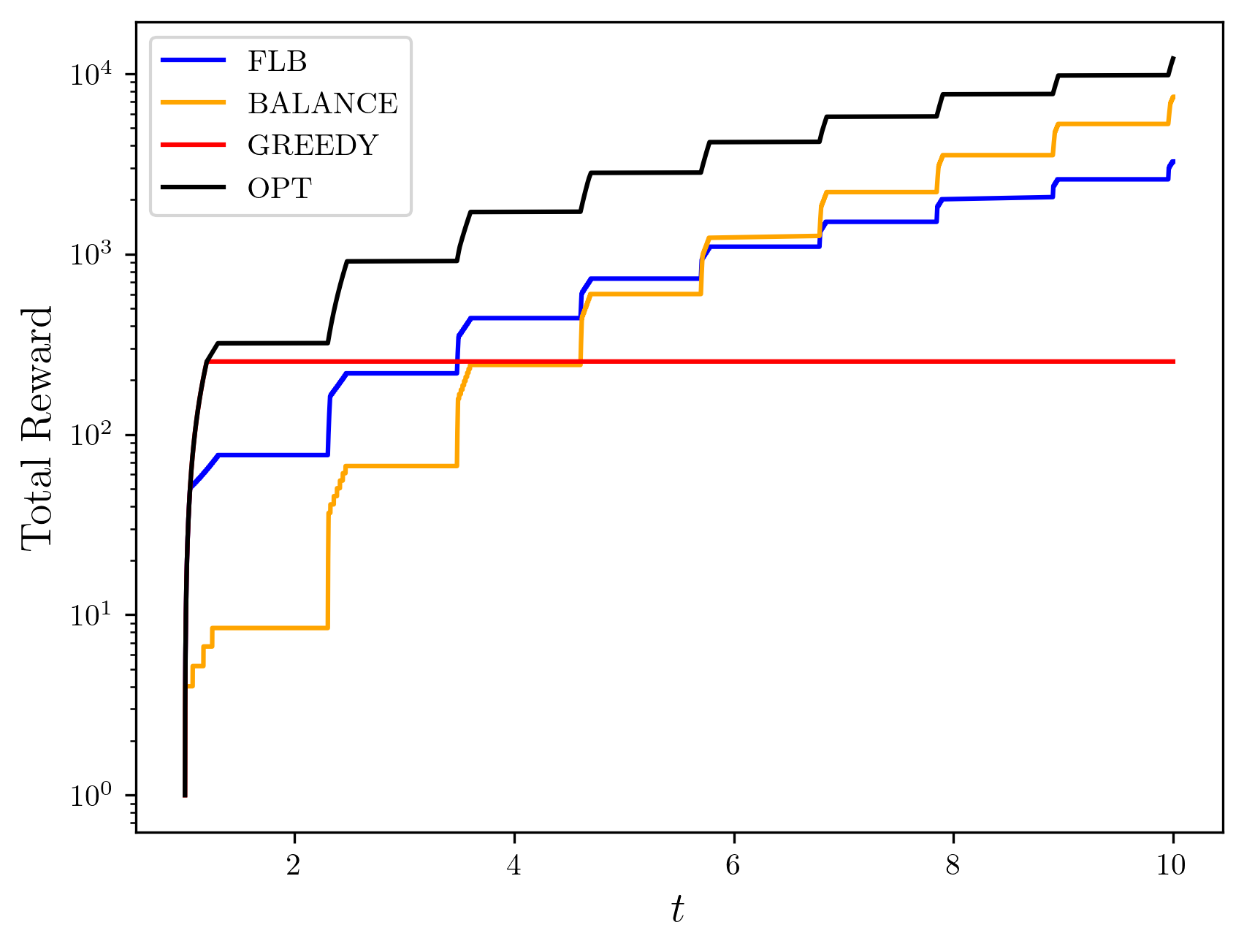}
    \caption{\centering The left panel illustrates the performance ratio of the algorithms relative to the optimal offline solution, with the solid black line indicating the tight theoretical guarantee $y = \frac{1}{\ln(\maxreward\maxduration)} \approx 0.217$. The right panel presents the total reward for the instance with $m=1000$. The observed jumps result from the discrete nature of durations, which take integer values.}
    \label{fig:badExampleNumerical}
\vspace{-7mm}
\end{figure}

\xhdr{Discussion.} The performance of the algorithms compared to the optimal offline solution is shown in \Cref{fig:badExampleNumerical}. As observed, {\GRD} accepts all jobs until reaching full capacity at job number $200$, after which it loses competitiveness. In contrast, {\BALANCE} initially assigns very conservatively due to the high penalty required for feasibility, resulting in a poor performance ratio if the problem instance has fewer than approximately 500 jobs. However, {\FLB} consistently maintains its performance ratio above the theoretical bound $\frac{1}{\ln(\maxreward\maxduration)}$, as established theoretically in \Cref{sec:competitive ratio analysis}.

\subsection{Against a Randomized Practical Instance} \label{sec:RandomInstance}
\begin{figure}
    \centering
    \includegraphics[width=0.45\linewidth]{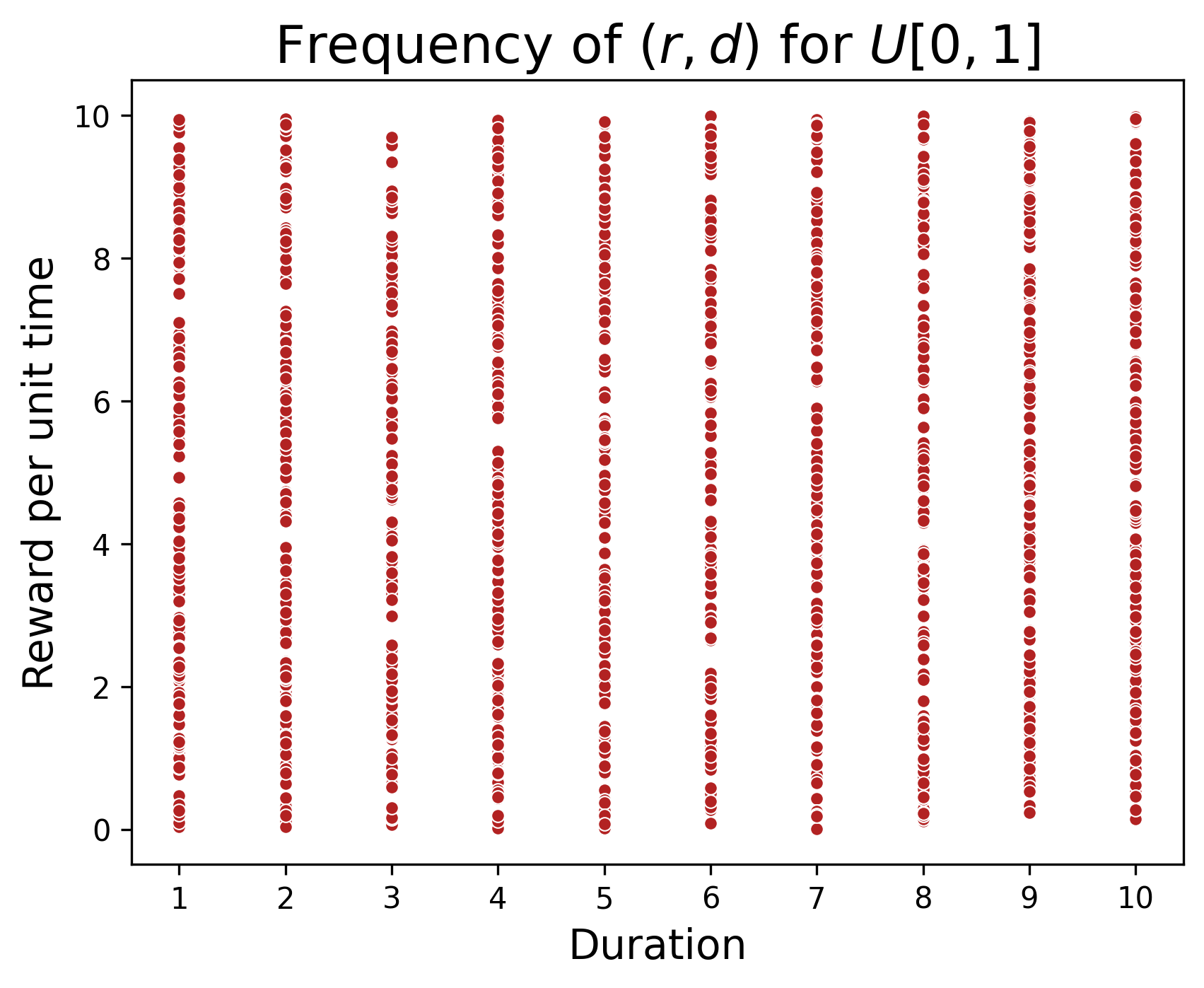}
    \includegraphics[width=0.45\linewidth]{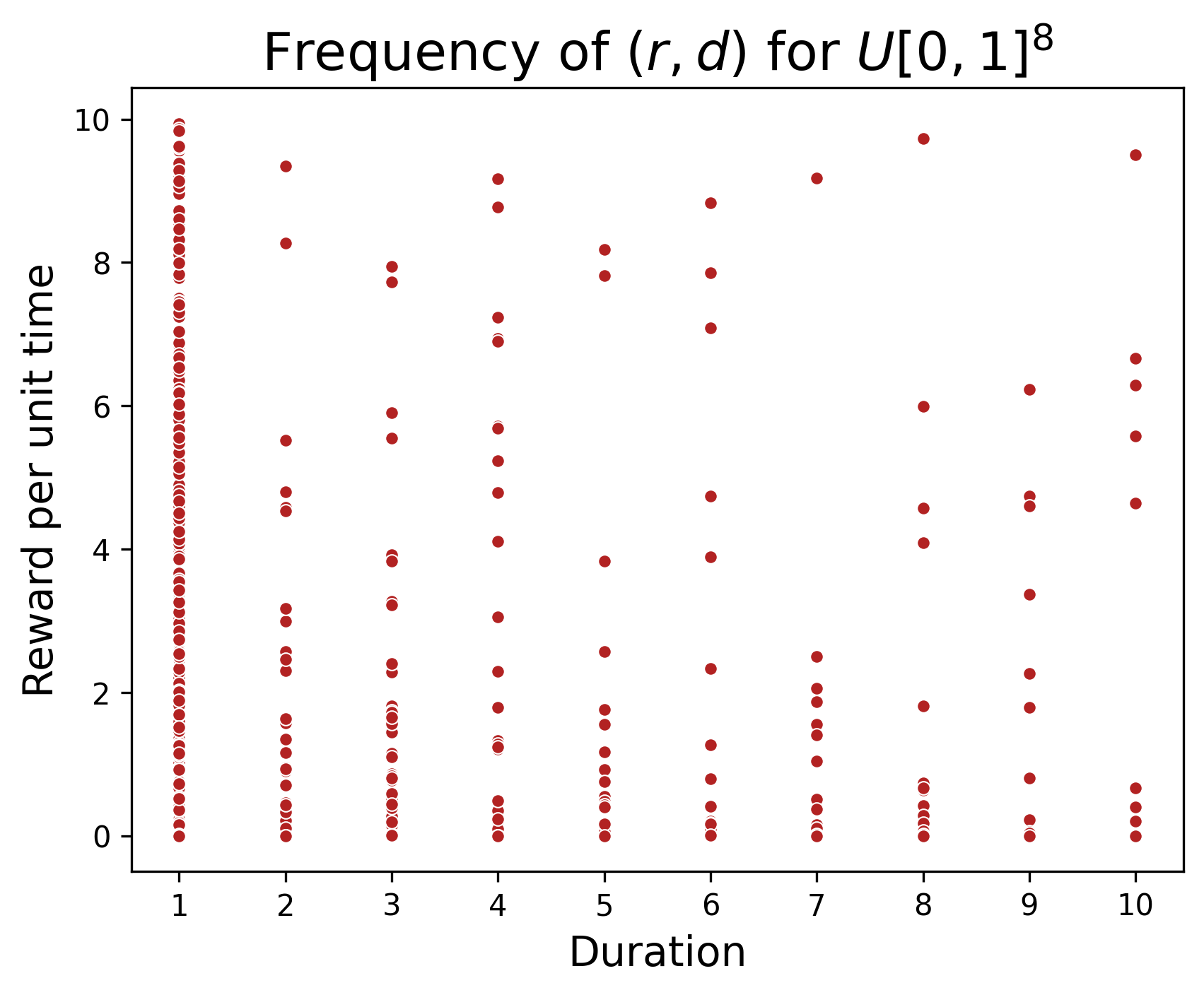}
    \caption{\revcolor{Scatter plots comparing the joint distribution of rewards and durations $(\reward, \duration)$ for $m = 500$ jobs. The left panel shows jobs sampled from a standard uniform distribution $\text{Beta}(1,1)$, yielding a homogeneous spread. The right panel demonstrates the heavily skewed $\text{Beta}(0.125, 1)$ distribution used in our experiments, which generates a dense cluster of low-value, short-duration jobs interspersed with rare, high-value opportunities.}}
\vspace{-8mm}
    \label{fig:scatter}
\end{figure}

\begin{figure}
    \centering
    \includegraphics[width=0.45\linewidth]{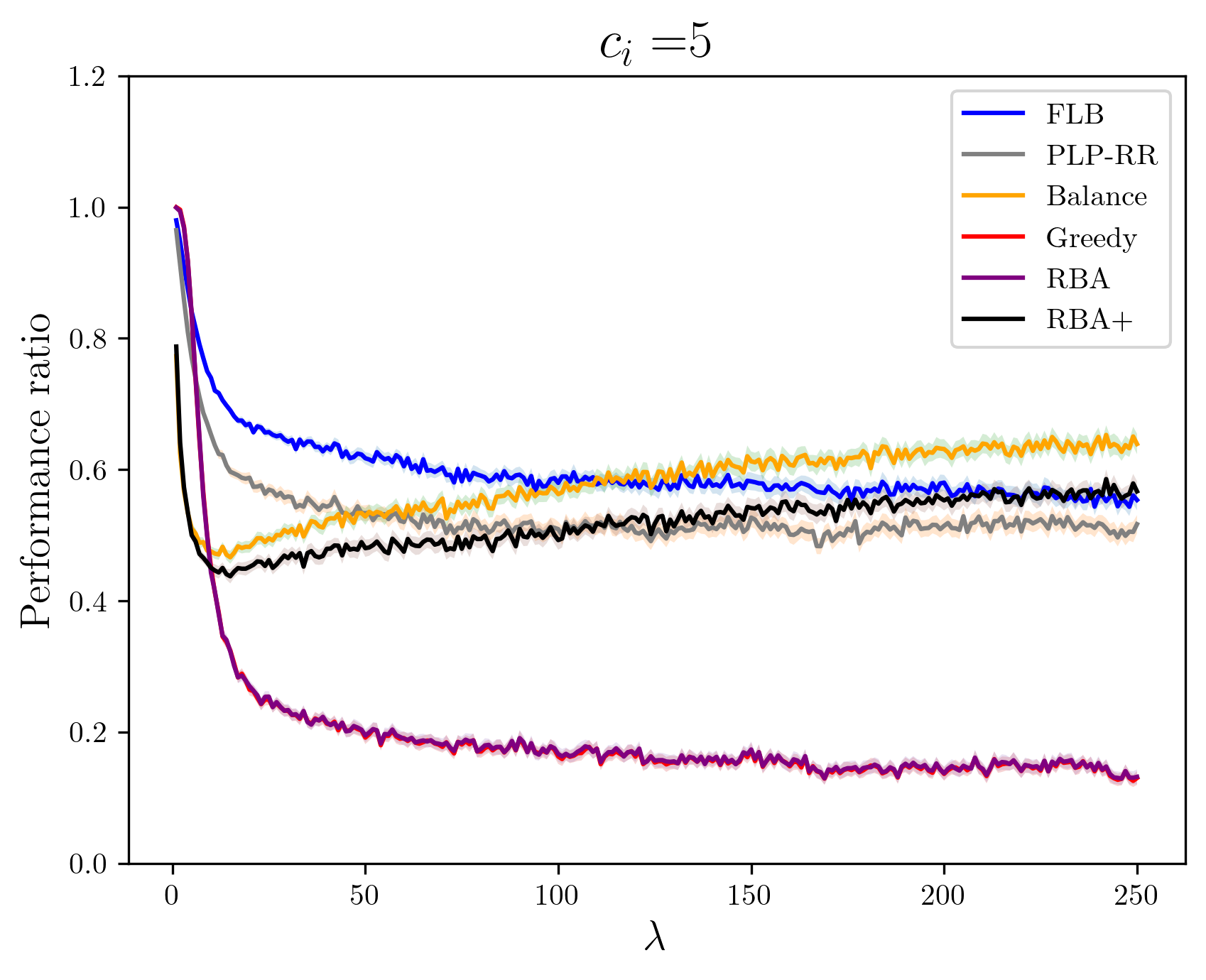}\includegraphics[width=0.45\linewidth]{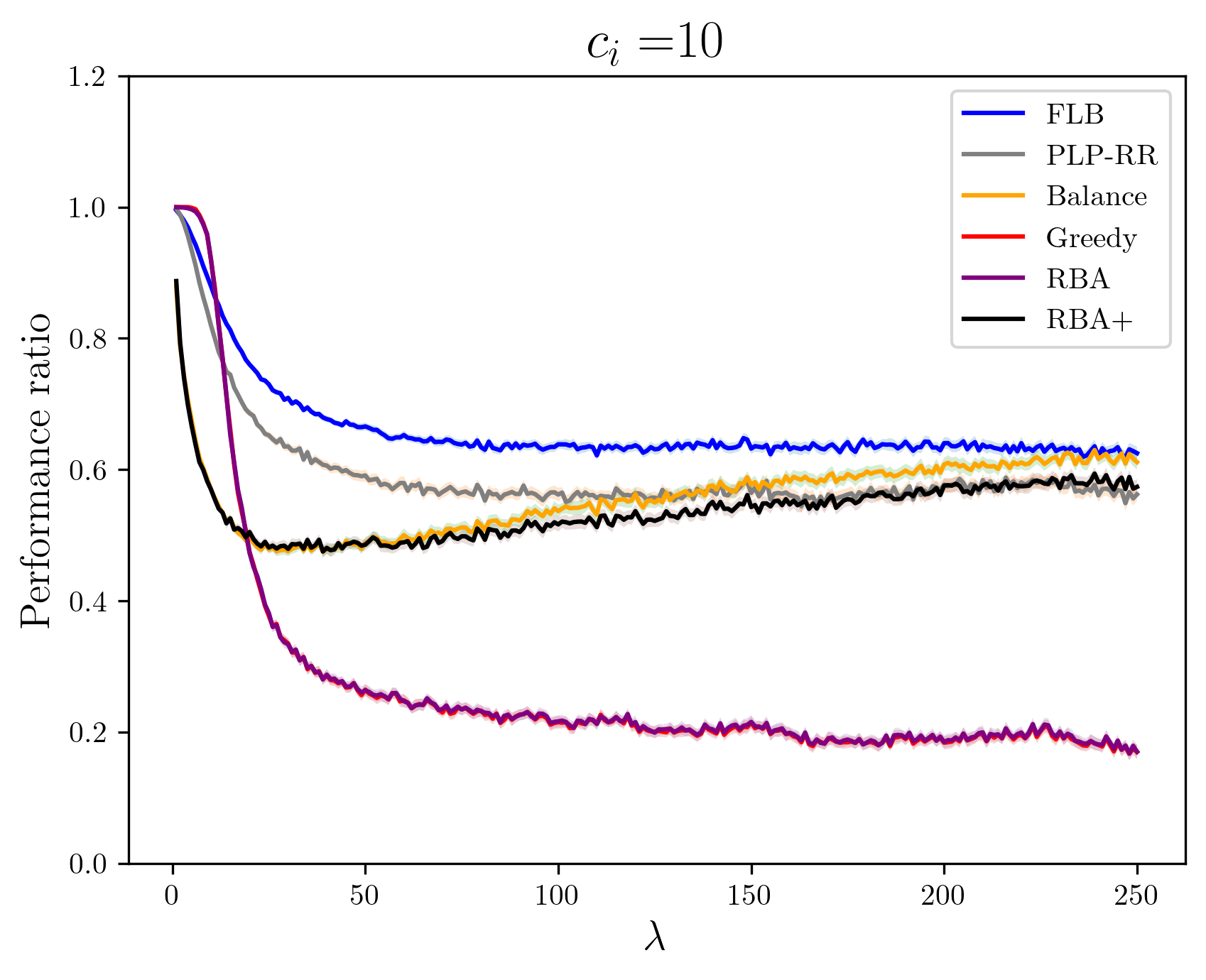}
    \\ \includegraphics[width=0.45\linewidth]{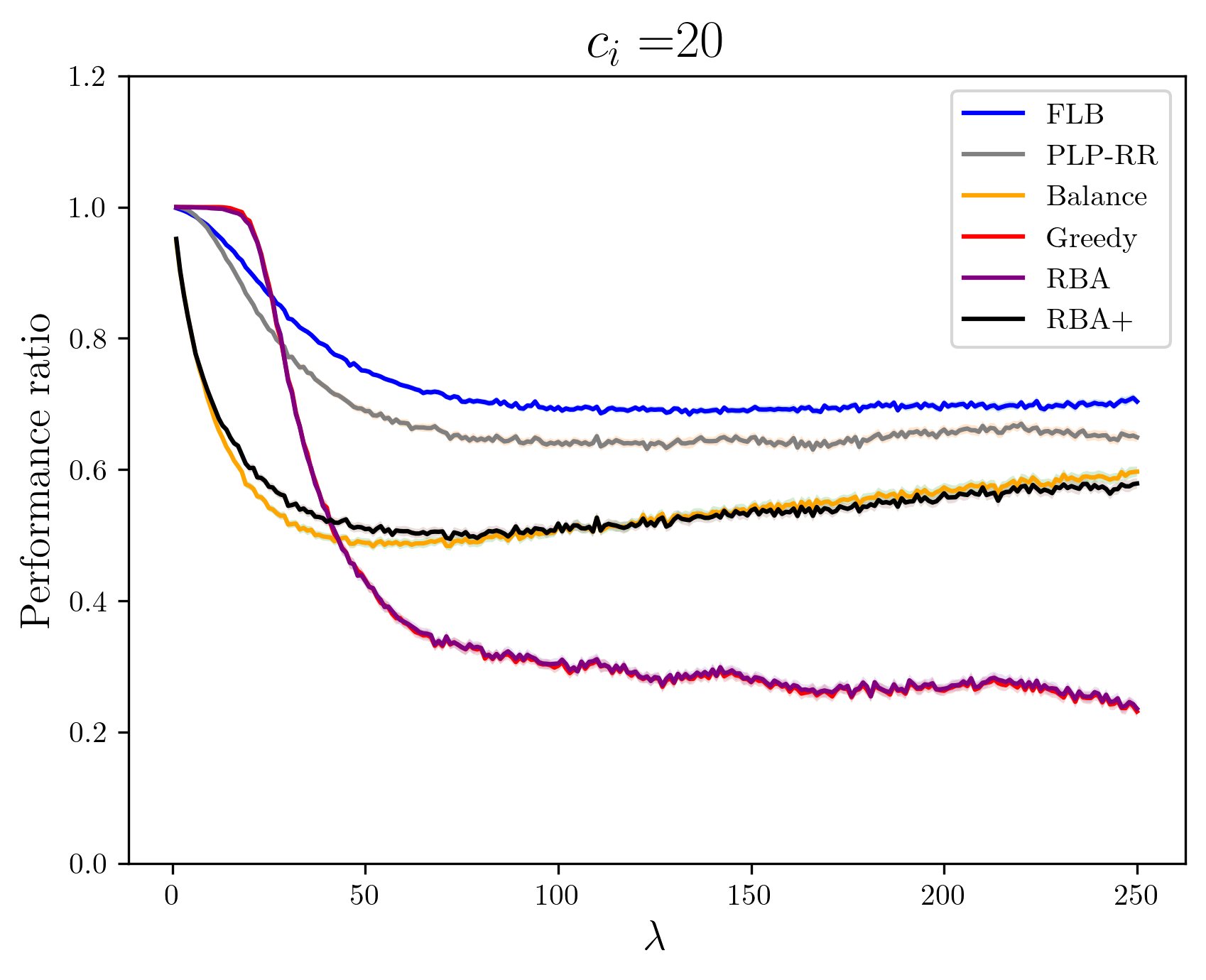}\includegraphics[width=0.45\linewidth]{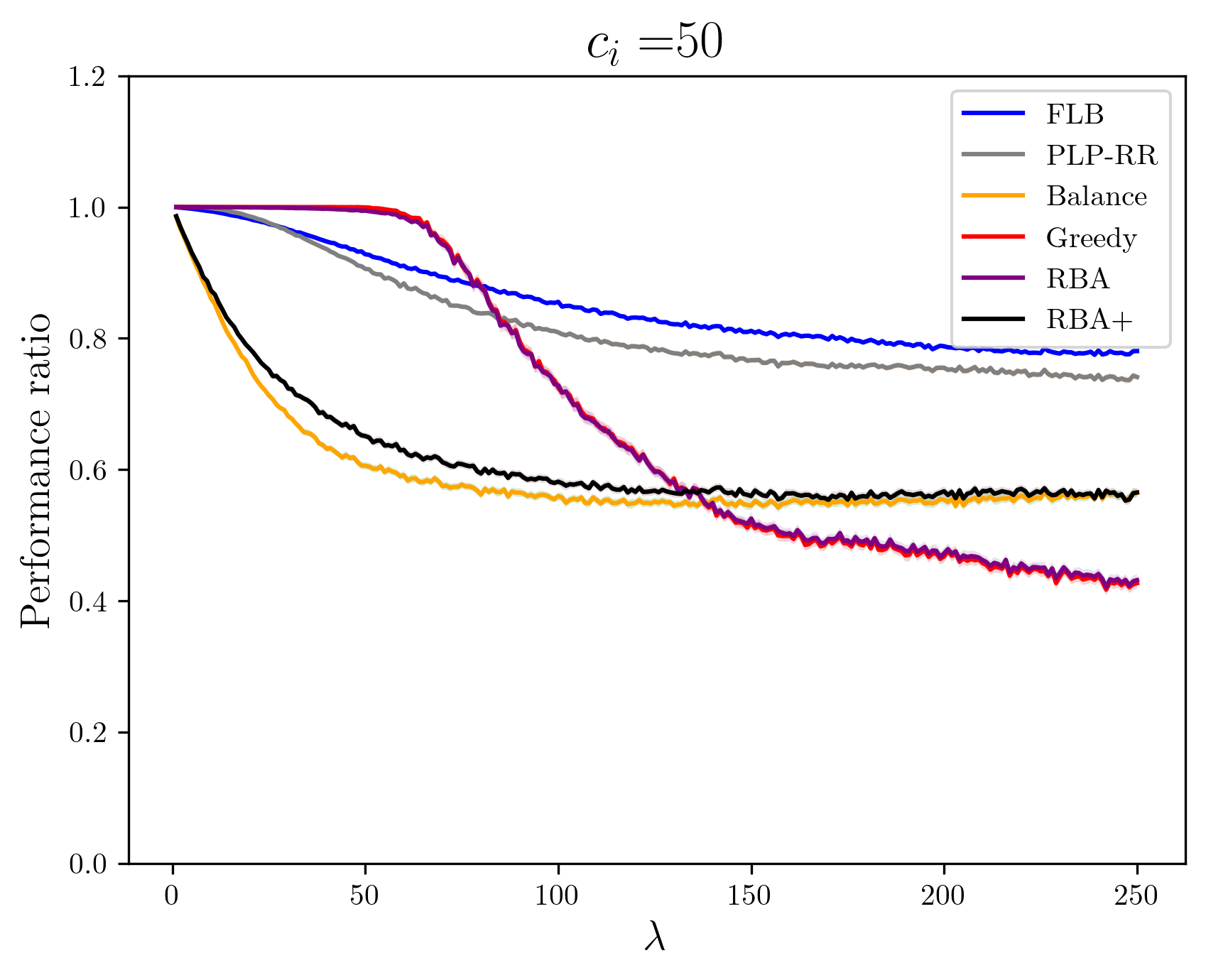}
    \caption{Performance comparison and confidence intervals for \FLB\ and alternative algorithms on the random instance for different values of $c_i$.
    Parameters: $n = 3, \maxreward = \maxduration = 10, m = 500, \inventory_1=\inventory_2=\inventory_3$.}
    \label{fig:Numerics}
\vspace{-8mm}
\end{figure}

\xhdr{Setting.} \revcolor{We consider $n=3$ servers, each sharing a symmetric capacity $c \in \{5,10,20,50\}$. A total of $m=500$ jobs arrive according to a Poisson process with rate $\lambda$. To evaluate the algorithms under varying levels of congestion, we vary $\lambda$ from 1 to 250, generating 100 independent random instances for each rate. Setting the maximum parameters to $\maxreward=\maxduration=10$, job rewards are drawn independently and identically from a right-skewed scaled Beta distribution. Specifically, rewards equal $\maxreward \cdot X$ where $X \sim \text{Beta}(1/8, 1)$ (equivalently generated via $\maxreward \cdot U^8$, with $U \sim \text{Uniform}(0,1)$). Job durations are strictly positive integers, determined by taking the ceiling of independent samples drawn from an identically shaped distribution scaled by $\maxduration$. In \Cref{fig:Numerics}, we report the mean performance of each algorithm across the tested values of $\lambda$ for each capacity setting.\\
This heavily skewed distribution is specifically selected rather than a standard uniform distribution ($\text{Beta}(1, 1)$) to create a non-trivial admission control environment. Under a uniform distribution, the $(r_j, d_j)$ pairs are relatively homogeneous, meaning simple non-idling policies perform well on average. In contrast, the $\text{Beta}(0.125, 1)$ distribution models a more realistic and challenging scenario: the vast majority of arriving requests are jobs with low rewards and short durations, while highly profitable, long-duration jobs arrive only occasionally (see \Cref{fig:scatter}). This structure rigorously tests an algorithm's ability to resist prematurely exhausting its capacity on low-value requests in order to preserve availability for rare but high reward future arrivals.}

\xhdr{Benchmarks.}
\revcolor{Beyond the previously discussed {\FLB}, {\BALANCE}, {\GRD}, and {\sf OPT}, we evaluate two additional algorithms from the literature, adapted to our duration-aware matching environment:
\begin{itemize}
    \item \underline{{\sf PLP-RR}:} The Protection Level Policy for Reusable Resources (\citealp{HC-22}) is extended to our matching instances by assigning each arriving request to the server $i$ that maximizes the net reward, defined as $r_id_i - R_{i,t}$, where $R_{i,t}$ represents the rejection price defined in their paper.
    \item \underline{{\sf RBA}:} We modify the Rank-Based Allocation algorithm (\citealp{GIU-20}) to account for heterogeneous usage durations. Under this adaptation, {\sf RBA} routes each job to the server $\hat{i}$ according to the rule:
    \begin{align}\label{eq:RBA}\tag{RBA}
        \hat{i}\in\argmax_i \;\;d_ir_i \left(1-e^{-\frac{z_i(t)}{c_i}}\right)
    \end{align}
    Because of the positive scaling factor in \eqref{eq:RBA}, {\GRD} and {\sf RBA} share a structural similarity in our setting: both algorithms will inherently admit an arriving job whenever capacity is available. 
    \item \underline{{\sf RBA+}:} An alternative approach to modifying the Rank-Based Allocation algorithm (\citealp{GIU-20}) is to incorporate a penalty term subtracted from the reward, conceptually similar to {\BALANCE}. Under this formulation, {\sf RBA+} assigns each incoming job to a server $\hat{i}$ by the following criteria:
    \begin{align}\label{eq:RBAPlus}\tag{RBA+}
        \hat{i}\in\argmax_i \;\;d_ir_i -\frac{RD}{e-1}(e^{1-\frac{z_i(t)}{c_i}}-1)
    \end{align}
    As formulated, {\BALANCE} and {\sf RBA+} exhibit a clear structural resemblance, differing primarily in how they account for server availability.
\end{itemize}
}

\xhdr{Discussion.}
\revcolor{
As illustrated in \Cref{fig:Numerics}, {\FLB} and {\sf PLP-RR} demonstrate superior performance across the majority of the parameter space, while {\BALANCE}, {\GRD}, {\sf RBA}, and {\sf RBA+} only prove effective in specific extreme regimes. 

For small values of $\lambda$ and high capacity $c$, the system is underutilized, making active capacity management less critical. As a result, non-idling policies like {\GRD} and {\sf RBA} outperform the other algorithms by naturally accepting nearly all profitable arrivals. In contrast, {\BALANCE} and {\sf RBA+} perform poorly in this regime; their highly conservative penalty structure leads to the unnecessary rejection of a significant number of jobs that could have easily been accommodated given the low arrival rate.

Conversely, for exceptionally large $\lambda$, the system becomes highly congested. The high volume of arrivals means most capacity is exhausted very early, effectively reducing the problem to a non-reusable setting and diminishing the performance gap between {\FLB} and both {\BALANCE} and {\sf RBA+}. Furthermore, because the reward and duration distributions are heavily skewed toward zero, {\GRD} and {\sf RBA} perform arbitrarily poorly in this congested regime. They greedily exhaust their capacity on the earliest arriving jobs, which are statistically highly likely to have short durations and low rewards (see \Cref{fig:scatter}).

Finally, in the intermediate utilization regime---where arrivals are high enough to necessitate careful capacity management, but not so overwhelming that the temporal aspect of reusability becomes irrelevant---{\FLB} (followed by {\sf PLP-RR}) significantly outperforms all other benchmarks.}

\section{Conclusion \& Future Directions}
Motivated by applications in cloud computing marketplaces, we introduced the online job assignment problem. We proposed {\FLBLong} ({\FLB}) and characterized its asymptotic competitive ratio in various settings using a novel primal-dual analysis, obtaining optimal or nearly optimal results. Although closely related to extensively studied models such as online packing/covering and reusable-resource matching, existing algorithms from these literature do not provide strong guarantees due to fundamental features in our model, such as continuous arrivals and heterogeneous job durations and rewards. We developed new techniques to address these challenges.

\xhdr{Future research.} Several open questions remain. First, while this work mainly focuses on the large-capacity regime, deriving nearly optimal competitive ratios under small capacity would also be interesting; even for the special case $\maxreward = \maxduration = 1$, the best known competitive ratio is $0.589 < 1 - 1/e$ in that regime~\citep{DFNSU-22}. Second, in certain practical settings, the platform may allow only discrete sets of possible rewards and job durations~\citep{MS-20}. Can we apply \FLB\ to obtain more refined competitive-ratio guarantees depending explicitly on these discrete sets? Third, platforms might leverage machine-learning predictions of future job arrivals. Given the forward-looking nature of {\FLB}, could such forecasts be incorporated into the algorithm? As another research question, some platforms allow deferring jobs if all servers are occupied through wait lists, or they allow for advanced booking. It would be interesting to analyze the performance of {\FLB} or develop new algorithms for these variants. Finally, one can consider the variant of the problem with machine-learning-based ``advice,'' similar to \cite{choo2025learning,mahdian2007allocating}, and study the characterization of the robustness-consistency tradeoff.

\newcommand{\newblock}{}
\setlength{\bibsep}{0.0pt}
\bibliographystyle{plainnat}
{\footnotesize
\bibliography{refs}}

\renewcommand{\theHchapter}{A\arabic{chapter}}
\renewcommand{\theHsection}{A\arabic{section}}

\newpage
\clearpage
\normalsize
\pagestyle{ECheadings}%
\ECHowEquations
\ECHowSections
\setcounter{figure}{0}%
\renewcommand\thefigure{EC.\@arabic\c@figure}%
\setcounter{table}{0}%
\renewcommand\thetable{EC.\@arabic\c@table}%
\setcounter{page}{1}\def\thepage{ec\arabic{page}}%
\hspace*{1em}

\ECDisclaimer

\section{Further Related Work}
\label{apx:further related work}
\xhdr{Online resource allocation of non-reusable resources.}
There is a \revcolor{vast} literature \revcolor{on} online resource allocation, where the consumer sequence is determined by an adversary, known as the adversarial arrival setting. In the basic model of online bipartite matching, \citet{KVV-90} introduce the RANKING algorithm and demonstrate its optimality with a competitive ratio of $\frac{e}{e - 1}$. \citet{KP-00} study the online bipartite b-matching where each offline node has a capacity (inventory) constraint. The authors propose the {\BALANCE} algorithm and show it also achieves the optimal $\frac{e}{e-1}$-competitive under a large inventory assumption. Later work generalizes the {\BALANCE} algorithm to various other models, including the AdWords problem \citep{MSVV-05}, the online assortment problem \citep{GNR-14}, batch arrival \citep{FN-21}, and unknown capacity \citep{MRS-22}. See the recent survey \cite{niazadeh2026modernprimaldualframeworkspriorfree} for more details. All aforementioned works assume that each resource has an identical reward that is independent of the consumer. For online resource allocation of non-reusable resources with heterogeneous rewards, \citet{BQ-09} investigate the same model as ours but with the restriction that all resources are non-reusable. They introduce a ``protection level'' policy and demonstrate its asymptotic optimality with a competitive ratio of \revcolor{$\log \maxreward +1$} under a large inventory assumption \revcolor{(in fact, this paper shows a more refined instance-dependent competitive ratio given a finite set of types, which becomes $\log \maxreward +1$ with infinitely many types)}. A similar result is also found in \citet{BN-09} and subsequent works \citep[e.g.,][]{Aza-16} that study the online packing/covering problem. \citet{MS-20} study the same problem,  present a generalization of the {\BALANCE} algorithm, and show it achieves a more refined instance-dependent competitive ratio. Our results align with the competitive ratio results established in \citet{KP-00,BQ-09,MS-20}.

\xhdr{Online resource allocation of reusable resources.} 
Several studies have been conducted on online resource allocation of reusable resources in the adversarial arrival setting. \citet{GGISUW-22} study the adversarial setting with identical rental fees and stochastic i.i.d.\ rental duration and show that the greedy algorithm achieves a competitive ratio of 2. \citet{SZZ-22} introduce a more general model with decaying rental fees and rental duration distributions. \citet{GIU-20} study the same model as \citet{GGISUW-22} and design a $\frac{e}{e-1}$ competitive algorithm under a large inventory assumption. \revcolor{Notably, in this model, the durations are assumed to be stochastic and i.i.d. from a known distribution, but their optimal competitive algorithm (under large budgets) does not require knowing the realization of the durations at the time of allocation.\citet{FNS-19} study the same model but with deterministic durations. Under a large inventory assumption, \citet{FNS-19} shows that the {\BALANCE} algorithm is $\frac{e}{e-1}$ competitive. \citet{DFNSU-22} studies the integral online bipartite matching of reusable resources with identical rental fees and identical rental durations, and design a $1.98$-competitive algorithm \emph{without} a large inventory assumption.} All the aforementioned previous works consider identical rental fees and durations, without considering heterogeneity. In contrast, our paper addresses the issue of heterogeneous rental fees and durations. Notably, all algorithms in these prior works are not forward-looking, i.e., the allocation decision is made based on the current inventory without an eye toward the anticipated inventory level in the future. Our result recovers the result in \citet{FNS-19}. Finally, we note the growing body of work that considers the learning problem in the allocation and pricing of reusable resources~\cite{JSS-24,FDJQ-24}. This differs from our work, as we focus on the adversarial setting rather than a stochastic learning framework.  

A recent closely related work is \citet{HC-22}. This paper shares the same model as ours \revcolor{(where the durations and rewards are fully heterogeneous, but durations are known to the online algorithm at the time of allocation)} but introduces the additional restriction of having only a single resource. They further make three types of monotonicity assumptions regarding consumers' personalized rental fees and durations. They generalize the protection level policy proposed by \citet{BQ-09}. Under a large inventory assumption, they prove a competitive ratio of $\phi\cdot \ln (\heterosize)$, $\phi\cdot\maxreward\ln (\heterosize)$, and  $\phi\cdot\maxduration\ln (\heterosize)$ for three types of monotonicity assumptions, respectively, where $\phi\in[1, 2)$ is a constant that depends on the consumer sequence. In comparison to \citet{HC-22}, our paper considers multiple reusable resources without imposing any monotonicity assumptions and achieves an asymptotically optimal competitive ratio of $\ln(\heterosize)$. Therefore, the competitive ratio of $\ln(\heterosize)$ in our paper not only improves the coefficient in a more general model but also represents an exponential improvement compared to \citet{HC-22}. Finally, it should be noted that \citet{HC-22} uses the standard LP as the benchmark for the analysis of competitive ratios. Consequently, their primal-dual proof requires the aforementioned monotonicity assumption, and the dual assignment construction necessitates the utilization of a novel, albeit complex, auxiliary algorithm. In contrast, our paper considers the competitive ratio with respect to the configuration LP, enabling a relatively simpler primal-dual analysis and achieving an asymptotically optimal competitive ratio without any monotonicity assumptions.

\xhdr{Online covering/packing LPs under adversarial arrivals.} Another closely related line of work is the classic literature in computer science on online covering and online packing linear programs under adversarial arrivals~\citep{BN-09}, and the related models such as online routing~\citep{AAP-93}, online load balancing~\citep{AKPPW-97}, interval scheduling~\citep{LT-94}, and combinatorial auction~\citep{BG-15}. By using non-trivial arguments, one might be able to show that special cases of the online job assignment problem with discrete time, integer durations, and single server can be reduced to these models, resulting in (somewhat complicated) algorithms that obtain $\mathcal{O}(\log (R\cdot D))$ competitive ratios in these special cases (with highly sub-optimal asymptotic logarithmic constants). This is completely in contrast to our simple and practical FLB algorithm, which obtains the optimal asymptotic logarithmic constant of $\logcons^*=1$ in the most \emph{general} case of the online job assignment problem with continuous time, real durations, and multiple servers. \revcolor{A related extension of the online packing LP setting under adversarial arrival is online conic optimization~\citep{EF-16} and, in particular, the PSD-cone resource-allocation model of \citet{ESF-18}. This latter model contains online budgeted allocation/AdWords as a special case, but not the online job assignment; in fact, it extends the classical budgeted non-reusable resource model in a different direction, by allowing conic objectives and constraints over monotone PSD resource consumption, but does not capture the reusable resource feature central to online job assignment: an accepted job occupies capacity only over a future interval and then returns it, so such a feasibility constraint cannot be encoded by a single cumulative budget.}


\revcolor{\xhdr{Online LP and convex programming under stochastic arrivals.}
A large literature studies online LP and convex programming under random or distributional input models, rather than adversarial arrivals. In random-order and i.i.d.\ unknown-distribution models, each request reveals a column or action. Besides primal-only methods, a powerful algorithmic framework in these settings is primal-dual, where an online algorithm learns dual prices (i.e., updates resource multipliers) to guide its primal allocations. Representative models and results include random-permutation AdWords and dynamic online LP algorithms~\revcitep{DevHayes-09,AWY-14}, stochastic AdWords and prior-robust resource allocation~\revcitep{DSA-12,NKBC-19}, random-order packing LPs with improved dependence on the capacity ratio~\revcitep{MR-14,KRTV-18}, expert-learning and mirror-descent methods for online packing and allocation~\revcitep{GM-16,BLM-23}, online stochastic convex programming~\revcitep{AD-15}, and sharper regret and dual-convergence guarantees for online LP~\revcitep{LiYe-22}. Recent dynamic matching and allocation models with post-allocation service also use dual-learning and Lyapunov/online-learning ideas to handle dynamic resources modeling congestion in queues~\revcitep{BLMNP-25}.}

\revcolor{This literature shares with our work the broad primal-dual philosophy: a dual object---such as bid prices, multipliers, or convex dual variables---guides online primal decisions through approximate complementary slackness or KKT conditions. The key difference is how this dual object is obtained. Under i.i.d.\ arrivals or random order,  and with the large-budget assumption, the algorithm can learn or estimate the dual solution of an offline, fluid, or expected LP, for example by periodic re-solving or no-regret online learning/first-order methods. As this learning error vanishes, the induced primal decisions become near-optimal for the corresponding stochastic benchmark. In our prior-free adversarial model, this route is unavailable: arrivals, rewards, durations, and compatibilities may all be arbitrarily changing over time, and the dual variables could not be learned. However, they can be \emph{approximated}. Thus, {\FLB} can be viewed as an adversarial dual-fitting counterpart to the same primal-dual recipe. Rather than learning the optimal dual, it constructs, along the realized instance, a problem-specific approximately feasible dual certificate for the configuration LP whose objective value equals the value earned by the algorithm. This certificate justifies the reward adjustment in {\FLB} and yields the sharp $\ln(\heterosize)$ asymptotic constant in a heterogeneous, multi-server reusable-resource setting; to the best of our knowledge, no off-the-shelf application of online learning or first-order dual updates could construct such an approximately optimal dual solution.}

\revcolor{We should also highlight that a parallel stream of work studies Bayesian or known-distribution models, where arrivals are drawn from known, possibly time-varying distributions, or where sampling access is available. These works typically use an expected-LP or ex-ante relaxation as the benchmark, and implement it through randomized routing, contention resolution or discarding, simulation-based control, or sample-based pricing policies~\revcitep{AHL-12,DHKN-21,BM-22,FNS-22,BBP-24,GSW-25}. This is again different from our setting, where no arrival distribution or sampling oracle is available to guide the online algorithm.}

Here is a summary of further related prior work in the literature directly related to our work.

\yfdelete{
\xhdr{Online resource allocation of non-reusable resources.}
There is a long literature about online resource allocation, where the consumer sequence is determined by an adversary, known as the adversarial arrival setting. In the basic model of online bipartite matching, \citet{KVV-90} introduce the RANKING algorithm and demonstrate its optimality with a competitive ratio of $\frac{e}{e - 1}$. \citet{KP-00} study the online bipartite b-matching where each offline node has a capacity (inventory) constraint. The authors propose the BALANCE algorithm and show it also achieves the optimal $\frac{e}{e-1}$-competitive under a large inventory assumption. Later work generalizes the BALANCE algorithm to various other models, including the AdWords problem \citep{MSVV-05}, the online assortment problem \citep{GNR-14}, batch arrival \citep{FN-21}, and unknown capacity \citep{MRS-22}. All aforementioned works assume that each resource has an identical reward that is independent of the consumer. For online resource allocation of non-reusable resources with heterogeneous rewards, \citet{BQ-09} investigate the same model as ours but with the restriction that all resources are non-reusable. They introduce a ``protection level'' policy and demonstrate its asymptotic optimality with a competitive ratio of $O(\log \maxreward)$ under a large inventory assumption. A similar result is also found in \citet{BN-09} and subsequent works \citep[e.g.,][]{Aza-16} that study the online packing/covering problem. \citet{MS-20} study the same problem,  present a generalization of the BALANCE algorithm, and show it achieves a more refined instance-dependent competitive ratio. Our results align with the competitive ratio results established in \citet{KP-00,BQ-09,MS-20}.

\xhdr{Online resource allocation of reusable resources.} 
Several studies have been conducted on online resource allocation of reusable resources in the adversarial arrival setting. \citet{GGISUW-22} study the adversarial setting with identical rental fees and stochastic i.i.d.\ rental duration and show that the greedy algorithm achieves a competitive ratio of 2. \citet{SZZ-22} introduce a more general model with decaying rental fees and rental duration distributions. \citet{GIU-20} study the same model as \citet{GGISUW-22} and design a $\frac{e}{e-1}$ competitive algorithm under a large inventory assumption. \citet{FNS-19} and \citet{DFNSU-22} study the online bipartite matching of reusable resources with identical rental fees and identical rental durations. Under a large inventory assumption, \citet{FNS-19} show that the BALANCE algorithm is $\frac{e}{e-1}$ competitive. \citet{DFNSU-22} design a $1.98$-competitive algorithm without a large inventory assumption. All the aforementioned previous works consider identical rental fees and durations, without considering heterogeneity. In contrast, our paper addresses the issue of heterogeneous rental fees and durations. Notably, all algorithms in these prior works are not forward-looking, i.e., the allocation decision is made based on the current inventory without an eye toward the anticipated inventory level in the future. Our result recovers the result in \citet{FNS-19}. Finally, we note the growing body of work that considers the learning problem in the allocation and pricing of reusable resources~\cite{JSS-24,FDJQ-24}. This differs from our work, as we focus on the adversarial setting rather than a stochastic learning framework.  
}

\xhdr{Job assignment in cloud computing platforms.} In the cloud computing setting specifically, prior work uses assumptions such as preemptable jobs \cite{LMNY-13,CI-95,AMX-23} and delayed commitments \cite{LMNY-13,ZG-12,AKLMNY-15}, neither of which our approach requires. 
Other work additionally includes pricing under various assumptions, in some cases taking a Bayesian approach~\citep[e.g.,][]{ZWLW-15,WNLL-13,NKBC-19,HXT-11,ZG-12,ZJLLVL-16,WMQCHL-15,KKS-19,ZMQSL-17,CDHKMS-17}, but lacks our nearly-optimal worst-case guarantees for the assignment problem. 

\yfdelete{
A recent closely related work is \citet{HC-22}. This paper shares the same model as ours but introduces the additional restriction of having only a single resource. They further make three types of monotonicity assumptions regarding consumers' personalized rental fees and durations. They generalize the protection level policy proposed by \citet{BQ-09}. Under a large inventory assumption, they prove a competitive ratio of $\phi\cdot \ln (\heterosize)$, $\phi\cdot\maxreward\ln (\heterosize)$, and  $\phi\cdot\maxduration\ln (\heterosize)$ for three types of monotonicity assumptions respectively, where $\phi\in[1, 2)$ is a constant that depends on the consumer sequence. In comparison to \citet{HC-22}, our paper considers multiple reusable resources without imposing any monotonicity assumptions and achieves an asymptotically optimal competitive ratio of $\ln(\heterosize)$. Therefore, the competitive ratio of $\ln(\heterosize)$ in our paper not only improves the coefficient in a more general model but also represents an exponential improvement compared to \citet{HC-22}. Finally, it should be noted that \citet{HC-22} uses the standard LP as the benchmark for the analysis of competitive ratios. Consequently, their primal-dual proof requires the aforementioned monotonicity assumption, and the dual assignment construction necessitates the utilization of a novel, albeit complex, auxiliary algorithm. In contrast, our paper considers the competitive ratio with respect to the configuration LP, enabling a relatively simpler primal-dual analysis and achieving an asymptotically optimal competitive ratio without any monotonicity assumptions.
}

\xhdr{Primal-dual analysis with non-standard LP.} 
The concept of configuration LP relaxation was initially introduced to design polynomial time approximation algorithms for the offline combinatorial optimization problems. It has been applied to various problems such as the cutting stock problem \citep{Eis-57}, the facility location problem \citep{JMMSV-03}, the bin packing problem \citep{BCS-06}, combinatorial auctions \citep{ABDR-12}, and the scheduling problem \citep{VW-14}. In many of these problems, the configuration LP achieves a smaller integrality gap compared to the standard LP relaxation \citep[see e.g.,][]{VW-14}. In online algorithm design, there are also recent works using the configuration LP. \citet{Tha-20} studies the online packing/covering problem and designs competitive algorithms with respect to the configuration LP. \citet{HZZ-20} study the Adwords problem and design an online algorithm whose competitive ratio against the configuration LP beats the greedy algorithm. \citet{HZ-20} and \citet{HJSSWZ-23} study  online bipartite matching with stochastic rewards. They use a randomized primal-dual framework to prove the competitive ratio guarantees of their algorithms with respect to the configuration LP. 

In addition to the configuration LP, primal-dual analysis with other non-standard LPs has been studied in the field of online matching. \citet{GU-23} introduce a path-based formulation LP and an LP-free primal-dual analysis framework for online bipartite matching with stochastic rewards. This technique has been subsequently applied to online bipartite matching with stochastic i.i.d.\ reusable resources \citep{GIU-20}, and AdWords with unknown budgets \citep{Udw-21,MRS-22}. \citet{HS-21} propose a natural LP for online stochastic bipartite matching and achieved a better competitive ratio than previous results, while \citet{AM-22} propose a truncated LP with a similar idea for a new variant model involving correlated arrivals.

\section{Competitive Ratio Analysis for Real-valued Durations}
\label{apx:FLB analysis real duration}

In this subsection, we present the formal analysis for the competitive ratio upper bound of {\FLB} in \Cref{thm:FLB competitive ratio general}. 

\crgeneral*

Similar to the integer-duration environment studied in \Cref{sec:competitive ratio analysis}, in \Cref{apx:FLB feasibility real duration}, we first characterize a sufficient condition for the capacity feasibility of {\FLB}. We then upper bound the competitive ratio of capacity feasible {\FLB} as a function of its parameters $(\inspecfreq,\penscalarTwo,\penscalar)$ in \Cref{apx:FLB primal dual competitive ratio}. Finally, in \Cref{apx:FLB parameter optimize}, we show an appropriate selection of parameters that obtains the competitive ratio upper bound in \Cref{thm:FLB competitive ratio general}. 

\subsection{Sufficient Condition for the Capacity Feasibility}
\label{apx:FLB feasibility real duration}
We present the following sufficient condition for the capacity feasibility of {\FLB}.

\PrimalFeasibilityReal*
Similar to \Cref{prop:FLB feasibility integer duration} following from \Cref{lem:inventory invariant integer duration}, we prove the following invariant for the real-valued case and we use the same machinery to prove \Cref{lem:FLB feasibility real duration}.
\begin{lemma}
\label{lem:inventory invariant real duration}
\yfedit{Consider a hypothetical scenario in which {\FLB} can assign a job to a server that has no available capacity (which could result in a negative {\estimated} capacity level). 
With integer-valued inspection-frequency scalar $\inspecfreq\in\naturals$,
the {\estimated} available capacity under {\FLB} satisfies the following property: for every server $i\in[n]$, time points $t,\tpre\in\reals_+$ such that $\inspecfreq (\tpre - t) \in \naturals_+$ and index $k\in\naturals$:}
\yfdelete{
In the capacity-unconstrained environment, the {\estimated} capacity level in {\FLB} with positive integer inspection-frequency scalar $\inspecfreq$ satisfies that for every server $i\in[n]$, time point $t,\tpre\in\reals_+$ such that $\inspecfreq (\tpre - t) \in \naturals_+$ and real duration $\duration\in\reals$:
}

    \begin{align*}
         \curinventory_{i,t\rightarrow \tau + \frac{k}{\inspecfreq}} - \curinventory_{i,t\rightarrow \tau} \leq 
         \begin{cases} 
      -\frac{\ln \left(1-\frac{\maxreward}{\maxreward + \penscalarTwo \inspecfreq + \penscalarTwo} \left(\frac{\maxreward + \penscalarTwo \inspecfreq + \penscalarTwo}{\maxreward + \penscalarTwo \inspecfreq} \right)^k -\epsilon\right)}{\ln \left(\penscalar \right)} & k \leq \inspecfreq \\
      -\frac{\ln \left(\Lambda \right) + \ln \left(\rho_{\frac{\maxreward}{\penscalarTwo\inspecfreq+\maxreward}}(k) \right)}{\ln \left(\penscalar \right)}  & k \geq \inspecfreq + 1
   \end{cases}
    \end{align*}
where $\ln (\Lambda) = \ln\left(1+ \left(\inspecfreq +\frac{\maxreward}{\penscalarTwo} \right) (1-\epsilon)
         -
         \frac{\maxreward}{\penscalarTwo} \left(\frac{\maxreward + \penscalarTwo \inspecfreq + \penscalarTwo}{\maxreward + \penscalarTwo \inspecfreq} \right)^{\inspecfreq}\right) - \ln\left(\frac{\inspecfreq + 1}{\frac{\penscalarTwo \inspecfreq}{\penscalarTwo\inspecfreq + \maxreward}}\rho_{\frac{\maxreward}{\penscalarTwo\inspecfreq+\maxreward}}(\inspecfreq + 1)\right)$, and $\epsilon = \frac{(\penscalarTwo \inspecfreq + \maxreward)\ln \left(\penscalar \right)}{\maxreward\inventory_i}$,
         and $\rho_{\frac{\maxreward}{\penscalarTwo\inspecfreq+\maxreward}}(k) = \prod\nolimits_{ \ell \in[k]}\left(1-\frac{\maxreward}{\ell (\maxreward+\inspecfreq\penscalarTwo
        )}\right)$.
        %
\end{lemma}
\begin{proofof}{Proof of \revcolor{\Cref{lem:inventory invariant real duration}}}
    This holds initially at time $0$ as we have $\alpha_{i,0 \rightarrow t} = 1$ for all $t$. Assume this holds true until arrival time of customer $j$ at $t_j$. We will prove this inequality holds true after the possible new assignment. The only interesting and non-trivial case is when the algorithm decides to assign job $j$ to server~$i$ and $t_j \leq \revcolor{\tau} < t_j + \durationij \leq \revcolor{\tau} + \frac{k}{\inspecfreq}$ which means $\maxreward \durationij - \sum\nolimits_{k'\in[0:\lceil \inspecfreq \durationij \rceil - 1]}
    \pen\left(\curinventory_{i,t_j\rightarrow t_j+\frac{k'}{\inspecfreq}}\right)
     > 0$ or: 
    \begin{align*}
        \frac{\maxreward \durationij}{\penscalarTwo}
         & > 
         \frac{1}{\penscalarTwo}\sum\nolimits_{k'\in[0:\lceil \inspecfreq \durationij \rceil - 1]}
    \pen\left(\curinventory_{i,t_j\rightarrow t_j+\frac{k'}{\inspecfreq}}\right)\\
         & = 
         \frac{1}{\penscalarTwo}\sum\nolimits_{k'\in[0:\inspecfreq (\tau-t_j)]}
    \pen\left(\curinventory_{i,t_j\rightarrow t_j+\frac{k'}{\inspecfreq}}\right) +
         \frac{1}{\penscalarTwo}\sum\nolimits_{k'\in[\lceil \inspecfreq \durationij \rceil + \inspecfreq (t_j - \tau) - 1]}
    \pen\left(\curinventory_{i,t_j\rightarrow \tau + \frac{k'}{\inspecfreq}} \right)\\
         & \geq 
         \left(1 + \inspecfreq (\revcolor{\tau}-t_j) \right) \left(\penscalar^{(1-\curinventory_{i,t_j\rightarrow \tau})} -1 \right) +
         \sum\nolimits_{k'\in[\lceil \inspecfreq \durationij \rceil + \inspecfreq (t_j - \revcolor{\tau}) - 1]}
         \left(\penscalar^{\left(1-\curinventory_{i,t_j\rightarrow \tau + \frac{k'}{\inspecfreq}}\right)}-1\right)
         \end{align*}
    Case (i): $t_j + \durationij \leq \revcolor{\tau}+ 1$:
         \begin{align*}
         & \overset{(a)}{\geq} - \inspecfreq \durationij + \left(1 + \inspecfreq (\revcolor{\tau}-t_j) \right) \penscalar^{(1-\curinventory_{i,t_j\rightarrow \tau})} +
         \sum\nolimits_{k'\in[\lceil \inspecfreq \durationij \rceil + \inspecfreq (t_j - \revcolor{\tau}) - 1]}
         \penscalar^{1-\curinventory_{i,t_j\rightarrow \tau}+\frac{\ln \left(1-\frac{\maxreward}{\maxreward + \penscalarTwo \inspecfreq + \penscalarTwo} \left(\frac{\maxreward + \penscalarTwo \inspecfreq + \penscalarTwo}{\maxreward + \penscalarTwo \inspecfreq} \right)^{k'} -\epsilon \right)}{\ln \left(\penscalar \right)}}
         \\
         & =  - \inspecfreq \durationij + \penscalar^{(1-\curinventory_{i,t_j\rightarrow \tau})} \left( 1 + \inspecfreq (\revcolor{\tau}-t_j)  +
         \sum\nolimits_{k'\in[\lceil \inspecfreq \durationij \rceil + \inspecfreq (t_j - \revcolor{\tau}) - 1]}
         \left(1-\frac{\maxreward}{\maxreward + \penscalarTwo \inspecfreq + \penscalarTwo} \left(\frac{\maxreward + \penscalarTwo \inspecfreq + \penscalarTwo}{\maxreward + \penscalarTwo \inspecfreq} \right)^{k'} -\epsilon \right) \right)\\
         & \geq  - \inspecfreq \durationij + \penscalar^{(1-\curinventory_{i,t_j\rightarrow \tau})} \left(\inspecfreq \durationij (1-\epsilon) - 
         \frac{\maxreward}{\maxreward + \penscalarTwo \inspecfreq + \penscalarTwo} \sum\nolimits_{k'\in[\lceil \inspecfreq \durationij \rceil + \inspecfreq (t_j - \revcolor{\tau}) - 1]}
         \left(\frac{\maxreward + \penscalarTwo \inspecfreq + \penscalarTwo}{\maxreward + \penscalarTwo \inspecfreq} \right)^{k'} \right)\\
         & \geq  - \inspecfreq \durationij  + \penscalar^{(1-\curinventory_{i,t_j\rightarrow \tau})} \left(\inspecfreq \durationij (1-\epsilon) - 
         \frac{\maxreward \durationij}{\maxreward + \penscalarTwo \inspecfreq + \penscalarTwo} \sum\nolimits_{k' \in [k - 1]}
         \left(\frac{\maxreward + \penscalarTwo \inspecfreq + \penscalarTwo}{\maxreward + \penscalarTwo \inspecfreq} \right)^{k'}  \right) \\
         & = - \inspecfreq \durationij + \penscalar^{(1-\curinventory_{i,t_j\rightarrow \tau})} \left(\inspecfreq \durationij (1-\epsilon) - 
         \frac{\maxreward \durationij}{\maxreward + \penscalarTwo \inspecfreq + \penscalarTwo} \frac{\left(\frac{\maxreward + \penscalarTwo \inspecfreq + \penscalarTwo}{\maxreward + \penscalarTwo \inspecfreq} \right)^{k}-\frac{\maxreward + \penscalarTwo \inspecfreq + \penscalarTwo}{\maxreward + \penscalarTwo \inspecfreq}}{\frac{\penscalarTwo}{\maxreward + \penscalarTwo \inspecfreq}} \right) \\
         & = - \inspecfreq \durationij + \penscalar^{(1-\curinventory_{i,t_j\rightarrow \tau})} \left(\left(\inspecfreq+\frac{\maxreward}{\penscalarTwo}\right) \durationij -  \inspecfreq \durationij \epsilon - 
         \frac{\maxreward \durationij}{\maxreward + \penscalarTwo \inspecfreq + \penscalarTwo} \frac{\left(\frac{\maxreward + \penscalarTwo \inspecfreq + \penscalarTwo}{\maxreward + \penscalarTwo \inspecfreq} \right)^{k}}{\frac{\penscalarTwo}{\maxreward + \penscalarTwo \inspecfreq}} \right) \\
    \end{align*}
    \revcolor{where inequality~(a) holds due to the induction hypothesis.}
    Hence:
    \begin{align*}
         &1 \geq \penscalar^{(1-\curinventory_{i,t_j\rightarrow \tau})} 
         \left(1 - \frac{\penscalarTwo \inspecfreq}{\penscalarTwo \inspecfreq + \maxreward}\epsilon - 
         \frac{\maxreward}{\maxreward + \penscalarTwo \inspecfreq + \penscalarTwo} \left(\frac{\maxreward + \penscalarTwo \inspecfreq + \penscalarTwo}{\maxreward + \penscalarTwo \inspecfreq} \right)^{k}\right)
         \\
         \iff& -\frac{\ln \left(1 - \frac{\penscalarTwo \inspecfreq}{\penscalarTwo \inspecfreq + \maxreward}\epsilon - 
         \frac{\maxreward}{\maxreward + \penscalarTwo \inspecfreq + \penscalarTwo} \left(\frac{\maxreward + \penscalarTwo \inspecfreq + \penscalarTwo}{\maxreward + \penscalarTwo \inspecfreq} \right)^{k} \right)}{\ln \left(\penscalar \right)} \geq 1-\estimateinventoryij
    \end{align*} 
    On the other hand:
    \begin{align*}
         - \frac{\ln \left(1 - \epsilon - 
         \frac{\maxreward}{\maxreward + \penscalarTwo \inspecfreq + \penscalarTwo} \left(\frac{\maxreward + \penscalarTwo \inspecfreq + \penscalarTwo}{\maxreward + \penscalarTwo \inspecfreq} \right)^{k} \right)}{\ln \left(\penscalar \right)} & \geq 
         - \frac{\ln \left(1 - \frac{\penscalarTwo \inspecfreq}{\penscalarTwo \inspecfreq + \maxreward}\epsilon - 
         \frac{\maxreward}{\maxreward + \penscalarTwo \inspecfreq + \penscalarTwo} \left(\frac{\maxreward + \penscalarTwo \inspecfreq + \penscalarTwo}{\maxreward + \penscalarTwo \inspecfreq} \right)^{k} \right) - \frac{\maxreward}{\penscalarTwo \inspecfreq + \maxreward}\epsilon}{\ln \left(\penscalar \right)} \\
         & \geq 1-\estimateinventoryij + \frac{1}{\inventory_i}
    \end{align*}
    Case (ii): $t_j + \durationij > \revcolor{\tau}+ 1$:
    \begin{align*}
         \overset{(a)}{\geq} - \inspecfreq \durationij + \left(1 + \inspecfreq (\revcolor{\tau}-t_j) \right) \penscalar^{(1-\curinventory_{i,t_j\rightarrow \tau})}
         & + \sum\nolimits_{k' \in [\inspecfreq]}
         \penscalar^{1-\curinventory_{i,t_j\rightarrow \tau}+\frac{\ln \left(1-\frac{\maxreward}{\maxreward + \penscalarTwo \inspecfreq + \penscalarTwo} \left(\frac{\maxreward + \penscalarTwo \inspecfreq + \penscalarTwo}{\maxreward + \penscalarTwo \inspecfreq} \right)^{k'} -\epsilon \right)}{\ln \left(\penscalar \right)}} \\
         & + \sum\nolimits_{k' \in [\inspecfreq + 1:\lceil \inspecfreq \durationij \rceil + \inspecfreq (t_j - \revcolor{\tau}) - 1]}
         \penscalar^{1-\curinventory_{i,t_j\rightarrow \tau}+\frac{\ln \left(\Lambda \right) + \ln \left(\rho_{\frac{\maxreward}{\penscalarTwo\inspecfreq+\maxreward}}(k')\right)}{\ln \left(\penscalar \right)}}
    \end{align*}
    \begin{align*}
         \geq &- \inspecfreq \durationij + \penscalar^{(1-\curinventory_{i,t_j\rightarrow \tau})} \left(1+\inspecfreq  \left( 1 -\epsilon \right)
         -
         \frac{\maxreward}{\maxreward + \penscalarTwo \inspecfreq + \penscalarTwo} {\left(\left(\frac{\maxreward + \penscalarTwo \inspecfreq + \penscalarTwo}{\maxreward + \penscalarTwo \inspecfreq} \right)^{\inspecfreq + 1}-\frac{\maxreward + \penscalarTwo \inspecfreq + \penscalarTwo}{\maxreward + \penscalarTwo \inspecfreq}\right)}{\frac{\maxreward + \penscalarTwo \inspecfreq}{\penscalarTwo}}\right)
         \\
         &\qquad\quad+\penscalar^{(1-\curinventory_{i,t_j\rightarrow \tau})} \left( \Lambda\sum\nolimits_{k' \in [\inspecfreq + 1:\lceil \inspecfreq \durationij \rceil - 1]}
         \rho_{\frac{\maxreward}{\penscalarTwo\inspecfreq+\maxreward}}(k') \right)\\
         = &- \inspecfreq \durationij + \penscalar^{(1-\curinventory_{i,t_j\rightarrow \tau})} \left(1+\inspecfreq  \left( 1-\epsilon \right)
         +\frac{\maxreward}{\penscalarTwo} -
         \frac{\maxreward}{\penscalarTwo} \left(\frac{\maxreward + \penscalarTwo \inspecfreq + \penscalarTwo}{\maxreward + \penscalarTwo \inspecfreq} \right)^{\inspecfreq}
         + \Lambda\sum\nolimits_{k' \in [\inspecfreq + 1:\lceil \inspecfreq \durationij \rceil - 1]}
         \rho_{\frac{\maxreward}{\penscalarTwo\inspecfreq+\maxreward}}(k') \right)\\
         \overset{(b)}{=} &- \inspecfreq \durationij + \penscalar^{(1-\curinventory_{i,t_j\rightarrow \tau})} \left(1+ \inspecfreq (1-\epsilon)
         +\frac{\maxreward}{\penscalarTwo} -
         \frac{\maxreward}{\penscalarTwo} \left(\frac{\maxreward + \penscalarTwo \inspecfreq + \penscalarTwo}{\maxreward + \penscalarTwo \inspecfreq} \right)^{\inspecfreq} 
         \right)
         \\
         &\qquad\quad+ \penscalar^{(1-\curinventory_{i,t_j\rightarrow \tau})}\Lambda
         \left(\frac{\lceil \inspecfreq \durationij \rceil}{\frac{\penscalarTwo \inspecfreq}{\penscalarTwo\inspecfreq + \maxreward}}\rho_{\frac{\maxreward}{\penscalarTwo\inspecfreq+\maxreward}}(\lceil \inspecfreq \durationij \rceil) - \frac{\inspecfreq + 1}{\frac{\penscalarTwo \inspecfreq}{\penscalarTwo\inspecfreq + \maxreward}}\rho_{\frac{\maxreward}{\penscalarTwo\inspecfreq+\maxreward}}(\inspecfreq + 1) \right) \\
         \geq & - \inspecfreq \durationij + \penscalar^{(1-\curinventory_{i,t_j\rightarrow \tau})}\left( \Lambda
         \left(\frac{\durationij}{\frac{\penscalarTwo}{\penscalarTwo\inspecfreq + \maxreward}}\rho_{\frac{\maxreward}{\penscalarTwo\inspecfreq+\maxreward}}(k) \right) +\frac{\maxreward}{\penscalarTwo}\epsilon \right)
    \end{align*}
    which implies:
    \begin{align*}
        1 \geq \penscalar^{(1-\curinventory_{i,t_j\rightarrow \tau})}\left( \Lambda
         \rho_{\frac{\maxreward}{\penscalarTwo\inspecfreq+\maxreward}}(\lceil \inspecfreq \durationij \rceil) +\frac{\maxreward}{\penscalarTwo\inspecfreq+\maxreward}\epsilon \right) \iff -\frac{\ln \left(\Lambda
         \rho_{\frac{\maxreward}{\penscalarTwo\inspecfreq+\maxreward}}(\lceil \inspecfreq \durationij \rceil) +\frac{\maxreward}{\penscalarTwo\inspecfreq+\maxreward}\epsilon\right)}{\ln (\penscalar)}\geq 1-\curinventory_{i,t_j\rightarrow \tau}
    \end{align*}
    Finally we get:
    \begin{align*}
         -\frac{\ln \left(\Lambda \right) +\ln \left(
         \rho_{\frac{\maxreward}{\penscalarTwo\inspecfreq+\maxreward}}(\lceil \inspecfreq \durationij \rceil) \right)}{\ln (\penscalar)} 
         \geq 
         -\frac{\ln \left(\Lambda
         \rho_{\frac{\maxreward}{\penscalarTwo\inspecfreq+\maxreward}}(\lceil \inspecfreq \durationij \rceil) +\frac{\maxreward}{\penscalarTwo\inspecfreq+\maxreward}\epsilon\right)-\frac{\maxreward}{\penscalarTwo \inspecfreq + \maxreward}\epsilon}{\ln (\penscalar)}\geq 1-\curinventory_{i,t_j\rightarrow \tau} + \frac{1}{\inventory_i}
    \end{align*}
\revcolor{Both} inequalities (a) holds due to the induction hypothesis and equality (b) holds due to \Cref{lem: rho lemma}
\end{proofof}

\subsection{Primal-Dual Analysis of the Competitive Ratio}
\label{apx:FLB primal dual competitive ratio}

We now present the following upper bound of capacity feasible {\FLB}'s competitive ratio.

\FLBCompetitiveRatioReal*

\begin{proofof}{Proof of \Cref{lem:FLB competitive ratio real duration}}
In this proof, we upper bound the competitive ratio of capacity feasible {\FLB} 
with inspection-frequency scalar $\inspecfreq \in (0, 1]$ and penalty parameters $\penscalarTwo > 0, \penscalar \geq e$. To simplify the notation, we rewrite the inspection time subset 
$\timesetij(\inspecfreq)$ as $\timesetij$.

Recall the dual program
of the configuration LP~\ref{eq:opt lp},
\begin{align*}
    \arraycolsep=5.4pt\def\arraystretch{2}
    &\begin{array}{llllll}
    &
    &
    &
    \min\limits_{\boldsymbol{\probdual,\inventorydual}\geq \mathbf 0} &
    \displaystyle\sum\nolimits_{j\in[m]}
    \onlinedual(j)
    +
    \displaystyle\sum\nolimits_{i\in[n]}
    \inventory_i
    \offlinedual(i)
    & \text{s.t.} \\
    & 
     & 
     &
     &
     \displaystyle\sum\nolimits_{j\in\config}
     \onlinedual(j)
     +
     \offlinedual(i)
     \geq 
     \displaystyle\sum\nolimits_{j\in\config}
     \rewardij \durationij
     &
     i\in[n],\config\in\configspace_i
    \end{array}
\end{align*}
We construct a dual solution based on 
the assignment decision made in FLB as follows.
First, set $\onlinedual(j) \gets 0$ and $\offlinedual(i) \gets 0$
for every job $j\in[m]$ and server $i\in[n]$.
Now consider every assignment of FLB.
For each job $j\in[m]$,
if FLB assigns
    server $i$ to job $j$, 
    update the dual variables as follows:
\begin{align*}
    \onlinedual(j) &\gets
    \rewardij\durationij
    -
    \sum\nolimits_{\tpre \in \timesetij}
 	\pen\left(
 	\estimateinventoryi
 	\right)
\\
    \offlinedual(i) &\gets
    \offlinedual(i) 
    +
    \sum\nolimits_{\tpre \in \timesetij}
 	\left(
    \pen\left(
 	\estimateinventoryij- \frac{1}{\inventory_i}
 	\right)
  -
    \pen\left(
 	\estimateinventoryi
 	\right)
    \right)
\end{align*}
Note the dual solution construction is well-defined.
In particular,
since {\FLB} is capacity feasible,
when FLB assigns server $i$ to job $j$, 
there exists at least one available unit of server $i$.
Namely, $\curinventory_{i, t\rightarrow t} \geq \frac{1}{\inventory_i}$.
Moreover, for every $\tpre \in \timesetij$, 
we have
$\estimateinventoryij\geq \curinventory_{i, t\rightarrow t} \geq \frac{1}{\inventory_i}$,
since $\estimateinventoryij$ is increasing in $\tpre$ and $\tpre \geq t_j$.

The rest of the proof is done in two steps.

\smallskip
\noindent[\emph{Step i}]
\emph{Comparing objective values in primal and dual.}
Here we show that the total reward of FLB is a 
$\ln(\penscalar)(1 + \inspecfreq\penscalarTwo
(1+\penscalar)(\penscalar^{\frac{1}{\mininventory}}-1))$-approximation of 
the objective value of the constructed dual solution.
To show this, we analyze 
the increment of the reward in FLB
and the increment of the objective value of the dual solution
due to every assignment decision in FLB separately.

Suppose FLB assigns server $i$ to job $j$.
The increment of the total reward in FLB is 
\begin{align*}
    \Delta(\text{Primal}) = \rewardij \durationij
\end{align*}
and the increment of the objective value of the constructed dual solution
can be upper bounded as follows,
\begin{align*}
    \Delta(\text{Dual}) 
    &=
    \rewardij \durationij
    -
    \sum\nolimits_{\tpre \in \timesetij}
 	\pen\left(
 	{\estimateinventoryij}
 	\right)
  +
  \sum\nolimits_{\tpre \in \timesetij}
  \inventory_i
  \left(
    \pen\left(
    \estimateinventoryij- 
 	\frac{1}{\inventory_i}
 	\right)
  -
    \pen\left(
 	{\estimateinventoryij}
  \right)
  \right)
  \\
  &\overset{(a)}{\leq}
  \ln\left(\penscalar\right)
  \left(
  \rewardij \durationij
    -
    \sum\nolimits_{\tpre \in \timesetij}
 	\pen\left(
 	{\estimateinventoryij}
 	\right)
  \right)
  -
  \sum\nolimits_{\tpre \in \timesetij}
  \inventory_i
  \left(
    \pen\left(
 	{\estimateinventoryij}
  \right)
    -
    \pen\left(
    \estimateinventoryij- 
 	\frac{1}{\inventory_i}
 	\right)
  \right)
  \\
  &=
  \rewardij \durationij
  \ln\left(\penscalar\right)
  -
  \sum\nolimits_{\tpre \in \timesetij}
  \left(
  \ln\left(\penscalar\right)
  \pen\left(
 	{\estimateinventoryij}
 	\right)
  +
  \inventory_i
  \left(
    \pen\left(
 	{\estimateinventoryij}
  \right)
    -
    \pen\left(
    \estimateinventoryij- 
 	\frac{1}{\inventory_i}
 	\right)
  \right)
  \right)
  \\
  &\overset{(b)}{\leq}
  \rewardij \durationij
  \ln\left(\penscalar\right)
  -
  \sum\nolimits_{\tpre \in \timesetij}
  \left(
  \ln\left(\penscalar\right)
  \pen\left(
 	{\estimateinventoryij}
 	\right)
  +
    \pen'\left(
    \estimateinventoryij- 
 	\frac{1}{\inventory_i}
 	\right)
  \right)
  \\
  &\overset{(c)}{\leq}
  \rewardij \durationij
  \ln\left(\penscalar\right)
  +
  \sum\nolimits_{\tpre \in \timesetij}
 	\penscalarTwo
      \ln\left(\penscalar\right) 
  \left(
  1 + 
  \penscalar
    \left(
    \penscalar^{\frac{1}{\mininventory}
    }
    - 1
    \right)
  \right)
  \\
  &\overset{}{=}
  \rewardij \durationij
  \ln\left(\penscalar\right)
  +
  |\timesetij|
 	\penscalarTwo 
      \ln\left(\penscalar\right) 
  \left(
  1 + 
  \penscalar
    \left(
    \penscalar^{\frac{1}{\mininventory}
    }
    - 1
    \right)
  \right)
  \\
  &\overset{(d)}{\leq}
  \rewardij \durationij
  \ln\left(\penscalar\right)
  +
  \rewardij
  \inspecfreq\durationij\penscalarTwo
      \ln\left(\penscalar\right) 
  \left(
  1 + \penscalar
    \left(
    \penscalar^{\frac{1}{\mininventory}
    }
    - 1
    \right)
  \right)
  \\
  &=  \Delta(\text{Primal})\cdot 
  \ln\left(\penscalar\right)
  \left(1 + \inspecfreq\penscalarTwo
  \left(1+\penscalar
  \left(\penscalar^{\frac{1}{\mininventory}}-1\right)\right)\right)
\end{align*}
In the above derivation,
inequality~(a) holds since
$\ln(\penscalar) \geq 1$ for every $\penscalar \geq e$
and server~$i$'s reduced reward 
$\rewardij \durationij
    -
    \sum\nolimits_{\tpre \in \timesetij}
 	\pen\left(
 	{\estimateinventoryij}
 	\right) > 0$
  implied by the assumption that
  FLB assigns 
  server $i$ to job $j$.
  Inequality~(b) holds since penalty function $\pen$
  is convex and thus 
  $
  \inventory_i
  (
    \pen(
 	{\estimateinventoryij}
  )
    -
    \pen(
    \estimateinventoryij- 
 	\frac{1}{\inventory_i}
 	)
  )
  \geq 
  \inventory_i(\estimateinventoryij- (\estimateinventoryij- \frac{1}{\inventory_i})
  )
    \pen'(
    \estimateinventoryij- 
 	\frac{1}{\inventory_i}
 	))
  =
    \pen'(
    \estimateinventoryij- 
 	\frac{1}{\inventory_i}
 	))$.
Inequality~(c) holds by algebra as follows,
\begin{align*}
    &\ln\left(\penscalar\right) 
 	\pen\left(
 	{\estimateinventoryij}
 	\right)
    +
    \pen'\left(
    \estimateinventoryij-
    \frac{ 1}{\inventory_i}
    \right)
    \\
    =&
    \left(\ln\left(\penscalar\right) 
 	\pen\left(
 	{\estimateinventoryij}
 	\right)
    +
    \pen'\left(
    {\estimateinventoryij}{}
    \right)
    \right)
    +
    \left(
    \pen'\left(
    \estimateinventoryij-
    \frac{ 1}{\inventory_i}
    \right)
    -
    \pen'\left(
    {\estimateinventoryij}{}
    \right)
    \right)
    \\
    =&
 	-\penscalarTwo
      \ln\left(\penscalar\right)
  -
 	\penscalarTwo
  \ln\left(\penscalar\right) 
  \penscalar^{
  \left(1 - \frac{\estimateinventoryij}{\inventory_i}\right)}
    \left(
    \left(\penscalar\right)^{\frac{1}{\inventory_i}
    }
    - 1
    \right)
    \\
    \geq &
 	-\penscalarTwo
      \ln\left(\penscalar\right) 
  \left(
  1 + 
  \penscalar
    \left(
    \penscalar^{\frac{1}{\mininventory}
    }
    - 1
    \right)
  \right)
\end{align*}
Finally, inequality~(d) holds since 
term $|\timesetij|=\lfloor\inspecfreq\durationij\rfloor\leq \inspecfreq\durationij$
and $\rewardij \geq 1$

\smallskip
\noindent[\emph{Step ii}]
\emph{Check the approximate feasibility of dual.}
Now we show that the constructed dual solution is 
$(\frac{\inspecfreq}{\inspecfreq-1})$-approximately
satisfied,
i.e.,
\begin{align*}
    \frac{\inspecfreq}{\inspecfreq-1}
    \left(\sum\nolimits_{j\in\config}\onlinedual(j) + \offlinedual(i)\right) 
    \geq \sum\nolimits_{j\in\config}\rewardij\durationij
\end{align*}
Fix an arbitrary
server $i\in[n]$ and
configuration $\config\in\config_i$.
Note the construction of FLB and the 
dual solution construction guarantee that
$\onlinedual(j)$ 
is weakly larger than the reduced reward of server $i$
for every time point $j\in\config$
(i.e., $\onlinedual(j) \geq \rewardij \durationij -
    \sum\nolimits_{\tpre \in \timesetij}
    \pen\left(\estimateinventoryij\right)$);
    and $\onlinedual(j) \geq 0$.
Thus, 
the dual constraint associated with primal variable $\allocconfig$ 
is
$(\frac{\inspecfreq}{\inspecfreq-1})$-approximately
satisfied if the following inequality holds.
\begin{align*}
    \frac{\inspecfreq}{\inspecfreq-1}
    \offlinedual(i) 
    -
    \sum\nolimits_{j\in{\config}}
    \sum\nolimits_{\tpre\in\timesetij}
 	\pen\left(
 	{\estimateinventoryij}{}
 	\right) 
    \geq 0
\end{align*}
We prove this inequality
with a \emph{charging argument}.
At a high level, 
our goal is to identify values $\{\offlinedualdecompi\}_{j\in\config,\tpre \in \timesetij}$
such that 
\begin{align}
\label{eq:decomposed offline dual general per time feasibility}
    &\forall j\in\config, \tpre\in\timesetij:~\offlinedualdecompi -\pen(\estimateinventoryij) \geq 0
    \\
\label{eq:decomposed offline dual general summation}
    &\sum\nolimits_{j\in\config}\sum\nolimits_{\tpre\in\timesetij}\offlinedualdecompi \leq 
     \frac{\inspecfreq}{\inspecfreq-1}\offlinedual(i)
\end{align}
The actual charging argument is as follows.

With slight abuse of notation, 
we introduce an auxiliary notation 
$\timesetiS \triangleq
\{(j,\tpre): j\in \config, \tpre \in \timesetij\}$
for every server $i\in[n]$ and 
configuration $\config\in \configspace_i$.
Recall that for every configuration $\config\in\configspace_i$,
the durations of two different jobs $j_1, j_2\in\config$
do not overlap and thus
$\timeset_{ij_1} \cap \timeset_{ij_2} = \emptyset$.
Therefore,
for every two $(j_1,\tpre_1), (j_2,\tpre_2) \in \timesetiS$,
if $\tpre_1 \not = \tpre_2$, then $j_1 \not = j_2$.

We further partition set $\timesetiS$
into $\timesetiS^{(a)}\sqcup \timesetiS^{(b)}$ as follows,
\begin{align*}
    \timesetiS^{(a)} &\triangleq \timesetiS \backslash \timesetiS^{(a)}
    \\
    \timesetiS^{(b)} &\triangleq
    \left\{
    (j,\tpre) \in \timesetiS:
    \tpre = \max \timesetij \text{ and }
    |\timesetij| \geq 2
    \right\}
\end{align*}
In words, set $\timesetiS^{(b)}$ contains all pairs of $(j,\tpre)$
if $\tpre$ is the last inspection time from $\timesetij$
and $|\timesetij| \geq 2$.
As a sanity check, for every $j\in\config$, there exists at most one $\tpre$ such that $(j,\tpre) \in \timeset^{(b)}(i, t)$.

We construct a one-to-many \emph{correspondence} $\chargemapping$
mapping from $\timesetiS^{(a)}$ 
to $[m]\times [T]$ that satisfies the following three properties.
\begin{itemize}
    \item[-] {\cseparation:}
    for every two different $(j_1,\tpre_1), (j_2,\tpre_2)\in \timesetiS$,
     $\chargemapping(j_1,\tpre_1) \cap \chargemapping(j_2,\tpre_2) = \emptyset$.
    
    \item[-] {{\cfeasibilitya:}}
    for every $(j,\tpre)\in\timesetiS$, 
    $|\chargemapping(j,\tpre)| = \inventory_i - \inventory_i\estimateinventoryij$.
    Moreover, for every two different $(j_1,\tpre_1), (j_2,\tpre_2) \in \chargemapping(j,\tpre)$,
    $j_1 \not=j_2$.
    
    \item[-] {{\cfeasibilityb:}}
    for every $(j,\tpre)\in\timesetiS$, 
    and every $(j',\tpre') \in \chargemapping(j,\tpre)$,
    $t' < t$,
    $\tpre' \leq \tpre$,
    $\tpre'\in\timeset_{ij'}$,
    and 
    a unit of server $i$ is assigned to 
    job $j'$
    with   duration $\duration_{ij'} > \tpre - t_{j'}$.
\end{itemize}
Before showing the existence of the desired 
correspondence $\chargemapping$,
we first illustrate how to construct 
values $\{\offlinedualdecompi\}_{(j,\tpre) \in \timesetiS}$
given correspondence $\chargemapping$.
We first construct $\offlinedualdecompi$
for every $(j,\tpre) \in \timesetiS^{(a)}$.
Specifically, for every $(j,\tpre) \in \timesetiS^{(a)}$, define
\begin{align*}
    \offlinedualdecompi \triangleq 
    \sum\nolimits_{(j',\tpre')\in\chargemapping(j,\tpre)} 
    \left(
    \pen\left(
    \curinventory_{i, t_{j'}\rightarrow\tpre'}
    - \frac{1}{\inventory_i}
    \right)
    -
    \pen\left(
    \curinventory_{i, t_{j'}\rightarrow\tpre'}
    \right)
    \right)
\end{align*}
Note that property~\cseparation\ and property~\cfeasibilityb\ of 
correspondence $\chargemapping$ guarantee 
that
\begin{align}
\label{eq:decomposed offline dual general summation set a}
    \sum\nolimits_{(j,\tpre)\in\timesetiS^{(a)}}
    \offlinedualdecompi \leq \offlinedual(i)
\end{align}
To argue condition~\eqref{eq:decomposed offline dual general per time feasibility}
for every $(j,\tpre) \in \timesetiS^{(a)}$,
note that
\begin{align*}
    \offlinedualdecompi-
    \pen(\estimateinventoryij)
    &=
    \sum\nolimits_{(j',\tpre')\in\chargemapping(j,\tpre)} 
    \left(
    \pen\left(
    \curinventory_{i, t_{j'}\rightarrow\tpre'}
    - \frac{1}{\inventory_i}
    \right)
    -
    \pen\left(
    \curinventory_{i, t_{j'}\rightarrow\tpre'}
    \right)
    \right)
    -
    \pen(\estimateinventoryij)
    \\
    &\overset{(a)}{\geq}
    \sum\nolimits_{(j',\tpre')\in\chargemapping(j,\tpre)} 
    \left(
    \pen\left(
    \curinventory_{i, t_{j'}\rightarrow\tpre}
    - \frac{1}{\inventory_i}
    \right)
    -
    \pen\left(
    \curinventory_{i, t_{j'}\rightarrow\tpre}
    \right)
    \right)
    -
    \pen(\estimateinventoryij)
    \\
    &\overset{(b)}{\geq}
    \sum\nolimits_{\ell \in [\inventory_i - \inventory_i\estimateinventoryij]}
    \left(
    \pen\left(
    \frac{\inventory_i - \ell}{\inventory_i}
    \right)
    -
    \pen\left(
    \frac{\inventory_i - \ell+1}{\inventory_i}
    \right)
    \right)
    -
    \pen\left(\estimateinventoryij\right)
    \\
    &\overset{(c)}{=}
    -\pen(1)\overset{(d)}{=}
    0
\end{align*}
In the above derivation, 
inequality~(a) holds since {\estimated} capacity level $\curinventory_{i,t_{j'}\rightarrow\tpre'}$
is weakly increasing in $\tpre'$,
and property~\cfeasibilityb\ of correspondence $\chargemapping$
ensures $\tpre' \leq \tpre$,
and penalty function $\pen$ is convex.
For inequality~(b), note 
property~\cfeasibilitya\ and property~\cfeasibilityb\ of correspondence $\chargemapping$
guarantee the fact that there exists $s \triangleq \inventory_i - \inventory_i\estimateinventoryij$
jobs $j_1 < j_2 <  \dots <  j_s < j$,
each of which is assigned 
a different unit of server $i$
and holds this unit at time point $\tpre$.
This fact further implies  that 
for each $\ell\in[s]$,
$\curinventory_{i,t_{j_\ell}\rightarrow\tpre} \leq \frac{\inventory_i - \ell + 1}{\inventory_i}$.
Invoking the convexity of penalty function $\pen$
completes the argument for inequality~(b).
Finally,
inequality~(c) holds by algebra
and inequality~(d) holds since $\pen(1) = 0$.

Next we construct $\offlinedualdecompi$
for every $(j,\tpre) \in \timesetiS^{(b)}$
with $\{\offlinedualdecompi\}_{(j,\tpre)\in\timesetiS^{(a)}}$ constructed above.
Specifically, for every $(j,\tpre) \in \timesetiS^{(b)}$, define
\begin{align*}
    \offlinedualdecompi \triangleq
    \sum\nolimits_{\tpre' \in \timesetij \backslash \{\tpre\}}
    \frac{
    \offlinedual_{(j, \tpre')}(i)}
    {|\timesetij \backslash \{\tpre\}|}
\end{align*}
Note the above construction is well-defined. 
In particular, $|\timesetij\backslash\{\tpre\}| \geq 1$ due 
to the definition of set $\timesetiS^{(b)}$.

To check the constructed values $\{\offlinedualdecompi\}_{(j,\tpre)\in \timeset}$
satisfies condition~\eqref{eq:decomposed offline dual general summation},
note that
\begin{align*}
\begin{split}
    \sum\nolimits_{(j,\tpre)\in\timesetiS}
    \offlinedualdecompi
    &=
    \sum\nolimits_{(j,\tpre)\in\timesetiS^{(a)}}
    \offlinedualdecompi
    +
    \sum\nolimits_{(j,\tpre)\in\timesetiS^{(b)}}
    \offlinedualdecompi
    \\
    &=
    \sum\nolimits_{(j,\tpre)\in\timesetiS^{(a)}}
    \offlinedualdecompi
    +
    \sum\nolimits_{(j,\tpre) \in \timesetiS^{(b)}}
    \sum\nolimits_{\tpre' \in \timesetij \backslash \{\tpre\}}
    \frac{
    \offlinedual_{(j, \tpre')}(i)}
    {|\timesetij \backslash \{\tpre\}|}
    \\
    &\overset{(a)}{\leq}
    \sum\nolimits_{(j,\tpre)\in\timesetiS^{(a)}}
    \offlinedualdecompi
    +
    \frac{1}{\inspecfreq - 1}
    \sum\nolimits_{(j,\tpre) \in \timesetiS^{(b)}}
    \sum\nolimits_{\tpre' \in \timesetij \backslash \{\tpre\}}
    \offlinedual_{(j, \tpre')}(i)
    \\
    &\overset{(b)}{\leq}
    \frac{\inspecfreq}{\inspecfreq-1} 
    \sum\nolimits_{(j,\tpre) \in \timesetiS^{(a)}} 
    \offlinedualdecompi
    \\
    &\overset{(c)}{\leq}
    \frac{\inspecfreq}{\inspecfreq-1}
    \offlinedual(i)
\end{split}
\end{align*}
where inequality~(a) holds since $|\timesetij\backslash\{\tpre\}| \geq \inspecfreq - 1$ due to the definition of 
inspection time subset $\timesetij$ parameterized by inspection-frequency scalar $\inspecfreq$;
inequality~(b) holds since for every $j\in\config$
there exists at most one $\tpre$ such that $(j,\tpre) \in \timeset^{(b)}(i, t)$;
and inequality~(c) holds due to inequality~\eqref{eq:decomposed offline dual general summation set a}.

To argue condition~\eqref{eq:decomposed offline dual general per time feasibility}
for every $(j,\tpre)\in \timesetiS^{(b)}$, note that
\begin{align*}
\offlinedualdecompi-
    \pen(\estimateinventoryij)
    &
    \overset{(a)}{\geq}
    \sum\nolimits_{\tpre' \in \timesetij \backslash \{\tpre\}}
    \frac{
    \offlinedual_{(j, \tpre')}(i)}
    {|\timesetij \backslash \{\tpre\}|}
    -
    \sum\nolimits_{\tpre' \in \timesetij \backslash \{\tpre\}}
    \frac{
    \pen(\curinventory_{i, t_j\rightarrow\tpre'})
    }
    {|\timesetij \backslash \{\tpre\}|}
    \\
    &=
    \sum\nolimits_{\tpre' \in \timesetij \backslash \{\tpre\}}
    \frac{
    \offlinedual_{(j, \tpre')}(i)
    -
    \pen(\curinventory_{i, t_j\rightarrow\tpre'})
    }
    {|\timesetij \backslash \{\tpre\}|}
    \overset{(b)}{\geq} 0
\end{align*}
where inequality~(a) holds due to the fact $\pen(\estimateinventoryij)\leq \pen(\curinventory_{i, t\rightarrow\tpre'})$
implied by $\tpre' < \tpre$
and the definition of $\offlinedualdecompi$
for $(j,\tpre)\in \timesetiS^{(b)}$;
and inequality~(b) holds since condition~\eqref{eq:decomposed offline dual general per time feasibility}
is shown to be satisfied for $(j, \tpre') \in \timesetiS^{(a)}$.

Now we show the existence of correspondence $\chargemapping$
with properties~\cseparation, \cfeasibilitya\ and \cfeasibilityb\
by the following explicit construction of $\chargemapping$.
Fix an arbitrary $(j,\tpre)\in\timesetiS$.
The definition of {\estimated} capacity level $\estimateinventoryij$ implies that
there are $s \triangleq \inventory_i - \inventory_i\estimateinventoryij$
jobs $j_1 < j_2 <  \dots <  j_s < j$,
each of which is assigned 
a different unit of server $i$
and holds this unit at time point $\tpre$.
Namely, for every $\ell\in[s]$, 
$t_{j_\ell} + \duration_{ij_\ell} \geq \tpre$.
Moreover, there exists unique $\tpre_\ell\in\timeset_{ij_\ell}$
such that $\tpre_\ell \leq \tpre < \tpre_\ell + \frac{1}{\inspecfreq}$.
By setting $\chargemapping(j,\tpre) \triangleq 
\{(j_\ell, \tpre_\ell)\}_{\ell\in[m]}$,
properties \cfeasibilitya\ and \cfeasibilityb\
are satisfied straightforwardly.
To show property \cseparation, 
fix arbitrary two different $(j',\tpre'), (j'', \tpre'') \in \timesetiS^{(a)}$ such that $j' \leq j''$.
It is sufficient to argue 
    $|\tpre'' - \tpre'| \geq \frac{1}{\inspecfreq}$.
We consider two cases separately.
If $j' = j''$, for $\tpre' \not = \tpre''$,
the construction of inspection time subset $\timeset_{ij'}$ ensures
$|\tpre'' - \tpre'| \geq \frac{1}{\inspecfreq}$ as desired.
If $j' < j''$, since both $j', j''\in \config$,
inspection time subsets
$\timeset_{ij'}$ and $\timeset_{ij''}$
do not overlap, i.e., $t_{j'} + \duration_{ij'} \leq t_{j''}$,
and thus 
$\tpre' \leq t_{j'} + \duration_{ij'} \leq t_{j''}
 \leq \tpre''$.
Moreover, the construction of $\timesetiS^{(a)}$
ensures that
$\tpre' 
\leq t_{j'} + \duration_{ij'} - \frac{1}{\inspecfreq}$.
Hence,
$|\tpre'' - \tpre'| = \tpre'' - \tpre'
\geq \frac{1}{\inspecfreq}$ as desired.

\smallskip
\noindent
Finally, 
since 
the total reward of FLB is 
$\ln(\penscalar)(1 + \inspecfreq\penscalarTwo
(1+\penscalar)(\penscalar^{\frac{1}{\mininventory}}-1))$-approximation
of the objective value of the constructed dual solution (Step i)
and
the constructed dual solution is $(\frac{\inspecfreq}{\inspecfreq-1})$-approximately feasible (Step ii),
invoking
the LP weak duality concludes the proof.
\end{proofof}

\subsection{Proof of Theorem~\ref{thm:FLB competitive ratio general}}
\label{apx:FLB parameter optimize}

Similar to the discussion in \Cref{sec:competitive ratio analysis}, we can formulate the task of optimizing parameters $(\inspecfreq,\penscalarTwo,\penscalar)$ of {\FLB} for its competitive ratio as the following program:
\begin{align}
\tag{$\FLBOPTReal{\maxreward,\maxduration,\mininventory}$}
\label{eq:FLB parameter real duration}
    &\begin{array}{lll}
    \min\nolimits_{\inspecfreq,\penscalarTwo,\penscalar}~ &
        \frac{\inspecfreq}{\inspecfreq - 1}\cdot \ln(\penscalar)\cdot \left(1 + \inspecfreq\penscalarTwo\left(1 + \penscalar\left(\penscalar^{\frac{1}{\mininventory}}-1\right)\right)\right)
    &
    \text{s.t.}
    \vspace{5pt}
    \\
    &\ln(\penscalar) \geq - \ln\left(\prod\nolimits_{k \in[\lceil\inspecfreq\maxduration\rceil]}\left(1-\frac{\maxreward}{k(\maxreward+\inspecfreq\penscalarTwo
        )}\right)  
        \right)&
        \\
        &\qquad\qquad\qquad-
        \ln\left(
        1+\left(\inspecfreq+\frac{\maxreward}{\penscalarTwo} \right)\left(1 - 
        \frac{(\penscalarTwo \inspecfreq + \maxreward)\ln \left(\penscalar \right)}{\maxreward\mininventory}
        \right) - \frac{\maxreward}{\penscalarTwo} \left(1+\frac{\penscalarTwo}{\maxreward+\inspecfreq\penscalarTwo}\right)^\inspecfreq
        \right)
        \\
        &\qquad\qquad\qquad+
        \ln\left(\frac{(\inspecfreq+1)(\maxreward+\inspecfreq\penscalarTwo)}{\inspecfreq\penscalarTwo}\right)
        +
        \ln\left(\prod\nolimits_{k \in[\inspecfreq+1]}\left(1-\frac{\maxreward}{k(\maxreward+\inspecfreq\penscalarTwo
        )}\right)  
        \right)
    \vspace{5pt}
    \\
    &
    \inspecfreq \geq 2,
    \penscalarTwo > 0,\penscalar\geq e
    &
    \end{array}
\end{align}
where the first constraint comes from the capacity feasibility condition in \Cref{lem:FLB feasibility real duration}, while the objective and second constraint come from \Cref{lem:FLB competitive ratio real duration}. Now we are ready to prove \Cref{thm:FLB competitive ratio general}.
\begin{proofof}{Proof of \Cref{thm:FLB competitive ratio general}}
    We consider two regimes based on the magnitude of $\maxreward,\maxduration$ separately.
    
    \vspace{5pt}
    \noindent\textsl{\underline{Regime (i)}:} Suppose $\maxreward\vee\maxduration = \omega(1)$. In this case,
    we assign $\inspecfreq = \lceil\ln(\maxreward\vee\maxduration)\rceil\vee2$, $\penscalarTwo = \left(\frac{1}{\ln(\maxreward\vee\maxduration)}\right)^2$. 
    We set $\penscalar$ such that the first constraint in program~$\FLBOPTReal{\maxreward,\maxduration,\infty}$ binds. We claim that $\ln(\penscalar) = \ln(\maxreward\maxduration) + 3\ln\ln(\maxreward\vee\maxduration) + O(1)$. To see this, consider each term on the right-hand side of the first constraint in $\FLBOPTReal{\maxreward,\maxduration,\infty}$:
    \begin{align*}
        - \ln\left(\prod\nolimits_{k \in[\inspecfreq\maxduration]}\left(1-\frac{\maxreward}{k(\maxreward+\inspecfreq\penscalarTwo
        )}\right)  
        \right) 
        &\overset{(a)}{=}
        \ln(\maxreward) + 2\ln\ln(\maxreward\vee\maxduration) + \ln(\maxduration) + O(1)
        \\
        -
        \ln\left(
        1+\inspecfreq+\frac{\maxreward}{\penscalarTwo}\left(1-\left(1+\frac{\penscalarTwo}{\maxreward+\inspecfreq\penscalarTwo}\right)^\inspecfreq\right)
        \right)
        &\leq 0
        \\
        \ln\left(\frac{(\inspecfreq+1)(\maxreward+\inspecfreq\penscalarTwo)}{\inspecfreq\penscalarTwo}\right)
        +
        \ln\left(1-\frac{\maxreward}{(\maxreward+\inspecfreq\penscalarTwo
        )}\right) 
        &=
        \ln\ln(\maxreward\vee\maxduration) + O(1)
        \\
        \ln\left(\prod\nolimits_{k \in[2:\inspecfreq+1]}\left(1-\frac{\maxreward}{k(\maxreward+\inspecfreq\penscalarTwo
        )}\right)  
        \right)
        & \leq 0
    \end{align*}
    where equality~(a) follows the same reason in \Cref{thm:competitive ratio integer duration}. Since $\maxreward\vee\maxduration = \omega(1)$, constraint $\penscalar\geq e$~from $\FLBOPTReal{\maxreward,\maxduration,\infty}$ is satisfied. Therefore, the competitive ratio of {\FLB} (with the aforementioned parameter values) under large capacity is at most 
    \begin{align*}
        \frac{\inspecfreq}{\inspecfreq - 1}\cdot \ln(\penscalar)\cdot \left(1 + \inspecfreq\penscalarTwo\right)
        =
        \ln(\maxreward\maxduration) + 3\ln\ln(\maxreward\vee\maxduration) + O(1)
    \end{align*}
    
    \noindent\textsl{\underline{Regime (ii)}:} Suppose $\maxreward\vee\maxduration = O(1)$. In this case,
    we assign $\inspecfreq = 2$, $\penscalarTwo = \frac{\maxreward}{e - 1}$. Note that with $\penscalarTwo = \frac{\maxreward}{e - 1}$, all terms in the right-hand side of the first constraint from $\FLBOPTReal{\maxreward,\maxduration,\infty}$ is independent of $\maxreward$. Moreover, it can be verified that the right-hand side is $O(1)$ since $\maxduration = O(1)$. Hence, we set $\penscalar$ be the maximum between this right-hand side and $e$. To sum up, we obtain a feasible solution of $\FLBOPTReal{\maxreward,\maxduration,\infty}$ and the the competitive ratio of {\FLB} (with the aforementioned parameter values) is $\frac{\inspecfreq}{\inspecfreq - 1}\cdot \ln(\penscalar)\cdot \left(1 + \inspecfreq\penscalarTwo\right) = O(\maxreward) = O(1)$.
    
\vspace{10pt}
\noindent
    Finally, the special case of $\maxreward=\maxduration=1$ is shown in \Cref{subsec:CR optimization} using program~$\FLBOPTInt{\maxreward,\maxduration,\infty}$ for integer-valued durations.
\end{proofof}

\subsection{Improved Competitive Ratios for Homogeneous Rewards}
Similar to \Cref{prop:CR identical reward integer duration}, we can also derive a tighter bound on the competitive ratio for instances with heterogeneous durations but homogeneous rewards (i.e., $\maxduration \geq 1$ but $\maxreward = 1$). For this result, we extend the definition of {\FLB} with finite inspection-frequency scalar $\inspecfreq\in[1,\infty)$ to infinite inspection-frequency scalar $\inspecfreq = \infty$. Specifically, under $\inspecfreq = \infty$, we generalize the definition of inspection-time subset $\timesetij(\infty)\triangleq [t_j, t_j + \durationij)$ and (projected-utilization-based) reduced reward as 
\begin{align*}
    \rewardij \durationij -
    \int\nolimits_{\tpre \in \timesetij(\infty)}
    \pen\left(\estimateinventoryij\right)\cdot \text{d} \tpre
\end{align*}
The improved competitive ratio is as follows.

\begin{proposition}[Competitive Ratio of {\FLB} for Homogeneous Rewards]
\label{prop:CR identical reward real duration}
\yfedit{For instances with homogeneous rewards (i.e., $\maxreward = 1$), 
there exists a choice of parameters $(\penscalarTwo^*,\penscalar^*)$ such that the asymptotic competitive ratio of {\FLB}, with inspection-frequency scalar $\inspecfreq^* = \infty$ and penalty parameters $(\penscalarTwo^*,\penscalar^*)$, is at most
    $\ln(\maxduration) + 4$.
    Moreover, there exist instances involving a single server with $\maxreward=1,\maxduration \geq 1$, for which the asymptotic competitive ratio of any online algorithm (possibly fractional or randomized) against the optimal offline benchmark is at least $\ln(\maxduration) + 2$. 
    }
\yfdelete{
For instances with homogeneous rewards (i.e., $\maxreward = 1$), under large capacity, (i) there exists $(\inspecfreq,\penscalarTwo,\penscalar)$ such that the competitive ratio of the {\FLB} with $(\inspecfreq,\penscalarTwo,\penscalar)$ is at most $\ln(\maxduration) + 4$. (ii) The competitive ratio of an (possibly fractional and randomized) online algorithm against the optimal offline benchmark is at least $\ln(\maxduration) + 2$.}
\end{proposition}

\begin{proofof}{Proof of \Cref{prop:CR identical reward real duration}}
    (i) Let $\pen(x) = \penscalarTwo \left(e^{(1-x)}-1 \right) dx$ and $\inspecfreq = \infty$, the capacity feasibility condition (refer to \Cref{lem: invariang infinity inspectifreq}) simplifies to 
\begin{align*}
        1 \geq -\ln \left(1 - \frac{1}{1+\penscalarTwo}e^{\frac{\penscalarTwo}{1+\penscalarTwo}}\right)
+\frac{1}{1+\penscalarTwo}\ln\left(\maxduration \right)
\end{align*}
We will prove $\penscalarTwo = \ln (\maxduration) + 3$ satisfies this inequality:
\begin{align*}
        \frac{4}{\ln (\maxduration) + 4} \geq 
        -\ln \left(1 - \frac{1}{\ln (\maxduration) + 4}e^{\frac{\ln (\maxduration) + 3}{\ln (\maxduration) + 4}}\right) 
        \iff& e^{-\frac{4}{\ln (\maxduration) + 4}} \leq \left(1 - \frac{1}{\ln (\maxduration) + 4}e^{\frac{\ln (\maxduration) + 3}{\ln (\maxduration) + 4}}\right)
\end{align*}
Defining $\zeta = \frac{1}{\ln(D)+4}$ we need to prove $0 \leq e^{\zeta} - e^{-3\zeta} - e\zeta$ for $\zeta \leq \frac{1}{4}$. Taking derivative of the right hand side we will have $e^{\zeta} + 3e^{-3\zeta} - e$ which is increasing in $\zeta$ and hence less than $e^{0.25} + 3e^{-0.75} - e < 0$. This means $e^{\zeta} - e^{-3\zeta} - e\zeta$ is decreasing in $\zeta$. Since The inequality holds for $\zeta = 0$ we are done. Finally our competitive ratio based on \Cref{lem:FLB competitive ratio real duration} will be $1+\penscalarTwo = \ln(\maxduration) + 4$.




(ii) We use proof by contradiction. Assume we have only one resource and Let $M > \inventory$ be a large and $\epsilon$ be a very small number. The bad example consists of two phases. In phase 1 there will be $M^2$ arrivals from length $1+M\epsilon$ to $1$ in a decreasing fashion and then jobs in phase $2$ will have an increasing length from $1$ to $\maxduration$.
    
    Phase 1 starts at time $0$. there will be $M$ jobs arriving with decreasing lengths for each number picked from the set $\{1+(M-i)\epsilon | \forall i \in [M]\}$. Since $M > \inventory$ then there will be a number $\delta = 1+(M-1-i^*)\epsilon$ such that the algorithm will reject that job. Let $\delta$ be the first among all of them. At that time adversary stops phase 1 and starts the second phase by sending a continuum ($M$ jobs of each length) of jobs at time $\delta$ with an increasing lengths.
    Notice that the algorithm accepted $i^*$ jobs in phase 1 so it has $\inventory-i^*$ left inventory. Assume the algorithm accepts job with lengths $\{y_1,...,y_\ell\}$ where $\ell \leq \inventory-i^*$. The moment before job $y_i$ we have:
    \begin{align*}
        \frac{\text{Online ALG}}{\text{Optimal Offline}} = \frac{i^* + \sum\nolimits_{k\in[i-1]} y_k}{\inventory(1+y_i)} > \frac{1}{\ln(\maxduration) + 2}
    \end{align*}
    This implies $y_1 < \frac{i^*(\ln(\maxduration) + 2)}{c} - 1$. Since $y_1 > 1$ then the last inequality also implies $i^* > \frac{\inventory}{\ln (\maxduration) + 2}$. Now we will prove by induction $y_i < \left(\frac{i^*(\ln(\maxduration) + 2)}{c} - 1\right)\left(1+\frac{(\ln(\maxduration) + 2)}{c} \right)^{(i-1)}$. For $i = 1$ it directly follows from the inequality above. For $i>1$ notice:
    \begin{align*}
        &1+y_i 
        \\< {}&
        \frac{\ln(\maxduration) + 2}{c}\left( i^* + \sum\nolimits_{k\in[i-1]} y_k\right) 
        \\
        <{} & \frac{\ln(\maxduration) + 2}{c}\left( i^* + \left(\frac{i^*(\ln(\maxduration) + 2)}{c} - 1\right)\frac{\left(1+\frac{(\ln(\maxduration) + 2)}{c} \right)^{(i-1)}-1}{\frac{(\ln(\maxduration) + 2)}{c}}
        \right)\\
         ={} & \left(\frac{i^*(\ln(\maxduration) + 2)}{c} - 1\right)\left(1+\frac{(\ln(\maxduration) + 2)}{c} \right)^{(i-1)} + 1
    \end{align*}
    Also at the end of the phase $2$ we have:
    \begin{align*}
        \frac{\text{Online ALG}}{\text{Optimal Offline}} = \frac{i^* + \sum_{k\in[\ell]} y_k}{\inventory(1+\maxduration)} > \frac{1}{\ln(\maxduration) + 2}
    \end{align*}
    Which means:
    \begin{align*}
        \maxduration <& \left(\frac{i^*(\ln(\maxduration) + 2)}{c} - 1\right)\left(1+\frac{(\ln(\maxduration) + 2)}{c} \right)^{\ell} \leq 
        \left(1+\frac{(\ln(\maxduration) + 2)}{\inventory} \right)^{\inventory-\frac{\inventory}{\ln(\maxduration) + 2}}\\
        =& \left(1+\frac{(\ln(\maxduration) + 2)}{\inventory} \right)^{ \frac{\inventory}{\ln(\maxduration) + 2}\ln(\maxduration)} \leq \maxduration
    \end{align*}
    The contradiction shows no algorithm can beat competitive ratio $\ln (\maxduration) + 2 $.
\end{proofof}

\yfdelete{\subsection{Invariant for \texorpdfstring{$\inspecfreq = \infty$}{gamma = infinite}}
When $\inspecfreq \rightarrow \infty$ then $\penscalarTwo$ needs to be infinitesimal and all the sums should become integrals.}
\begin{lemma}\label{lem: invariang infinity inspectifreq}
\yfedit{Consider a hypothetical scenario in which {\FLB} can assign a job to a server that has no available capacity (which could result in a negative {\estimated} capacity level). 
With integer-valued inspection-frequency scalar $\inspecfreq\in\naturals$,
the {\estimated} available capacity under {\FLB} satisfies the following property: for every server $i\in[n]$, time points $s,t\in\reals_+$ and real duration $\duration\in\reals$:}
    \yfdelete{During running algorithm \ref{alg:FLB} with $\inspecfreq = \infty$ and $\pen(x)=\penscalarTwo\left(\penscalar^{(1-x)} - 1\right)dx$ we will have:}
    \begin{align*}
         \curinventory_{i, s \rightarrow t+d} - \curinventory_{i, s \rightarrow t} \leq 
\begin{cases} 
      -\RHSlessOne & d \leq 1 \\
      -\RHSgreaterOne & d \geq 1
   \end{cases}
    \end{align*}
    where  $\epsilon = \frac{(\maxreward + \penscalarTwo)\ln \left(\penscalar \right)}{\maxreward\inventory_i}$ and $\delta = \frac{\maxduration \ln \left(\penscalar \right)}{\inventory_i}$.
\end{lemma}
\begin{proofof}{Proof of \Cref{lem: invariang infinity inspectifreq}}
    This holds trivially initially as $\alpha = 1$. Assume this holds true until arrival time of customer $j$ at $t_j$. We will prove this inequality holds true after the possible new assignment. The only interesting and non-trivial case is when $s = t_j \leq t < t_j + \durationij \leq t + d$ and the algorithm decides to assign which means
    $\maxreward \durationij + \int_{t_j} ^{t_j + \durationij}
    \pen\left(\estimateinventoryij\right) d\tau
     > 0$ or: 
    \begin{align*}
        \maxreward \durationij
         & > -
         \int_{t_j} ^{t_j + \durationij}
         \pen\left(\estimateinventoryij\right) d\tau \\
         & \geq - (t-t_j)\pen\left(\estimateinventoryijt\right)
         - \int_{t} ^{t_j + \durationij}
         \pen\left(\estimateinventoryij\right) d\tau\\
         & = - (t-t_j)\penscalarTwo \left(1-\penscalar^{(1-\estimateinventoryijt)}\right)
         - \penscalarTwo\int_{t} ^{t_j + \durationij}
         \left(1-\penscalar^{(1-\estimateinventoryij)}\right) d\tau
    \end{align*}
    Case (i): $t_j + \durationij \leq t+ 1$:
    \begin{align*}
         \frac{\maxreward \durationij}{\penscalarTwo} & \overset{(a)}{\geq} - \durationij 
         + \penscalar^{(1-\estimateinventoryijt)} \left(t-t_j
         + \int_{0} ^{t_j + \durationij - t}
         \penscalar^{\RHSlessOnetau}d\tau \right)  \\
         & = - \durationij 
         + \penscalar^{(1-\estimateinventoryijt)} \left(t-t_j + \int_{0} ^{t_j + \durationij - t}
         \left(1-\frac{\maxreward}{\maxreward+\penscalarTwo}e^{\penscalarTwo\frac{\tau}{\maxreward+\penscalarTwo}} - \epsilon \right) d\tau \right)\\
         & \geq - \durationij 
         + \penscalar^{(1-\estimateinventoryijt)}
         \left(\durationij(1 - \epsilon) -\frac{\maxreward}{\penscalarTwo}\left(e^{\penscalarTwo\frac{t_j + \durationij - t}{\maxreward+\penscalarTwo}}-1\right)\right)\\
         & \geq - \durationij 
         + \penscalar^{(1-\estimateinventoryijt)}
         \left(\durationij(1 - \epsilon) -\frac{\maxreward}{\penscalarTwo}\left(e^{\penscalarTwo\frac{d}{\maxreward+\penscalarTwo}}-1\right)\durationij\right)
    \end{align*}
        The last inequality came from $t_j-t+\durationij \leq d$, $\durationij \geq 1$ and $1-e^{\penscalarTwo\frac{\durationij}{\maxreward+\penscalarTwo}} \geq 1-e^{\penscalarTwo\frac{d}{\maxreward+\penscalarTwo}} \geq (1-e^{\penscalarTwo\frac{d}{\maxreward+\penscalarTwo}})\durationij$. Hence:
    \begin{align*}
         & 1 \geq \penscalar^{(1- \estimateinventoryijt)}
         \left(1 -\frac{\maxreward}{\maxreward+\penscalarTwo}e^{\penscalarTwo\frac{d}{\maxreward+\penscalarTwo}}-\frac{\penscalarTwo}{\maxreward+\penscalarTwo} \epsilon\right) \iff 
     -
\frac{\ln \left(1 - \frac{\maxreward}{\maxreward+\penscalarTwo}e^{\penscalarTwo\frac{d}{\maxreward+\penscalarTwo}}-\frac{\penscalarTwo}{\maxreward+\penscalarTwo} \epsilon \right) 
}{\ln \left(\penscalar \right)} \geq 1-\estimateinventoryijt
\end{align*} 

\begin{align*}\Rightarrow
-\RHSlessOne &\geq
-\frac{\ln \left(1 - \frac{\maxreward}{\maxreward+\penscalarTwo}e^{\penscalarTwo\frac{d}{\maxreward+\penscalarTwo}}-\frac{\penscalarTwo}{\maxreward+\penscalarTwo} \epsilon \right) - \frac{\maxreward}{\maxreward+\penscalarTwo} \epsilon
}{\ln \left(\penscalar \right)}
\\ & =
-\frac{\ln \left(1 - \frac{\maxreward}{\maxreward+\penscalarTwo}e^{\penscalarTwo\frac{d}{\maxreward+\penscalarTwo}}-\frac{\penscalarTwo}{\maxreward+\penscalarTwo} \epsilon \right) 
}{\ln \left(\penscalar \right)} + \frac{1}{\inventory_i} \geq 1-\estimateinventoryijt + \frac{1}{\inventory_i}
    \end{align*}
    Case (ii): $t_j + \durationij \geq t+ 1$:
    \begin{align*}
         \maxreward\frac{\durationij}{\penscalarTwo} \overset{(a)}{\geq} - \durationij 
         + \penscalar^{(1-\estimateinventoryijt)} \Biggr(t-t_j
         &+ \int_{0} ^{1}
         \penscalar^{\RHSlessOnetau} d\tau\\
         &+ \int_{1} ^{t_j + \durationij - t}
         \penscalar^{\RHSgreaterOnetau}d\tau\Biggr)
    \end{align*}
    \begin{align*}
         = & - \durationij 
         + \penscalar^{(1-\estimateinventoryijt)} \left(t-t_j
         + 1 - \epsilon - \frac{\maxreward}{\penscalarTwo}\left(e^{\frac{\penscalarTwo}{\maxreward+\penscalarTwo}}-1\right)
         + \left(1 - \frac{\maxreward}{\maxreward+\penscalarTwo}e^{\frac{\penscalarTwo}{\maxreward+\penscalarTwo}}-\epsilon \right) 
         (1-\delta)\int_{1} ^{t_j + \durationij - t}
         \tau^{-\frac{\maxreward}{\maxreward+\penscalarTwo}}\right) \\
         \geq & - \durationij 
         + \penscalar^{(1-\estimateinventoryijt)} \left(
         \frac{\maxreward+\penscalarTwo}{\penscalarTwo} - \epsilon - \frac{\maxreward}{\penscalarTwo}e^{\frac{\penscalarTwo}{\maxreward+\penscalarTwo}}
         + \left(1 - \frac{\maxreward}{\maxreward+\penscalarTwo}e^{\frac{\penscalarTwo}{\maxreward+\penscalarTwo}}-\epsilon \right) 
         (1-\delta)\int_{1} ^{\durationij}
         \tau^{-\frac{\maxreward}{\maxreward+\penscalarTwo}}\right) \\
         = & - \durationij 
         + \penscalar^{(1-\estimateinventoryijt)} \left(\frac{\maxreward+\penscalarTwo}{\penscalarTwo} - \epsilon - \frac{\maxreward}{\penscalarTwo}e^{\frac{\penscalarTwo}{\maxreward+\penscalarTwo}}
         + \left(\frac{\maxreward+\penscalarTwo}{\penscalarTwo}(1 - \epsilon) - \frac{\maxreward}{\penscalarTwo}e^{\frac{\penscalarTwo}{\maxreward+\penscalarTwo}}\right) 
         (1-\delta)\left(\durationij^{\frac{\penscalarTwo}{\maxreward+\penscalarTwo}}-1\right)\right)\\
         \geq & - \durationij 
         + \penscalar^{(1-\estimateinventoryijt)} \left(
         \frac{\maxreward+\penscalarTwo}{\penscalarTwo}(1 - \epsilon) - \frac{\maxreward}{\penscalarTwo}e^{\frac{\penscalarTwo}{\maxreward+\penscalarTwo}}\right) 
         \left((1-\delta)\durationij^{\frac{\penscalarTwo}{\maxreward+\penscalarTwo}} + \delta\right)
    \end{align*}
    Hence:
    \begin{align*}
         1 \geq \penscalar^{(1-\estimateinventoryijt)} \left(
         1 - \epsilon - \frac{\maxreward}{\maxreward+\penscalarTwo}e^{\frac{\penscalarTwo}{\maxreward+\penscalarTwo}}\right)
         \left((1-\delta)d^{-\frac{\maxreward}{\maxreward+\penscalarTwo}}+\delta d^{-1}\right) 
         \\
         \Rightarrow -\RHSgreaterOne \geq 1-\estimateinventoryij
    \end{align*}
    Since $(1-\delta)d^{\frac{-\maxreward}{\maxreward+\penscalarTwo}} + \delta d ^ {-1} < 1$ we know:
    \begin{align*}
        \ln \left( (1-\delta)d^{\frac{-\maxreward}{\maxreward+\penscalarTwo}} + \delta d ^ {-1} \right) - \ln \left( (1-\delta)d^{\frac{-\maxreward}{\maxreward+\penscalarTwo}} \right) \geq \delta d ^ {-1} \geq \frac{\ln \left(\penscalar \right)}{\inventory_i}
    \end{align*}
    Then:
    \begin{align*}
        &-\RHSgreaterOne \\
        \geq &-{\frac{\ln \left(1 - \frac{\maxreward}{\maxreward+\penscalarTwo}e^{\penscalarTwo\frac{1}{\maxreward+\penscalarTwo}}-\epsilon\right)
         + \ln \left((1-\delta)d^{-\frac{\maxreward}{\maxreward+\penscalarTwo}}+\delta d^{-1}\right)
         }{\ln \left(\penscalar \right)}} + \frac{1}{\inventory_i} \geq 1-\estimateinventoryij + \frac{1}{\inventory_i}
    \end{align*}
inequalities ~(a) come from the induction hypothesis. These two cases show the induction hypothesis remains true after any assignment.
\end{proofof}

\section{Omitted Proofs}
\label{apx:proofs}
In this section, we provide all missing proofs in the main text.
\subsection{Proof of Lemma~\ref{lem: rho lemma}}
\label{apx: rho lemma}
\rholemma*
\begin{proofof}{\revcolor{Proof} of \Cref{lem: rho lemma}}
    Let us prove this by induction. For $d = 1$ the equation holds trivially. For $d>1$:
    \begin{align*}
    \sum\nolimits_{\ell \in[0:d-1]} \prod\nolimits_{k\in [\ell]} (1-\frac{z}{k}) &= \frac{d-1}{1-z}\prod\nolimits_{k\in [d-1]} (1-\frac{z}{k}) +\prod\nolimits_{k\in [d-1]} (1-\frac{z}{k})\\
    & = \frac{d-z}{1-z} \prod\nolimits_{k\in [d-1]} (1-\frac{z}{k}) =
    \frac{d}{1-z}\prod\nolimits_{k\in [d]} (1-\frac{z}{k})
    \end{align*}
    which finishes the proof of the lemma as desired.
\end{proofof}

\subsection{Proof of Lemma~\ref{lem:relaxation}}
\label{apx:relaxation}
\lemrelax*
\begin{proofof}{\revcolor{Proof} of \Cref{lem:relaxation}}
For each server $i\in[n]$, unit $k\in[\inventory_i]$, and job subset $\config\in\configspace_i$, let $\event(i, k, \config)$ represent the indicator whether unit $k$ of server $i$ is assigned exclusively to jobs in $\config$ in the optimal offline benchmark.
Since the optimal offline benchmark is feasible, it is evident that for every job $j\in [m]$, 
\begin{align*}
 \sum\nolimits_{i\in[n]}
 \sum\nolimits_{k\in[\inventory_i]}
\sum\nolimits_{\config\in\configspace_i:j\in \config}
     \indicator{\event(i, k, \config)} \leq 1
\end{align*}
and for every server $i\in[n]$ and its unit $k\in[\inventory_i]$,
\begin{align*}
    \sum\nolimits_{\config\in\configspace_i}
     \indicator{\event(i, k, \config)} \leq 1
\end{align*}
Therefore, $\allocconfig = \sum_{k\in[\inventory_i]}\event(i, k, \config)$ results in a feasible assignment for program~\ref{eq:opt lp}. Moreover, the objective value under this assignment will be equal to the total reward of the optimal offline benchmark, thereby concluding the proof.
\end{proofof}

\subsection{Proof of Lemma~\ref{lem: bounding theta}}
\label{apx: bounding theta}

\boundingtheta*

\begin{proofof}{\revcolor{Proof} of \Cref{lem: bounding theta}}

    Notice that:
    \begin{align*}
        1 = \sum\nolimits_{k \in[\maxduration]} \ln \left(\frac{k}{k-\lambda_D}\right) = \sum\nolimits_{k \in[\maxduration]} \ln \left(1+ \frac{\lambda_D}{k-\lambda_D}\right) \leq& \sum\nolimits_{k \in[\maxduration]} \frac{\lambda_D}{k-\lambda_D} \leq \lambda_D\left(\frac{1}{1-\lambda_D} + H(D) \right) 
    \end{align*}
    solving the quadratic equation:
    \begin{align*}
        \frac{1}{\lambda_D} \leq \frac{2H(D)}{H(D)+2-\sqrt{(H(D))^2+4}} \leq H(D) + 2
    \end{align*}
    The last inequality is equivalent to $(H(D)+2)\sqrt{(H(D))^2+4} \leq (H(D))^2+2H(D)+4$ and By squaring both sides of the inequality we get $0 \leq 4(H(D))^2$
\end{proofof}


\revcolor{\section{Extensions}}
\label{apx:extension}
\revcolor{In this section, we generalize our {\FLB} algorithm to multiple extension models. All three subsections focus on the generalization of the baseline model from different perspectives.

\subsection{Network Revenue Management}
\label{apx:NRM}

In this section, we extend our baseline model and {\FLBLong} ({\FLB}) to a network revenue management (NRM) setting, where a job may consume multiple units of capacity from one or more servers (e.g., a cloud job requiring CPUs, GPUs, and memory). We first describe the generalized model, and then present the extension of {\FLB}. 
Finally, we discuss capacity feasibility as well as the competitive ratio analysis.\footnote{\label{footnote:focus-integer-duration}\revcolor{For the ease of presentation, we only consider the capacity feasibility and competitive ratio under the integer-duration environment. The extension to the real-duration environment follows the same approach.}}

\xhdr{Model.} We consider the following generalization of the online job assignment problem. For each job $j\in[m]$, the platform selects a meta-assignment $\jobconfig$ from a finite set of feasible meta-assignments $\jobconfigspace_j$. Each meta-assignment $\jobconfig\in\jobconfigspace_j$ specifies, for every server $i\in[n]$, a binary demand $\consump_{i\jobconfig j}\in\{0, 1\}$ representing the capacity of server $i$ consumed by job $j$ under meta-assignment $\jobconfig$.
The job occupies these units for a duration $\duration_{\jobconfig j}\in[1,\maxduration]$, and the platform earns a per-period reward $\reward_{\jobconfig j}\in[1,\maxreward]$ for each occupied unit. Thus, the total reward obtained by assigning job $j$ to meta-assignment $\jobconfig$ is\footnote{\revcolor{As a sanity check, this extension recovers our baseline model when each feasible meta-assignment $\jobconfig\in\jobconfigspace_j$ encodes a single compatible server $i\primed\in\neighbor(j)$. In this special case, we set $\consump_{i\jobconfig j} \triangleq \indicator{i = i\primed}$ for each $i\in[n]$, $\duration_{\jobconfig j} \triangleq \duration_{i\primed j}$, and $\reward_{\jobconfig j}\triangleq \reward_{i\primed j}$.}}
\begin{align*}
    \sum\nolimits_{i\in[n]} \consump_{i\jobconfig j}\cdot \reward_{\jobconfig j}\duration_{\jobconfig j}.
\end{align*}
Let the number of legs $\maxleg\geq 1$ denote the maximum of the total demand across all servers, over all feasible meta-assignments, i.e.,
\begin{align*}
\maxleg \triangleq \max\nolimits_{j\in[m]}\max\nolimits_{\jobconfig\in\jobconfigspace_j}~
\sum\nolimits_{i\in[n]}\consump_{i\jobconfig j}
\end{align*}

The capacity constraints are enforced as follows. For each server $i\in[n]$, let $\inventory_i\in\naturals$ denote its initial capacity. At any time $t$, the total demand assigned to server $i$ by jobs that are still active at time $t$ must not exceed $\inventory_i$. The offline benchmark and the online algorithm proceed as in the baseline model: jobs arrive sequentially, and upon the arrival of job $j$, the algorithm learns the feasible set $\jobconfigspace_j$ and the associated parameters $\{(\consump_{i\jobconfig j},\duration_{\jobconfig j},\reward_{\jobconfig j})_{i\in[n]}\}_{\jobconfig\in\jobconfigspace_j}$ and must immediately and irrevocably assign the job to a meta-assignment $\jobconfig\in\jobconfigspace_j$ (or reject it). 


\xhdr{Extension of {\FLB}.}
We generalize the {\FLB} algorithm to the NRM setting by considering a meta-assignment-level projected-utilization-based reduced reward defined as follows:
for each server $i\in[n]$ and time points $t,\tau\in[0,T]$ with $t\le \tau$, we define the (normalized) projected available capacity $\estimateinventoryij\in[0,1]$ exactly as in the baseline model (see \Cref{def:projected available capacity}). For a meta-assignment $\jobconfig\in\jobconfigspace_j$, the inspection-time subset is defined with respect to the meta-assignment duration $\duration_{\jobconfig j}$:
\begin{align*}
    \timeset_{\jobconfig j}(\inspecfreq)\triangleq \left\{\tpre 
    \in [t_j, t_j + \duration_{\jobconfig j}):
    \exists \ell\in\naturals
    \text{ s.t. }
    \tpre = 
    t_j + \tfrac{\ell}{\inspecfreq}
    \right\}.
\end{align*}
The reduced reward for meta-assignment $\jobconfig$ is then defined as\footnote{\label{footnote:NRM:reduced-reward}\revcolor{We remark that the exact definition of the reduced reward in Eqn.~\eqref{eq:reduced-reward-NRM} is carefully designed to ensure that both the capacity feasibility condition and the competitive ratio guarantee extend to this setting. Alternative definitions, such as $\sum\nolimits_{i\in[n]} \consump_{i\jobconfig j}\cdot \reward_{\jobconfig j}\duration_{\jobconfig j} - \sum\nolimits_{i\in[n]}\sum\nolimits_{\tpre \in \timeset_{\jobconfig j}(\inspecfreq)} \consump_{i\jobconfig j}\cdot \pen\left(\estimateinventoryij\right)$, do not admit an extension of the capacity feasibility condition, nor yield a competitive ratio guarantee consistent with the literature. See below for further discussion.}}
\begin{align}
\label{eq:reduced-reward-NRM}
    \reducedreward(j, \jobconfig)\triangleq
        \sum\nolimits_{i\in[n]} \consump_{i\jobconfig j}\cdot \reward_{\jobconfig j}\duration_{\jobconfig j}
    -
    \maxleg
    \cdot
    \left(\sum\nolimits_{i\in[n]}\consump_{i\jobconfig j}\cdot \sum\nolimits_{\tpre \in \timeset_{\jobconfig j}(\inspecfreq)}
    \pen\left(\estimateinventoryij\right)\right),
\end{align}
where $\pen(\cdot)$ is the exponential penalty function parameterized by $(\eta,\beta)$ as in the baseline model. This expression naturally generalizes the single-unit reduced reward: each unit of demand on server $i$ is penalized by the aggregate penalty evaluated at the projected available capacities of all servers with positive demand over the inspection times.

Upon the arrival of job $j$, the algorithm computes the reduced reward for every feasible meta-assignment $\jobconfig\in\jobconfigspace_j$. It then assigns job $j$ to the meta-assignment $\jobconfig^*$ with the highest positive reduced reward; if all meta-assignments have non-positive reduced reward, the job is rejected (i.e., allocated with fraction zero). 




\xhdr{Capacity feasibility.} The capacity feasibility characterization of {\FLB} extends to this NRM setting. For completeness, we include the formal proposition statement and its analysis below. 

\begin{proposition}[Capacity Feasibility of {FLB} under the NRM Setting]
\label{prop:NRM:capacity-feasibility}
In the integer-duration environment, the NRM extension of FLB (see Eqn.~\eqref{eq:reduced-reward-NRM}) with inspection-frequency scalar $\inspecfreq = 1$ is capacity feasible if the penalty parameters $(\penscalarTwo,\penscalar)$ satisfy
\begin{align*}
    \ln(\penscalar) \geq - \ln\left(\prod\nolimits_{k \in[\maxduration]}\left(1-\frac{\maxreward}{k(\maxreward+\penscalarTwo)}\right) 
    - \frac{(\maxreward+\penscalarTwo)\ln(\penscalar)}{\maxreward\mininventory}
    \right).
\end{align*}
\end{proposition}

\begin{proofof}{Proof}
The analysis closely follows that of \Cref{prop:FLB feasibility integer duration}. In particular, we show that \Cref{lem:inventory invariant integer duration} continues to hold for each server by the same induction argument. The formal and complete argument is given below.

We prove the statement of \Cref{lem:inventory invariant integer duration} in the NRM setting by an induction over the sequence of currently arrived jobs $[0:j]$.

\smallskip
\noindent\emph{Base case ($j = 0$):} Before the first job arrives, the {\estimated} available capacity $\curinventory_{i,0\rightarrow \tpre}$ is $1$ for all servers $i\in[n]$ and all future time points $\tpre \geq 0$ by definition. Therefore, the inequality in lemma statement holds trivially as the left-hand side is zero and the right-hand side is non-negative (for large enough $\mininventory$). 

\smallskip
\noindent\emph{Inductive step ($j \geq 1$):} \revcolor{Fix an arbitrary server $i\in[n]$ and suppose the inequality in the lemma statement holds for the arrival of first $j-1$ jobs. Now consider the arrival of job $j$ at time $t_j$.  We show the inequality holds after {\FLB} determines the assignment for job $j$. The only non-trivial case happens when {\FLB} assigns a meta-assignment $\jobconfig$ such that $t = t_j \leq \tau < t + \duration_{\jobconfig j} \leq t + d$ and the demand over server $i$ is positive, i.e.,  $\consump_{i \jobconfig j}>0$. 
In this case, we use the fact that the reduced reward $ \reducedreward(j, \jobconfig)$ of the assigned meta-assignment $\jobconfig$ is strictly positive. Thus, for this server $i$ with $\consump_{i \jobconfig j}=1$,
\begin{align*}
    0 < {} & \reducedreward(j, \jobconfig) 
    \overset{(a)}{\leq} 
    \sum\nolimits_{i\primed\in[n]} \consump_{i\primed\jobconfig j}\cdot \maxreward\duration_{\jobconfig j}
    -
    \maxleg
    \cdot
    \left(\sum\nolimits_{i\primed\in[n]}\consump_{i\primed\jobconfig j}\cdot \sum\nolimits_{\tpre \in \timeset_{\jobconfig j}(\inspecfreq)}
    \pen\left(\curinventory_{i\primed, t_j \rightarrow \tpre}\right)\right)
    \\
    \overset{(b)}{\leq} {} & 
    \maxleg\cdot \maxreward\duration_{\jobconfig j}
    -
    \maxleg
    \cdot
    \left(\sum\nolimits_{i\primed\in[n]}\consump_{i\primed\jobconfig j}\cdot \sum\nolimits_{\tpre \in \timeset_{\jobconfig j}(\inspecfreq)}
    \pen\left(\curinventory_{i\primed, t_j \rightarrow \tpre}\right)\right)
    \overset{(c)}{\leq} 
    \maxleg
    \cdot 
    \left(\maxreward\duration_{\jobconfig j}
    -
    \sum\nolimits_{\tpre \in \timeset_{\jobconfig j}(\inspecfreq)}
    \pen\left(\estimateinventoryij\right)
    \right),
\end{align*}
where inequality~(a) holds since $\reward_{\jobconfig j} \leq \maxreward$; inequality~(b) holds since $\sum\nolimits_{i\primed\in[n]} \consump_{i\primed\jobconfig j} \leq \maxleg$ by the definition of $\maxleg$; and inequality~(c) holds since $\sum\nolimits_{\tpre \in \timeset_{\jobconfig j}(\inspecfreq)}
\pen\left(\curinventory_{i\primed, t_j \rightarrow \tpre}\right) \geq 0$ for every server $i\primed\in[n]$.\footnote{\revcolor{As discussed in \Cref{footnote:NRM:reduced-reward}, inequality~(b) relies on the specific definition of the reduced reward. The alternative definition in \Cref{footnote:NRM:reduced-reward} does not ensure inequality~(b), and thus the capacity feasibility characterization cannot be extended.}} Since $\maxleg \geq 1$, we obtain
\begin{align*}
    \maxreward\duration_{\jobconfig j} > \sum\nolimits_{\tpre \in \timeset_{\jobconfig j}(\inspecfreq)}
    \pen\left(\estimateinventoryij\right).
\end{align*}}
Therefore, 
\begin{align*}
    \frac{1}{\penscalarTwo}\maxreward \duration_{\jobconfig j}
     >{}  &
     \frac{1}{\penscalarTwo}\sum\nolimits_{k' \in [0:\duration_{\jobconfig j}   - 1]}
    \pen\left(\curinventory_{i,t_j\rightarrow t_j + k'}\right)
    \\
    & = {} 
    \frac{1}{\penscalarTwo}\sum\nolimits_{k' \in [0:\tau -t_j-1]}
     \pen\left(\curinventory_{i,t_j\rightarrow t_j + k'}\right) +
     \frac{1}{\penscalarTwo}\sum\nolimits_{k' \in [0:\duration_{\jobconfig j}   +   t_j - \tau- 1]}
     \pen\left(\curinventory_{i,t_j\rightarrow \tau + k'}\right)\\
     \geq {} &
     \left(\tau -t_j \right) \left(\penscalar^{(1-\curinventory_{i,t_j\rightarrow \tau})} -1 \right) +
     \sum\nolimits_{k' \in [0:\duration_{\jobconfig j}   +   t_j - \tau- 1]}
     \left(\penscalar^{\left(1-\curinventory_{i,t_j\rightarrow \tau + k'}\right)}-1\right)\\
     \overset{(a)}{\geq} {} & -   \duration_{\jobconfig j} + \left(\tau - t_j \right) \penscalar^{\left(1-\curinventory_{i,t_j\rightarrow \tau}\right)}
     + \sum\nolimits_{k' \in [0:\duration_{\jobconfig j} + t_j - \tau- 1]}
     \penscalar^{1-\curinventory_{i,t_j\rightarrow \tau}+ \frac{\ln\left(\prod\nolimits_{k \in[k']}\left(1-\frac{\maxreward}{k(\maxreward+\penscalarTwo
    )}\right)  
    - \frac{(\maxreward+\penscalarTwo)\ln(\penscalar)}{\maxreward\mininventory}
    \right)}{\ln (\penscalar)}}\\
     \overset{\revcolor{(b)}}{\geq} {} &-   \duration_{\jobconfig j} + \penscalar^{(1-\curinventory_{i,t_j\rightarrow \tau})} \sum\nolimits_{k' \in [0:\duration_{\jobconfig j}   - 1]}
     \left(\prod\nolimits_{k \in[k']}\left(1-\frac{\maxreward}{k(\maxreward+\penscalarTwo
    )}\right)  - 
    \frac{(\maxreward+\penscalarTwo)\ln(\penscalar)}{\maxreward\mininventory}
    \right) \\
     \overset{(c)}{=}{} &-   \duration_{\jobconfig j} + \penscalar^{(1-\curinventory_{i,t_j\rightarrow \tau})} 
     \left(\frac{\duration_{\jobconfig j}}{\frac{\penscalarTwo}{\maxreward+ \penscalarTwo}}\prod\nolimits_{k \in[\duration_{\jobconfig j}]}\left(1-\frac{\maxreward}{k(\maxreward+\penscalarTwo)}\right)  
     - \frac{\duration_{\jobconfig j}(\maxreward+\penscalarTwo)\ln(\penscalar)}{\maxreward\mininventory}\right)
\end{align*}
where inequality~(a) holds due to the induction hypothesis and the fact that $\tpre - \revcolor{t_j} \leq \duration_{\jobconfig j}$, and inequality~\revcolor{(b)} follows by lower-bounding the scalar $\tau - t_j$ via $\tau - t_j$ individual terms in the summation, each of which is bounded above by one. Finally, equality~\revcolor{(c)} holds due to the identity in \Cref{lem: rho lemma} with $d = \duration_{\jobconfig j}, \ell = k'$ and $z = \frac{\maxreward}{\maxreward + \penscalarTwo}$.
After moving $\duration_{\jobconfig j}$ to the left-hand side and canceling $\left(\frac{\maxreward}{\penscalarTwo} + 1\right) \duration_{\jobconfig j}$ from both sides, and then taking a logarithm from both sides and rearranging the terms, we have:
\begin{align*}
        -\ln \left(\prod\nolimits_{k \in[\duration_{\jobconfig j}]}\left(1-\frac{\maxreward}{k(\maxreward+\penscalarTwo)}\right)  
         - \frac{\penscalarTwo\ln(\penscalar)}{\maxreward\mininventory}\right)\geq \left(1-\curinventory_{i,t_j\rightarrow \tau}\right) \cdot \ln (\penscalar)~.
    \end{align*}
and consequently,
    \begin{align*}
        &
        - \ln\left(\prod\nolimits_{k \in[\duration]}\left(1-\frac{\maxreward}{k(\maxreward+\penscalarTwo
        )}\right)  -\frac{(\maxreward+\penscalarTwo) \ln(\penscalar)}{\maxreward\mininventory}
        \right) 
        \overset{(a)}{\geq} 
        -\ln \left(\prod\nolimits_{k \in[\duration_{\jobconfig j}]}\left(1-\frac{\maxreward}{k(\maxreward+\penscalarTwo)}\right)  
         - \frac{\penscalarTwo\ln(\penscalar)}{\maxreward\mininventory}\right) + \frac{\ln(\penscalar)}{\mininventory}\\
         \overset{(b)}{\geq}& \left(1-\left(\curinventory_{i,t_j\rightarrow \tau}-\frac{1}{\mininventory}\right)\right) \cdot \ln (\penscalar)
         \overset{(c)}{\geq}{} 
         {\left(\curinventory_{i,t\rightarrow \tau + \duration}-\left(\curinventory_{i,t\rightarrow \tau}-\frac{1}{\inventory_i}\right)\right) \cdot \ln (\penscalar)}~.
    \end{align*}
where inequality~(a) holds due to the concavity of the logarithm function and the case assumption that $\duration \geq \duration_{\jobconfig j}$, inequality~(b) holds as we argued above, and inequality~(c) holds since $\curinventory_{i,t\rightarrow \tau + \duration} \leq 1$ and $\inventory_i \geq \mininventory$.
Note that terms $\curinventory_{i,t\rightarrow \tau + \duration}$ and $\curinventory_{i,t\rightarrow \tau}-\frac{1}{\inventory_i}$ above are the updated {\estimated} available capacities after the assignment of job $j$ to server $i$ in {\FLB}. Therefore, we have shown the induction hypothesis after {\FLB} determines the assignment for job $j$, which finishes the proof of the induction. 
\end{proofof}

\xhdr{Competitive ratio guarantee.} We present the following competitive ratio guarantee of the {\FLB}.
\begin{proposition}
\label{prop:NRM:competitive-ratio}
In the integer-duration environment, for every $\penscalarTwo > 0$ and $\penscalar \geq e$, the competitive ratio of a capacity-feasible NRM extension of FLB (see Eqn.~\eqref{eq:reduced-reward-NRM}) with inspection-frequency scalar $\inspecfreq = 1$ is at most 
\begin{align*}
\maxleg \cdot \ln(\penscalar)\cdot \left(1 + \penscalarTwo\left(1 + \penscalar\left(\penscalar^{\frac{1}{\mininventory}}-1\right)\right)\right)~.
\end{align*}
Therefore, there exists a choice of parameters $(\penscalarTwo^*,\penscalar^*)$ such that the asymptotic competitive ratio of NRM extension of {\FLB}, with inspection-frequency scalar $\inspecfreq^* = 1$ and penalty parameters $(\penscalarTwo^*,\penscalar^*)$, is at most
    \begin{align*}
        \maxleg\cdot \left(\ln(\maxreward\maxduration) + \ln\ln(\maxreward\vee\maxduration) + O(1)\right)~.
    \end{align*}
\end{proposition}
We remark that linear dependence on $\maxleg$ (i.e., the number of legs) also appears in similar online NRM settings \citep[e.g.,][]{MRST-20,BM-22,MMZ-26}. The proof of \Cref{prop:NRM:competitive-ratio} closely follows that of \Cref{prop:FLB competitive ratio integer duration}, and then applies the same argument as in \Cref{thm:competitive ratio integer duration} (see \Cref{subsec:CR optimization}). The key change is that we replace the primal-dual analysis based on the configuration LP with an LP-free analysis similar to \citet{GU-23,GIU-20}, while adapting the detailed arguments from \Cref{prop:FLB competitive ratio integer duration}.

\begin{proofof}{Proof of \Cref{prop:NRM:competitive-ratio}}
In this analysis, our goal is to construct a dual solution $\{\onlinedual(j), \offlinedual(i)\}_{i\in[n],j\in[m]}$ such that summation of constructed solution (i.e., $\sum\nolimits_{j\in[m]}\onlinedual(j) + \sum\nolimits_{i\in[n]}\offlinedual(i)$) is at most $\Gamma$ times the total reward collected by {\FLB}, and at least the total reward of the optimal offline benchmark, where $\Gamma = \maxleg\cdot \ln(\penscalar)\cdot \left(1 + \penscalarTwo\left(1 + \penscalar\left(\penscalar^{\frac{1}{\mininventory}}-1\right)\right)\right)$ presented in the proposition statement. We first present the construction and then establish the corresponding upper and lower bounds.

\noindent\textsl{[Step i] Solution construction.} We now construct a solution based on the execution of the {\FLB}: initially, set $\onlinedual(j) \gets 0$ and $\offlinedual(i) \gets 0$ for each job $j\in[m]$ and server $i\in [n]$. Now consider all assignments made by {\FLB}. For each job $j\in[m]$, if {\FLB} assigns job $j$ under meta-assignment $\jobconfig$, update the solutions as follows:\footnote{Compared with the dual solution construction in \Cref{prop:FLB competitive ratio integer duration}, we scale up the update to the dual variable $\offlinedual(i)$ by a multiplicative factor of $\inventory_i$, and later scale it down by the same factor in the analysis. We make this change to render the proof self-contained.}
\begin{align*}
    \onlinedual(j) &\gets
    \reducedreward(j, \jobconfig) \equiv
    \sum\nolimits_{i\in[n]}
    \consump_{i\jobconfig j}
    \cdot 
    \reward_{\jobconfig j}\duration_{\jobconfig j}
    -
    \maxleg\cdot 
    \left(\sum\nolimits_{i\in[n]}
    \consump_{i\jobconfig j}
    \cdot
    \sum\nolimits_{\tpre \in \timeset_{\jobconfig j}}
 	\pen\left(
 	\estimateinventoryij
 	\right)\right),
\intertext{and for every server $i\in[n]$ with positive demand $\consump_{i \jobconfig j} = 1$,}
    \offlinedual(i) &\gets
    \offlinedual(i) 
    +
    \maxleg \inventory_i\cdot \sum\nolimits_{\tpre \in \timeset_{\jobconfig j}}
 	\left(
    \pen\left(
 	\estimateinventoryij - \frac{1}{\inventory_i}
 	\right)
  -
    \pen\left(
 	\estimateinventoryij
 	\right)
    \right),
\end{align*}
where $\timeset_{\jobconfig j}$ is shorthand for the inspection-time subset $\timeset_{\jobconfig j}(\inspecfreq)$ with $\inspecfreq = 1$. Note that the above solution is well-defined (i.e., $\onlinedual(j) \geq 0$ and $\offlinedual(i) \geq 0$). 
To see this, note that for each online job $j$, by assigning meta-assignment $\jobconfig$, its reduced reward is positive, and thus $\onlinedual(j) = \reducedreward(j, \jobconfig) \geq 0$. For each server $i$ with positive demand $\consump_{i \config j} = 1$, since {\FLB} is capacity feasible, there is at least one available unit of server~$i$. Therefore, we have $\curinventory_{i, t_j \rightarrow t_j} \geq \frac{1}{\inventory_i}$. Additionally, for every $\tpre \in \timeset_{\jobconfig j}$, we have $\estimateinventoryij \geq \curinventory_{i, t_j\rightarrow t_j} \geq \frac{1}{\inventory_i}$, as $\estimateinventoryij$ is increasing in $\tpre$ and $\tpre \geq t_j$. Thus, each update to the dual variable $\offlinedual(i)$ is non-negative.

\smallskip
\noindent[\emph{Step ii}]
\emph{Comparing {\FLB} with the constructed solution.}
Here, we show that the total reward of {\FLB} is a $\Gamma$-approximation to the summation of the constructed solution, i.e., $\sum\nolimits_{j\in[m]}\onlinedual(j) + \sum\nolimits_{i\in[n]}\offlinedual(i)$, where $\Gamma = \maxleg\cdot \ln(\penscalar)\cdot \left(1 + \penscalarTwo\left(1 + \penscalar\left(\penscalar^{\frac{1}{\mininventory}}-1\right)\right)\right)$, as stated in \Cref{prop:NRM:competitive-ratio}.
To establish this, we analyze the increments of the reward in {\FLB} and the update of $\sum\nolimits_{j\in[m]}\onlinedual(j) + \sum\nolimits_{i\in[n]}\offlinedual(i)$ from each job assignment decision separately.

Suppose {\FLB} assigns job $j$, arriving at time $t_j$, under meta-assignment $\jobconfig$. The increment of the total reward in {\FLB} due to this assignment is $\Delta(\FLB) = \sum_{i\in[n]}\consump_{i\jobconfig j}\cdot \reward_{\jobconfig j}\duration_{\jobconfig j}$, while the increment of the summation over the constructed solution can be upper bounded as:
\begin{align*}
    &~\Delta(\textsc{Summation-Over-Solution}) 
    \\
    ={} &
    \left(\sum\nolimits_{i\in[n]}
    \consump_{i\jobconfig j}
    \cdot 
    \reward_{\jobconfig j}\duration_{\jobconfig j}
    -
    \maxleg\cdot \left(\sum\nolimits_{i\in[n]}
    \consump_{i\jobconfig j}
    \cdot
    \sum\nolimits_{\tpre \in \timeset_{\jobconfig j}}
 	\pen\left(
 	\estimateinventoryij
 	\right)\right)\right)
    \\
  &  \qquad\qquad
  +
  \sum\nolimits_{i\in[n]}\consump_{i\jobconfig j}\cdot \maxleg\inventory_i
  \cdot
  \sum\nolimits_{\tpre \in \timeset_{\jobconfig j}}
  \left(
    \pen\left(
    \estimateinventoryij - 
 	\frac{1}{\inventory_i}
 	\right)
  -
    \pen\left(
 	{\estimateinventoryij}
  \right)
  \right)
  \\
  \overset{(a)}{\leq} {} &
    \maxleg\cdot \left(\sum\nolimits_{i\in[n]}
    \consump_{i\jobconfig j}
    \cdot 
    \reward_{\jobconfig j}\duration_{\jobconfig j}
    -
    \sum\nolimits_{i\in[n]}
    \consump_{i\jobconfig j}
    \cdot
    \sum\nolimits_{\tpre \in \timeset_{\jobconfig j}}
 	\pen\left(
 	\estimateinventoryij
 	\right)\right)
    \\
  &  \qquad\qquad
  +\maxleg\cdot \left(
  \sum\nolimits_{i\in[n]} \consump_{i\jobconfig j}\cdot 
  \sum\nolimits_{\tpre \in \timeset_{\jobconfig j}}
  \inventory_i
  \left(
    \pen\left(
    \estimateinventoryij - 
 	\frac{1}{\inventory_i}
 	\right)
  -
    \pen\left(
 	{\estimateinventoryij}
  \right)
  \right)\right)
  \\
  \overset{(b)}{\leq}{}&
  \maxleg\cdot \sum\nolimits_{i\in[n]}\consump_{i\jobconfig j}\cdot 
  \left(
  \ln\left(\penscalar\right)
  \left(
  \reward_{\jobconfig j} \duration_{\jobconfig j}
    -
    \sum\nolimits_{\tpre \in \timeset_{\jobconfig j}}
 	\pen\left(
 	{\estimateinventoryij}
 	\right)
  \right)
  -
  \sum\nolimits_{\tpre \in \timeset_{\jobconfig j}}
  \inventory_i
  \left(
    \pen\left(
 	{\estimateinventoryij}
  \right)
    -
    \pen\left(
    \estimateinventoryij - 
 	\frac{1}{\inventory_i}
 	\right)
  \right)
  \right)
  \\
  ={} &
  \maxleg\cdot \sum\nolimits_{i\in[n]}\consump_{i\jobconfig j}\cdot 
  \left(\reward_{\jobconfig j} \duration_{\jobconfig j}
  \ln\left(\penscalar\right)
  -
  \sum\nolimits_{\tpre \in \timeset_{\jobconfig j}}
  \left(
  \ln\left(\penscalar\right)
  \pen\left(
 	{\estimateinventoryij}
 	\right)
  +
  \inventory_i
  \left(
    \pen\left(
 	{\estimateinventoryij}
  \right)
    -
    \pen\left(
    \estimateinventoryij - 
 	\frac{1}{\inventory_i}
 	\right)
  \right)
  \right)
  \right)
  \\
  \overset{(c)}{\leq}{} &
  \maxleg\cdot \sum\nolimits_{i\in[n]}\consump_{i\jobconfig j}\cdot 
  \left(
  \reward_{\jobconfig j} \duration_{\jobconfig j}
  \ln\left(\penscalar\right)
  -
  \sum\nolimits_{\tpre \in \timeset_{\jobconfig j}}
  \left(
  \ln\left(\penscalar\right)
  \pen\left(
 	{\estimateinventoryij}
 	\right)
  +
    \pen'\left(
    \estimateinventoryij - 
 	\frac{1}{\inventory_i}
 	\right)
  \right)
  \right)
  \\
  \overset{(d)}{\leq}{} &
  \maxleg\cdot \sum\nolimits_{i\in[n]}\consump_{i\jobconfig j}\cdot 
  \left(
  \reward_{\jobconfig j} \duration_{\jobconfig j}
  \ln\left(\penscalar\right)
  +
  \sum\nolimits_{\tpre \in \timeset_{\jobconfig j}}
 	\penscalarTwo 
      \ln\left(\penscalar\right) 
  \left(
  1 + 
  \penscalar
    \left(
    \penscalar^{\frac{1}{\mininventory}
    }
    - 1
    \right)
  \right)
  \right)
  \\
  \overset{}{=}{} &\maxleg\cdot \sum\nolimits_{i\in[n]}\consump_{i\jobconfig j}\cdot 
  \left(
  \reward_{\jobconfig j} \duration_{\jobconfig j}
  \ln\left(\penscalar\right)
  +
  |\timeset_{\jobconfig j}|
 	\penscalarTwo 
      \ln\left(\penscalar\right) 
  \left(
  1 + 
  \penscalar
    \left(
    \penscalar^{\frac{1}{\mininventory}
    }
    - 1
    \right)
  \right)
  \right)
  \\
  \overset{(e)}{\leq}{} &
  \maxleg\cdot \sum\nolimits_{i\in[n]}\consump_{i\jobconfig j}\cdot 
  \left(
  \reward_{\jobconfig j} \duration_{\jobconfig j}
  \ln\left(\penscalar\right)
  +
  \reward_{\jobconfig j}\duration_{\jobconfig j}
 	\penscalarTwo 
      \ln\left(\penscalar\right) 
  \left(
  1 + 
  \penscalar
    \left(
    \penscalar^{\frac{1}{\mininventory}
    }
    - 1
    \right)
  \right)
  \right)
  \\
  \overset{}{=}{}&
   \Gamma\cdot \sum\nolimits_{i\in[n]}\consump_{i\jobconfig j}\cdot 
  \reward_{\jobconfig j}\duration_{\jobconfig j}
  =  \Gamma\cdot\Delta(\textsc{primal})~.
\end{align*}
In the above derivation, inequality~(a) holds since $\maxleg \geq 1$.
Inequality~(b) holds since $\ln(\penscalar) \geq 1$ for all $\penscalar \geq e$, and the reduced reward of job $j$ under meta-assignment $\jobconfig$, is positive by the assumption that the {\FLB} algorithm assigns job $j$ under meta-assignment $\jobconfig$. Moreover, inequality~(c) holds since the penalty function $\pen$ is convex, and thus $\inventory_i \left(
\pen\left({\estimateinventoryij}\right) -
\pen\left(\estimateinventoryij - \frac{1}{\inventory_i}\right)\right)
\geq \pen'\left(\estimateinventoryij - \frac{1}{\inventory_i}\right).$ Also, inequality~(d) follows from algebraic simplifications.
Finally, inequality~(e) holds since 
$|\timeset_{\jobconfig j}| = \duration_{\jobconfig j}$ and $\reward_{\jobconfig j} \geq 1$.

\smallskip
\noindent[\emph{Step iii}]
\emph{Comparing the optimal offline benchmark with the constructed solution.} Next, we show that the total reward of the optimal offline benchmark is at most the summation of the constructed solution, i.e., $\sum\nolimits_{j\in[m]}\onlinedual(j) + \sum\nolimits_{i\in[n]}\offlinedual(i)$. 

First, due to the greedy-style selection rule in the {\FLB} and the construction of $\onlinedual(j)$ for each online job $j\in[m]$, we can lower bound $\onlinedual(j)$ as 
\begin{align*}
    \onlinedual(j) = \reducedreward(j, \jobconfig_j) \geq 
    \reducedreward(j, \jobconfig_j^*) = 
    \sum\nolimits_{i\in[n]}
    \consump_{i\jobconfig_j^* j}
    \cdot 
    \reward_{\jobconfig_j^* j}\duration_{\jobconfig_j^* j}
    -
    \maxleg\cdot 
    \left(\sum\nolimits_{i\in[n]}
    \consump_{i\jobconfig_j^* j}
    \cdot
    \sum\nolimits_{\tpre \in \timeset_{\jobconfig_j^* j}}
 	\pen\left(
 	\estimateinventoryij
 	\right)\right),
\end{align*}
where $\jobconfig$ and $\jobconfig^*$ are the meta-assignments chosen for online job $j$ by {\FLB} and the optimal offline benchmark, respectively (if job $j$ is rejected, we set the corresponding meta-assignment to $\emptyset$). 

For each server $i\in[n]$ and each of its unit $k\in[\inventory_i]$, 
let $\config^*_{ik}\subseteq[m]$ denote the subset of online jobs in the optimal offline benchmark that all occupy unit $k$ of server $i$ at different times under their corresponding meta-assignments $\jobconfig_j^*$. By definition, the total reward of the optimal offline benchmark can be expressed as
\begin{align*}
    \sum\nolimits_{i\in[n]}\sum\nolimits_{k\in[\inventory_i]}
    \sum\nolimits_{j \in \config^*_{ik}}
    \reward_{\jobconfig_j^* j}\duration_{\jobconfig_j^* j}
\end{align*}
Meanwhile, the summation of the constructed solution, i.e., $\sum\nolimits_{j\in[m]}\onlinedual(j) + \sum\nolimits_{i\in[n]}\offlinedual(i)$, can be lower bounded by the following expression:
\begin{align*}
    &\sum\nolimits_{j\in[m]}\onlinedual(j) + \sum\nolimits_{i\in[n]}\offlinedual(i)
    \\
    \overset{(a)}{\geq} &
    \sum\nolimits_{j\in[m]}
    \sum\nolimits_{i\in[n]}
    \consump_{i\jobconfig_j^* j}
    \cdot 
    \reward_{\jobconfig_j^* j}\duration_{\jobconfig_j^* j}
    -
    \maxleg\cdot 
    \left(\sum\nolimits_{i\in[n]}
    \consump_{i\jobconfig_j^* j}
    \cdot
    \sum\nolimits_{\tpre \in \timeset_{\jobconfig_j^* j}}
 	\pen\left(
 	\estimateinventoryij
 	\right)\right)
    + \sum\nolimits_{i\in[n]}\offlinedual(i)
    \\
    ={} &
    \sum\nolimits_{i\in[n]}\sum\nolimits_{k\in[\inventory_i]}
    \left(
    \sum\nolimits_{j \in \config^*_{ik}}
    \left(
    \reward_{\jobconfig_j^* j}\duration_{\jobconfig_j^* j}
    -
    \maxleg\cdot \sum\nolimits_{\tpre \in \timeset_{\jobconfig_j^* j}}
 	\pen\left(
 	\estimateinventoryij
 	\right)
    \right)
    +
    \frac{1}{\inventory_i}
    \cdot
    \offlinedual(i)
    \right)
\end{align*}
where the inequality holds due to lower bound of $\onlinedual(j)$ derived above, and the equality holds due to the definition of subset $\config^*_{ik}$.

Combining the above two expressions, it suffices the show that for each server $i\in[n]$ and each of its unit $k\in[\inventory_i]$, the following inequality holds 
\begin{align}
\label{eq:NRM:feasible}
    \frac{1}{\inventory_i}\cdot \offlinedual(i) 
    -
    \maxleg\cdot
    \sum\nolimits_{j\in{\config^*_{ik}}}
    \sum\nolimits_{\tpre\in\timeset_{\jobconfig_j^* j}}
 	\pen\left(
 	{\estimateinventoryij}{}
 	\right)
    \geq 0~.
\end{align}
The proof of inequality~\eqref{eq:NRM:feasible} is almost the same as the one for inequality~\eqref{eq:feasible}. We include it below for the completeness.

Fix any server $i\in[n]$ and any unit $k\in[\inventory_i]$. We abbreviate $\config^*_{ik}$ as $\config$.
We prove inequality~{\eqref{eq:NRM:feasible}} with a ``charging argument.'' At a high level, our goal is to identify values $\{\offlinedualdecompi\}_{j\in\config,\tpre \in \timeset_{\jobconfig_j^* j}}$
such that 
\begin{align}
\label{eq:NRM:decomposed offline dual divisible}
  \forall j\in\config, \tpre\in\timeset_{\jobconfig_j^* j}:~~ \offlinedualdecompi - \pen(\estimateinventoryij)  \geq 0
    \;\; 
    \text{and}
    \;\;
\maxleg\cdot \sum\nolimits_{j\in\config}\sum\nolimits_{\tpre\in\timeset_{\jobconfig_j^* j}}\offlinedualdecompi \leq \frac{1}{\inventory_i}\cdot \offlinedual(i)~,
\end{align}
which then implies inequality~\eqref{eq:NRM:feasible}. In the following, we provide details of this charging argument.

Let us introduce an auxiliary notation
$\timesetiS \triangleq
\{(j,\tpre): j\in \config, \tpre \in \timeset_{\jobconfig_j^* j}\}$
for each server $i\in[n]$ and subset $\config$.
Recall that subset $\config$ contains all online jobs such which all occupy unit $k$ of server $i$ at different times under their corresponding
meta-assignments $\jobconfig^*_j$ in the optimal offline benchmark. Hence, processing-time intervals of any two distinct jobs $j_1, j_2\in\config$ under their corresponding meta-assignments in the optimal offline benchmark
do not overlap, and thus
$\timeset_{\jobconfig_{j_1}^* j_1} \cap \timeset_{\jobconfig_{j_2}^* j_2} = \emptyset$.
Consequently,
for every pair $(j_1, \tpre_1), (j_2,\tpre_2) \in \timesetiS$,
if $\tpre_1 \not = \tpre_2$, then $j_1 \not = j_2$.
We now identify a one-to-many  correspondence $\chargemapping: \timesetiS\to[m]\times [T]$ that satisfies the following three properties.
\begin{itemize}
    \item[-] {\cseparation:}
    For every pair of distinct elements $(j_1, \tpre_1), (j_2,\tpre_2)\in \timesetiS$,
    it holds that $\chargemapping(j_1,\tpre_1) \cap \chargemapping(j_2,\tpre_2) = \emptyset$.
    
    \item[-] {{\cfeasibilitya:}}
    For every $(j, \tpre)\in\timesetiS$, 
    we have $|\chargemapping(j, \tpre)| = \inventory_i - \inventory_i\estimateinventoryij$.
    Moreover, for any two distinct elements $(j_1,\tpre_1), (j_2, \tpre_2) \in \chargemapping(j, \tpre)$,
    it holds that $j_1 \neq j_2$.
    
    \item[-] {{\cfeasibilityb:}}
    For every $(j, \tpre)\in\timesetiS$ 
    and every $(j', \tpre') \in \chargemapping(j, \tpre)$, 
    we have $j' < j$,
    $\tpre' \leq \tpre$,
    $\tpre'\in\timeset_{\jobconfig_{j'} j'}$,
    and a unit of server $i$ is assigned to 
    job $j'$ under meta-assignment $\jobconfig_{j'}$ in the {\FLB} with duration $\duration_{\jobconfig_{j'} j'} > \tpre - t_{j'}$.
\end{itemize}
Intuitively, the mapping $\sigma$ serves to synchronize the duration of different jobs to a common reference time frame. Before demonstrating the existence of such a correspondence $\chargemapping$, we first illustrate how to construct 
values $\{\offlinedualdecompi\}_{j\in\config,\tpre \in \timeset_{\jobconfig_j^* j}}$
that satisfy condition~\eqref{eq:NRM:decomposed offline dual divisible} given the correspondence $\chargemapping$. Specifically, for every $j\in\config$ and $\tpre\in\timeset_{\jobconfig_j^* j}$, define
\begin{align*}
    \offlinedualdecompi \triangleq 
    \sum\nolimits_{(j', \tpre')\in\chargemapping(j, \tpre)} 
    \left(
    \pen\left(
    \curinventory_{i, t_{j'}\rightarrow\tpre'}
    - \frac{1}{\inventory_i}
    \right)
    -
    \pen\left(
    \curinventory_{i, t_{j'}\rightarrow\tpre'}
    \right)
    \right)
\end{align*}
Note that property~\cseparation\ and property~\cfeasibilityb\ of the correspondence $\chargemapping$ guarantee the second half of condition~\eqref{eq:NRM:decomposed offline dual divisible}, i.e., 
$\sum_{j\in\config}\sum_{\tpre\in\timesetij}\offlinedualdecompi \leq \offlinedual(i)/\inventory_i$.
To establish the first half of condition~\eqref{eq:NRM:decomposed offline dual divisible}, i.e., 
$\offlinedualdecompi -\pen(\estimateinventoryij) \geq 0$, observe that
\begin{align*}
    \offlinedualdecompi - \pen(\estimateinventoryij)
    &=
    \sum\nolimits_{(j', \tpre')\in\chargemapping(j, \tpre)} 
    \left(
    \pen\left(
    \curinventory_{i, t_{j'}\rightarrow\tpre'}
    - \frac{1}{\inventory_i}
    \right)
    -
    \pen\left(
    \curinventory_{i, t_{j'}\rightarrow\tpre'}
    \right)
    \right)
    -
    \pen(\estimateinventoryij)
    \\
    &\overset{(a)}{\geq}
    \sum\nolimits_{(j', \tpre')\in\chargemapping(j, \tpre)} 
    \left(
    \pen\left(
    \curinventory_{i, t_{j'}\rightarrow\tpre}
    - \frac{1}{\inventory_i}
    \right)
    -
    \pen\left(
    \curinventory_{i, t_{j'}\rightarrow\tpre}
    \right)
    \right)
    -
    \pen(\estimateinventoryij)
    \\
    &\overset{(b)}{\geq}
    \sum\nolimits_{\ell \in [\inventory_i - \inventory_i\estimateinventoryij]}
    \left(
    \pen\left(
    \frac{\inventory_i - \ell}{\inventory_i}
    \right)
    -
    \pen\left(
    \frac{\inventory_i - \ell + 1}{\inventory_i}
    \right)
    \right)
    -
    \pen\left(\estimateinventoryij\right)
    \overset{(c)}{=}
    -\pen(1)
    \overset{(d)}{=}
    0.
\end{align*}
In the above derivation, inequality~(a) holds since the {\estimated} available capacity $\curinventory_{i,t_{j'}\rightarrow\tpre'}$ is weakly increasing in $\tpre'$, property~\cfeasibilityb\ of the correspondence $\chargemapping$ ensures $\tpre' \leq \tpre$, and the penalty function $\pen$ is convex. For inequality~(b), properties~\cfeasibilitya\ and~\cfeasibilityb\ of the correspondence $\chargemapping$ imply that there exist $s \triangleq \inventory_i - \inventory_i\estimateinventoryij$ jobs $j_1 < j_2 < \dots < j_s < j$, each of which is assigned to a different unit of server $i$ and occupies that unit at time $\tpre$. This further implies that, for each $\ell\in[s]$, $\curinventory_{i,t_{j_\ell}\rightarrow\tpre} \leq \frac{\inventory_i - \ell + 1}{\inventory_i}$. Applying the convexity of the penalty function $\pen$ completes the argument for inequality~(b). Finally, equality~(c) holds by simplifying algebra (noticing the telescopic summation), and equality~(d) follows since $\pen(1) = 0$.

Now we show the existence of the correspondence $\chargemapping$ satisfying properties~\cseparation, \cfeasibilitya, and \cfeasibilityb\ through the following explicit construction. Fix an arbitrary $(j,\tpre)\in\timesetiS$. As mentioned earlier, by definition of the {\estimated} available capacity $\estimateinventoryij$, there exist $s \triangleq \inventory_i - \inventory_i\estimateinventoryij$ jobs $j_1 < j_2 <  \dots <  j_s < j$, each of which is assigned to a distinct unit of server $i$ and occupies this unit at time $\tpre$. In other words, for every $\ell\in[s]$, we have $t_{j_\ell} + \duration_{\jobconfig_{j_\ell} j_\ell} \geq \tpre$. Moreover, there exists a unique $\tpre_\ell\in\timeset_{\jobconfig_{j_\ell} j_\ell}$ such that $\tpre_\ell \leq \tpre < \tpre_\ell + 1$. By defining 
\[
\chargemapping(j, \tpre) \triangleq 
\{(j_\ell, \tpre_\ell)\}_{\ell\in[s]},
\] 
properties~\cfeasibilitya\ and~\cfeasibilityb\ are straightforwardly satisfied.

To show property~\cseparation, fix two distinct elements $(j', \tpre'), (j'', \tpre'') \in \timesetiS$ such that $j' \leq j''$. It suffices to show that $|\tpre'' - \tpre'| \geq 1$. We consider two cases separately. If $j' = j''$, then for $\tpre' \neq \tpre''$, the construction of inspection-time subset $\timeset_{\jobconfig_{j'}^* j'}$ directly implies that $|\tpre'' - \tpre'| \geq 1$, as desired. If $j' < j''$, then since both $j', j''\in \config$, the inspection-time subsets $\timeset_{\jobconfig_{j'}^* j'}$ and $\timeset_{\jobconfig_{j''}^* j''}$ do not overlap, meaning that $t_{j'} + \duration_{ \jobconfig_{j'}^* j'} \leq t_{j''}$. Thus, we have
$
\tpre' \leq \max\timeset_{\jobconfig_{j'}^* j'} \leq t_{j'} + \duration_{\jobconfig_{j'}^* j'} \leq t_{j''} = \min\timeset_{\jobconfig_{j''}^* j''} \leq \tpre''.
$
Moreover, since all durations are integer-valued, we have
$\max \timeset_{\jobconfig_{j'}^* j'} = t_{j'} + \duration_{\jobconfig_{j'}^* j'} - 1$. Hence, 
$
|\tpre'' - \tpre'| = \tpre'' - \tpre' \geq t_{j''} - \max\timeset_{\jobconfig_{j'}^* j'} \geq 1,
$
as desired.

Finally, since the total reward of {\FLB} is a $\Gamma$-approximation of the summation over construction solution (Step ii), and the summation over construction solution is at least the total reward of the optimal offline benchmark (Step iii), we show the first guarantee in the proposition statement. The second guarantee follows the same analysis of \Cref{thm:competitive ratio integer duration} given in \Cref{subsec:CR optimization}.
\end{proofof}

\subsection{Multi-Unit Online Matching}
\label{apx:multi-unit}

The {\FLBLong} ({\FLB}) and its competitive ratio guarantees can be generalized to the extension model where assigning online job $j$ to server $i$ occupies a fraction $\consump_{ij}\in[0, 1]$ of server $i$'s units from time $t_j$ to $t_j + \durationij$ and generates a total reward of $\consump_{ij} \cdot \rewardij\durationij$.\footnote{\revcolor{Here we normalize demands so that $\consump_{ij}\in[0, 1]$ for all $i$ and $j$. This ensures consistency with the large-capacity regime and the definition of $\mininventory$.}}
In this setting, we generalize the {\FLB}'s construction as follows: for each job $j\in[m]$ and each compatible server $i\in \neighbor(j)$, the reduced reward is computed as 
    $\consump_{ij}\cdot \left( \rewardij \durationij -
    \sum\nolimits_{\tpre \in \timesetij(\inspecfreq)}
    \pen\left(\estimateinventoryij\right)
    \right)$.
Based on the reduced reward, {\FLB} makes the same greedy-style decision that assign job $j$ with server $i^*$ with the highest positive reduced reward.

It can be checked that the capacity feasibility characterizations established in the baseline model continue to hold exactly, due to the normalization $\consump_{ij}\in[0, 1]$. The competitive ratio guarantees also carry over unchanged. The proof follows by directly modifying the LP-free analysis in \Cref{prop:NRM:competitive-ratio} and adapting the detailed arguments from \Cref{prop:FLB competitive ratio integer duration,prop:NRM:competitive-ratio}.\textsuperscript{\ref{footnote:focus-integer-duration}}

Specifically, for the dual solution construction (Step~i in the proof of \Cref{prop:NRM:competitive-ratio}), we initialize all variables $\{\onlinedual(j),\offlinedual(i)\}$ to zero. Whenever {\FLB} assigns job $j$ to server $i$, we set $\onlinedual(j)$ to the reduced reward of server $i$, i.e., $\consump_{ij}\cdot ( \rewardij \durationij -
    \sum\nolimits_{\tpre \in \timesetij(\inspecfreq)}
    \pen(\estimateinventoryij)
    )$
and we increase $\offlinedual(i)$ by
$\consump_{ij}\inventory_i\cdot \sum_{\tpre\in\timesetij} (
\pen(
\estimateinventoryij - \frac{\consump_{ij}}{\inventory_i}
)
-
\pen(
\estimateinventoryij
)
)$.

To show that the total reward of {\FLB} is a $\Gamma$-approximation to $\sum_{i\in[n]}\offlinedual(i)+\sum_{j\in[m]}\onlinedual(j)$, where $\Gamma$ is as stated in \Cref{prop:FLB competitive ratio integer duration} (cf.\ Step~i in the proof of \Cref{prop:NRM:competitive-ratio}), the same analysis applies, except that both the primal and dual increments are scaled by $\consump_{ij}$.

To show that the total reward of the optimal offline benchmark is at most $\sum_{i\in[n]}\offlinedual(i)+\sum_{j\in[m]}\onlinedual(j)$ (cf.\ Step~iii in the proof of \Cref{prop:NRM:competitive-ratio}), we adapt the argument from \Cref{prop:NRM:competitive-ratio}: we first lower bound $\sum_{j\in[m]}\onlinedual(j)$ by the sum of reduced rewards obtained by assigning all jobs according to the optimal offline benchmark. We then argue that this lower bound plus $\sum_{i\in[n]}\offlinedual(i)$ is at least the optimal offline reward, by analyzing each server separately. For each server, we can partition its inventory into slices such that, within each slice, the assignments of jobs using that slice do not overlap. (In the baseline model and the NRM extension, unit-level slicing suffices because demands are binary; in this extension, the slices may be finer, but they still exist.) For each fixed slice of each fixed server, the remaining argument is identical to the charging argument in the baseline model and the NRM extension. We omit the formal proof to avoid repetition.

}

\subsection{Online Matching with Server Dependent Heterogeneity}
\label{apx:server-dependent}

The \FLBLong\ ({\FLB}) and its competitive ratio guarantees can be generalized to the extension model where different servers have different range of rewards and durations. In particular, consider the setting where each server $i\in[n]$ is associated with $(\minreward^{(i)}, \minduration^{(i)})$, and the range of reward (duration) between server $i$ and its compatible job $j\in N^{-1}(i)$ is $[\minreward^{(i)}, \maxreward\cdot\minreward^{(i)}]$ ($[\minduration^{(i)}, \maxreward\cdot\minduration^{(i)}]$). In this setting, we generalize \FLB's construction as follows: for each job $j\in[m]$ and each compatible server $i\in N(j)$, the inspection time subset $\timesetij$ given inspection-frequency scalar $\inspecfreq$ is 
    $\timesetij(\inspecfreq) \triangleq
    \left\{\tpre 
    \in [t_j, t_j + \durationij):
    \exists \ell\in\naturals
    \text{ s.t. }
    \tpre = 
    t_j + \tfrac{\ell}{\inspecfreq}\cdot 
    \minduration^{(i)}
    \right\}$
and the reduced reward is computed as 
    $\rewardij \durationij -
    \sum\nolimits_{\tpre \in \timesetij(\inspecfreq)}
    \pen\left(\estimateinventoryij\right)\cdot \minreward^{(i)}\minduration^{(i)}$.
Finally, \FLB \ makes the same greedy-style decision that assign job $j$ with server $i^*$ with the highest positive reduced reward.

For the capacity feasibility characterization and competitive ratio guarantees, it can be checked that all statements such as \Cref{thm:FLB competitive ratio general,thm:competitive ratio integer duration}, \Cref{prop:FLB feasibility integer duration}, \Cref{prop:CR identical reward integer duration,prop:CR identical reward real duration} as well as other technical lemmas continue to hold.


\revcolor{\section{The Impact of Parameter Knowledge}}
\label{apx:parameterKnowledge}
\revcolor{In this section, we investigate how restricted information regarding job and system parameters impacts the achievable competitive ratio. We approach this from two different scales. First, we establish a fundamental lower bound for algorithms that are entirely agnostic to the global system scales, $\maxreward$ and $\maxduration$. Second, we analyze the robustness of {\FLB}, when the specific characteristics of an arriving job, its individual reward $r_j$ and duration $d_j$, are subject to estimation error or noise.}

\newcommand{\overestimateduration}{\overline{d}}
\newcommand{\overestimatereward}{\overline{r}}

\newcommand{\underestimateduration}{\underline{d}}
\newcommand{\underestimatereward}{\underline{r}}

\newcommand{\estimaterewarderr}{\kappa_{\reward}}
\newcommand{\estimatedurationerr}{\kappa_{\duration}}

\newcommand{\noisyinstance}{\iota}

\revcolor{\subsection{Competitive Ratio Lower Bound of \texorpdfstring{$(\maxreward,\maxduration)$-Agnostic}{(\maxreward,\maxduration)-agnostic} Algorithms}\label{apx:agnostic}}
In this subsection, we provide hardness results of $(\maxreward,\maxduration)$-agnostic algorithms. Specifically, we say a $(\maxreward,\maxduration)$-agnostic algorithm \revcolor{has} a competitive ratio $O(f(\maxreward,\maxduration))$ for some function $f$ if for every $\maxreward\geq 1$ and $\maxduration \geq 1$, its competitive ratio is $O(f(\maxreward,\maxduration))$ among all instances with maximum reward $\maxreward$ and maximum duration $\maxduration$. We establish competitive ratio lower bounds for $(\maxreward,\maxduration)$-agnostic deterministic algorithms and $(\maxreward,\maxduration)$-agnostic random algorithms, respectively.

\begin{proposition}[Deterministic $(\maxreward,\maxduration)$-Agnostic Negative Result]
\label{prop:deterministic agnostic LB}
    There exists no $(\maxreward,\maxduration)$-agnostic deterministic integral online algorithm with competitive ratio $o(\maxreward\maxduration)$, even under large capacity.
\end{proposition}
\begin{proofof}{Proof of \revcolor{\Cref{prop:deterministic agnostic LB}}}
    Consider an example with only one server with capacity $c\in\naturals$. Assume the following instance: at time $0$, $c$ jobs arrive with with reward $r = \sqrt{\ell}$ and duration $d = \sqrt{\ell}$ for each $\ell\in[1:L]$ in an increasing order. Here $L$ is picked by the adversary. Since the algorithm is $(\maxreward,\maxduration)$-agnostic, for every $\ell$, the algorithm makes the same assignment decision for jobs with $r= d = \sqrt{\ell}$ for all examples with $L\geq \ell$. Thus, let $x_\ell\in\naturals$ be the number of jobs with $r=d=\sqrt{\ell}$ accepted by the algorithm. Notice that the capacity constraint says:
        $\sum\nolimits_{\ell=1}^\infty x_\ell \leq c$.
    This defines a game between the algorithm and the adversary. The algorithm picks $x_\ell$ and the adversary picks $L$. Since $x_\ell$ are non-negative integer, there is a constant $\hat{L}$ where $x_\ell = 0$ for all $\ell > \hat{L}$. Thus, if the adversary picks $L = \omega(\hat{L})$, the revenue of the algorithm is at most $O(c\hat{L}) = O(c)$, while the optimal offline benchmark is $cL$. Consequently, the competitive ratio is $\Omega(L)=\Omega(\maxreward\maxduration)$.
\end{proofof}
\begin{proposition}[$(\maxreward,\maxduration)$-agnostic Negative Result]
\label{prop:random agnostic LB}
     There exists no $(\maxreward,\maxduration)$-agnostic (possibly fractional or randomized) online algorithm with competitive ratio $O(\log(\maxreward\maxduration)\log\log(\maxreward\maxduration))$, even under large capacity.
\end{proposition}
\begin{proofof}{Proof \revcolor{of \Cref{prop:random agnostic LB}}}
    Consider the same example mentioned in \Cref{prop:deterministic agnostic LB}. We prove the proposition statement by contradiction. Fix an arbitrary constant $\alpha> 0$. Assume there exist a series of $x_\ell$ such that $\sum\nolimits_{r=\ell}^{\infty} x_\ell \leq c$ and for all $L \geq 2$:
    \begin{align*}
        \sum\nolimits_{\ell=1}^{L} \ell\cdot x_\ell \geq \frac{\alpha}{\ln(L)\ln\ln(L)}cL
    \end{align*}
    where the left-hand side is the revenue of the algorithm, and the right-hand side is $\frac{\alpha}{\ln(L)\ln\ln(L)}$-fraction of the revenue in the optimal offline benchmark.
    Multiplying above inequality by $\frac{1}{L}-\frac{1}{L+1} = \frac{1}{L(L+1)}$ and summing up for all $L$ gives us:
    \begin{align*}
        c \geq \sum\nolimits_{L=1}^{\infty} x_\ell \geq \alpha \sum\nolimits_{L=2}^\infty \frac{1}{(L+1)\ln(L)\ln\ln(L)}c
    \end{align*}
    which is a contradiction, since the series on right-hand side diverges. 
\end{proofof}

\revcolor{\subsection{Performance under Noisy Parameter Estimates}\label{apx:noisy}
While the previous subsection addresses uncertainty regarding the range of the instance, we now consider uncertainty regarding the specifics of the arrivals. In some real-world applications such as cloud scheduling, the exact duration $\duration_{ij}$ or reward $\reward_{ij}$ of a job may only be known through an imperfect estimate at the time of arrival. Instead, the algorithm only observes uncertainty intervals $[\underestimateduration_{ij}, \overestimateduration_{ij}]$ and $[\underestimatereward_{ij}, \overestimatereward_{ij}]$ containing the true values. We parameterize the sharpness of these estimates by bounded uncertainty ratios, where $\overestimateduration_{ij} / \underestimateduration_{ij} \leq \estimatedurationerr$ and $\overestimatereward_{ij} / \underestimatereward_{ij} \leq \estimaterewarderr$. To reflect the most restrictive setting, we assume that the decision-maker does not learn the true values of $\duration_{ij}$ and $\reward_{ij}$ until the job completely finishes its execution. In addition, for ease of presentation, we only consider the integer-duration environment for this extension. (The extension to the real-duration environment follows the same approach.)

Crucially, this formulation encapsulates our baseline setting as a special case: when $\estimatedurationerr = \estimaterewarderr = 1$, the uncertainty intervals collapse to singletons, and the analysis exactly recovers the performance guarantees of the fully-informed model. Here, we analyze the performance of {\FLB} when it must make routing decisions based on these uncertainty interval observations. We show that the algorithm's competitive ratio degrades gracefully as a function of the uncertainty magnitudes $\estimatedurationerr$ and $\estimaterewarderr$, demonstrating that {\FLB} remains robust to localized inaccuracies in job profiling.

To establish a general result for $\estimatedurationerr, \estimaterewarderr > 1$, we first consider the simpler case where uncertainty is isolated to the rewards and job durations are perfectly known ($\estimatedurationerr = 1$). While the objective value scales linearly with the noise in both reward and duration, noise in the duration parameters also directly interferes with capacity feasibility, introducing additional structural complexity.
\begin{proposition}[Noisy Reward Result]
\label{prop:NoisyR}
Consider the integer-duration environment.
When $\estimatedurationerr = 1$, the asymptotic competitive ratio of a conservative variant of {\FLB} (that makes all routing decisions based on the reward lower bounds $\{\underestimatereward_{ij}\}$) scales linearly in $\estimaterewarderr$. More specifically, it is at most
\begin{align*}
\estimaterewarderr\compratio(\maxreward,\maxduration),
\end{align*}
where 
$\compratio(\maxreward,\maxduration) = \ln(\maxreward\maxduration) + \ln\ln(\maxreward\vee\maxduration) + O(1)$.
\end{proposition}
\begin{proofof}{Proof of \Cref{prop:NoisyR}}
Let $\noisyinstance$ denote the true noisy problem instance, where the exact reward $\reward_{ij}$ of an arriving job is unknown but guaranteed to lie within the interval $[\underestimatereward_{ij}, \overestimatereward_{ij}]$. Because $\estimatedurationerr = 1$, job durations are perfectly known (i.e., $\duration_{ij} = \underestimateduration_{ij} = \overestimateduration_{ij}$). Let \FLB\ operates conservatively on $\noisyinstance$ by making all routing decisions based exclusively on the reward lower bounds $\underestimatereward_{ij}$.

To analyze this behavior, we construct an auxiliary, fully-informed baseline instance $\noisyinstance\primed$ where the true rewards are fixed exactly to these lower bounds ($\reward\primed_{ij} = \underestimatereward_{ij}$ for all $i,j$) and job durations remain unchanged. Let $\FLB(\cdot)$ and $\OPT(\cdot)$ denote the total rewards collected by the online algorithm and the offline optimal policy on a given instance, respectively. We establish the following chain of inequalities:
\begin{align*}
    \FLB(\noisyinstance) \geq \FLB(\noisyinstance\primed) \geq \frac{1}{\compratio(\maxreward,\maxduration)} \OPT(\noisyinstance\primed) \geq \frac{1}{\estimaterewarderr\compratio(\maxreward,\maxduration)} \OPT(\noisyinstance).
\end{align*}
These three steps are justified as follows:
\begin{enumerate}
    \item[(i)] $\FLB(\noisyinstance) \geq \FLB(\noisyinstance\primed)$: Because the true realized rewards in the noisy instance satisfy $\reward_{ij} \geq \underestimatereward_{ij} = \reward\primed_{ij}$, the algorithm collects a total reward on $\noisyinstance$ that is at least as large as what it earns on $\noisyinstance\primed$.
    \item[(ii)] $\FLB(\noisyinstance\primed) \geq \frac{1}{\compratio(\maxreward,\maxduration)} \OPT(\noisyinstance\primed)$: This follows directly from the performance guarantees established for \FLB\ under the fully-informed baseline model (\Cref{thm:competitive ratio integer duration,thm:FLB competitive ratio general}).
    \item[(iii)] $\OPT(\noisyinstance\primed) \geq \frac{1}{\estimaterewarderr} \OPT(\noisyinstance)$: By definition of the reward uncertainty ratio, the true rewards are bounded above by $\reward_{ij} \leq \overestimatereward_{ij} \leq \estimaterewarderr \underestimatereward_{ij} = \estimaterewarderr \reward\primed_{ij}$. Consequently, the total reward accumulated by any feasible offline schedule in $\noisyinstance$ can be at most $\estimaterewarderr$ times its reward in $\noisyinstance\primed$.
\end{enumerate}
Rearranging the combined inequality yields $\OPT(\noisyinstance) / \FLB(\noisyinstance) \leq \estimaterewarderr \compratio(\maxreward,\maxduration)$, which proves that the competitive ratio degrades by a factor of at most $\estimaterewarderr$. This bound is tight in the worst-case scenario where the algorithm only admits jobs yielding their lower-bound rewards $\underestimatereward_{ij}$, while the offline optimum selectively processes jobs achieving their upper-bound rewards $\overestimatereward_{ij}$.
\end{proofof}
Before proving the general result for $\estimatedurationerr, \estimaterewarderr > 1$, we illustrate the fundamental complexity introduced by duration uncertainty using a concrete example.

\begin{example}
Consider a setting with a single server ($n=1$) having capacity $\inventory_1$. At time $t = 0$, the first batch of $2\inventory_1$ jobs arrives. Every job $j$ in this set presents an identical reward $\reward_{1j} = 1$ and an identical observed duration interval of $[\underestimateduration_{1j}, \overestimateduration_{1j}] = [\maxduration-1, \maxduration]$. To ensure that the maximum duration parameter is formally well-defined, we assume a dummy job of duration $1$ also arrives at $t=0$. Crucially, the true durations of these first $2\inventory_1$ jobs are heterogeneous: exactly half ($\inventory_1$) have a true duration of $\maxduration-1$, while the remaining half have a true duration of $\maxduration$. Subsequently, at time $t = \maxduration - 1$, a second batch of $\inventory_1$ jobs arrives, each having a known true duration of $\maxduration$.

The optimum offline policy ($\OPT$) exploits its knowledge of the true durations and accepts the $\inventory_1$ jobs with duration $\maxduration-1$ in the first batch and the entire second batch of $\inventory_1$ jobs arriving at $t=\maxduration-1$, yielding a total reward of $\inventory_1(2\maxduration-1)$. In contrast, because the initial $2\inventory_1$ jobs appear completely identical to the online decision-maker at $t=0$, any online algorithm (including {\FLB}) is structurally incapable of distinguishing between them. In the worst-case adversarial realization, all jobs selected by the online algorithm turn out to be precisely those with the longer true duration $\maxduration$. As a result, the online algorithm accumulates a total reward of at most $\inventory_1\maxduration$.

Consequently, the online algorithm cannot achieve a competitive ratio better than $\frac{1}{2}$ of the optimal solution. This competitive gap represents an entirely new dimension of hardness that is completely absent under perfect duration information, demonstrating that duration uncertainty disrupts {\FLB} performance in a way that reward uncertainty does not.
\end{example}
Now, we turn to present and prove the main result of this section under joint uncertainty ($\estimatedurationerr, \estimaterewarderr > 1$). To accommodate interval uncertainty in both channels, we augment the assignment primitive of \FLB\ to incorporate robust, worst-case estimates. Specifically, upon the arrival of job $j$ at time $t_j$, the augmented algorithm routes the job to the server $i^*$ that maximizes a conservative surrogate welfare metric:
\begin{align}
\label{eq:robust-FLB-winner}
    i^* = \argmax_i \left\{ \underestimatereward_{ij}\underestimateduration_{ij} - \sum\nolimits_{\tpre \in \overline{\timeset}_{ij}(\inspecfreq)} \pen\left(\estimateinventoryij\right) \right\},
\end{align}
where the capacity penalty is evaluated over the maximum potential duration of the assignment, mapped by the discrete inspection set:
\begin{align}
\label{eq:robust-FLB-timeset}
    \overline{\timeset}_{ij}(\inspecfreq) \triangleq \left\{\tpre \in [t_j, t_j + \overestimateduration_{ij}): \exists \ell\in\naturals \text{ s.t. } \tpre = t_j + \tfrac{\ell}{\inspecfreq} \right\}.
\end{align}
Under this policy, \FLB\ simultaneously hedges against both dimensions of uncertainty. On the reward side, utilizing the lower bounds $\underestimatereward_{ij}$ and $\underestimateduration_{ij}$ protects the objective function against pessimistic, low-yielding realizations. On the capacity side, extending the penalty horizon to the upper bound $\overestimateduration_{ij}$ ensures that the algorithm remains highly risk-averse to severe, long-term capacity depletion from unexpectedly long-running jobs.
\begin{proposition}
[Noisy Reward \& Duration Result]
\label{prop:NoisyRD}
Consider the integer-duration environment.
The asymptotic competitive ratio of a robust variant of {\FLB} (see Eqn.~\eqref{eq:robust-FLB-winner} and \eqref{eq:robust-FLB-timeset}) 
is at most
\begin{align*}
\estimatedurationerr^2\estimaterewarderr\compratio(\maxreward,\maxduration),
\end{align*}
where 
$\compratio(\maxreward,\maxduration) = \ln(\maxreward\maxduration) + \ln\ln(\maxreward\vee\maxduration) + O(1)$.
\end{proposition}
\begin{proofof}{Proof of \Cref{prop:NoisyRD}} 
Given the pessimistic nature of the algorithm, capacity feasibility 
stays the same as proven in the baseline model. See \Cref{prop:FLB feasibility integer duration} for more details. 

For the competitive ratio analysis, we use the primal-dual analysis based on the configuration LP defined on the (unknown) true instance. Now consider every assignment of FLB.
For each job $j\in[m]$,
if FLB assigns
    server $i$ to job $j$, 
    update the dual variables as follows:
\begin{align*}
    \onlinedual(j) &\gets
    \underestimatereward_{ij}\underestimateduration_{ij}
    -
    \sum\nolimits_{\tpre \in \overline{\timeset}_{ij}}
 	\pen\left(
 	\estimateinventoryi
 	\right)
\\
    \offlinedual(i) &\gets
    \offlinedual(i) 
    +
    \sum\nolimits_{\tpre \in \overline{\timeset}_{ij}}
 	\left(
    \pen\left(
 	\estimateinventoryij- \frac{1}{\inventory_i}
 	\right)
  -
    \pen\left(
 	\estimateinventoryi
 	\right)
    \right)
\end{align*} 
The comparison between the primal and dual objective stays the same as the baseline model. (For more details see the proof of \Cref{prop:FLB competitive ratio integer duration}). 
The remaining step is to verify the dual feasibility of this constructed assignment. Below We demonstrate that the dual solution is $(\estimatedurationerr^2\estimaterewarderr)$-approximate feasible, which completes the proof.

Recall that for approximate dual feasibility, we must show that for every server $i \in [n]$ and all feasible configurations $\config$, the following inequality holds:
\begin{align*}
    \estimatedurationerr^2\estimaterewarderr \left(\sum\nolimits_{j\in\config}\onlinedual(j) + \offlinedual(i)\right) \geq \sum\nolimits_{j\in\config} \reward_{ij}d_{ij}
\end{align*}
Note that the job-side dual variable satisfies $\onlinedual(j) \geq 0$. Furthermore, by the greedy selection rule of {\FLB}, we can bound $\onlinedual(j)$ from below as follows:
\begin{align*}
    \onlinedual(j) &\geq 
        \underestimatereward_{ij}\underestimateduration_{ij} - \sum\nolimits_{\tpre \in \overline{\timeset}_{ij}(\inspecfreq)} \pen\left(\estimateinventoryij\right) \geq \underestimatereward_{ij}\underestimateduration_{ij} - \estimatedurationerr\sum\nolimits_{\tpre \in \underline{\timeset}_{ij}(\inspecfreq)} \pen\left(\estimateinventoryij\right)
\end{align*}
where the auxiliary inspection set over the lower-bound duration is defined as $\underline{\timeset}_{ij}(\inspecfreq) \triangleq \left\{\tpre \in [t_j, t_j + \underestimateduration_{ij}): \exists \ell\in\naturals \text{ s.t. } \tpre = t_j + \tfrac{\ell}{\inspecfreq} \right\}$. The second inequality stems from the property that the capacity penalty function $\pen\left(\estimateinventoryij\right)$ is monotonically decreasing in $\tau$. 

By scaling the joint dual expressions and leveraging the fact that the true reward and duration satisfy $\reward_{ij}\duration_{ij} \leq \estimaterewarderr \estimatedurationerr \underestimatereward_{ij}\underestimateduration_{ij}$, we can expand the aggregated dual variables over the configuration $\config$:
\begin{align*}
    \estimatedurationerr^2\estimaterewarderr \left(\sum\nolimits_{j\in\config}\onlinedual(j) + \offlinedual(i)\right) \geq{}& \estimatedurationerr\estimaterewarderr \sum\nolimits_{j\in\config}\onlinedual(j) + \estimatedurationerr^2\estimaterewarderr \offlinedual(i)
    \\ \geq{} & \sum\nolimits_{j\in\config} \left[\reward_{ij}\duration_{ij} -  \estimatedurationerr^2\estimaterewarderr \sum\nolimits_{\tpre \in \underline{\timeset}_{ij}(\inspecfreq)} \pen\left(\estimateinventoryij\right)\right] + \estimatedurationerr^2\estimaterewarderr \offlinedual(i)
\end{align*}
It remains to show that the accumulated server-side dual variable $\offlinedual(i)$ satisfies the following aggregated bound:
\begin{align*}
    \offlinedual(i) \geq& \sum\nolimits_{j\in\config} \sum\nolimits_{\tpre \in \underline{\timeset}_{ij}(\inspecfreq)} \pen\left(\estimateinventoryij\right).
\end{align*}
To establish this, we construct a charging argument by leveraging the same one-to-many correspondence $\sigma$ developed in our baseline analysis (\Cref{sec:competitive ratio analysis}). In this joint-uncertainty setting, the mapping $\sigma$ satisfies the following three relaxed properties:
\begin{itemize}
    \item[-] {\cseparation:}
    For every pair of distinct elements $(j_1, \tpre_1), (j_2,\tpre_2)\in \timesetiS$,
    it holds that $\chargemapping(j_1,\tpre_1) \cap \chargemapping(j_2,\tpre_2) = \emptyset$.
    \item[-] {{\cfeasibilitya:}}
    For every $(j, \tpre)\in\timesetiS$, 
    we have $|\chargemapping(j, \tpre)| \geq \inventory_i - \inventory_i\estimateinventoryij$.
    Moreover, for any two distinct elements $(j_1,\tpre_1), (j_2, \tpre_2) \in \chargemapping(j, \tpre)$,
    it holds that $j_1 \neq j_2$.
    \item[-] {{\cfeasibilityb:}}
    For every $(j, \tpre)\in\timesetiS$ 
    and every $(j', \tpre') \in \chargemapping(j, \tpre)$, 
    we have $j' < j$,
    $\tpre' \leq \tpre$,
    $\tpre'\in\timeset_{ij'}$,
    and a unit of server $i$ is assigned to 
    job $j'$ with maximum duration $\overline{\duration}_{ij'} > \tpre - t_{j'}$.
\end{itemize}
The existence proof for the mapping $\sigma$ proceeds analogously to the baseline deterministic setting, subject to two key structural modifications:
\begin{enumerate}
    \item[(i)] In Property~\cfeasibilitya, the cardinality condition holds with a weak inequality ($\geq$) because jobs may terminate strictly prior to their conservative upper bounds.
    \item[(ii)] In Property~\cfeasibilityb, the temporal boundary condition is governed by the maximum possible duration, yielding the inequality $\overestimateduration_{ij'} > \tpre - t_{j'}$. This directly reflects the algorithm's pessimistic design, where the server-side dual variable $\offlinedual(i)$ is proactively updated under the assumption that every accepted job will persist for its full upper-bound duration.
\end{enumerate}
To complete the proof, we execute a parallel charging argument by defining the incremental dual contribution mapped to a given job-time pair $(j, \tau)$ as follows:
\begin{align*}
    \offlinedualdecompi \triangleq 
    \sum\nolimits_{(j', \tpre')\in\chargemapping(j, \tpre)} 
    \left(
    \pen\left(
    \curinventory_{i, t_{j'}\rightarrow\tpre'}
    - \frac{1}{\inventory_i}
    \right)
    -
    \pen\left(
    \curinventory_{i, t_{j'}\rightarrow\tpre'}
    \right)
    \right)
\end{align*}
By evaluating this decomposition, we obtain:
\begin{align*}
    \offlinedualdecompi - \pen(\estimateinventoryij)
    &=
    \sum\nolimits_{(j', \tpre')\in\chargemapping(j, \tpre)} 
    \left(
    \pen\left(
    \curinventory_{i, t_{j'}\rightarrow\tpre'}
    - \frac{1}{\inventory_i}
    \right)
    -
    \pen\left(
    \curinventory_{i, t_{j'}\rightarrow\tpre'}
    \right)
    \right)
    -
    \pen(\estimateinventoryij)
    \\
    &\overset{(a)}{\geq}
    \sum\nolimits_{(j', \tpre')\in\chargemapping(j, \tpre)} 
    \left(
    \pen\left(
    \curinventory_{i, t_{j'}\rightarrow\tpre}
    - \frac{1}{\inventory_i}
    \right)
    -
    \pen\left(
    \curinventory_{i, t_{j'}\rightarrow\tpre}
    \right)
    \right)
    -
    \pen(\estimateinventoryij)
    \\
    &\overset{(b)}{\geq}
    \sum\nolimits_{\ell \in [\inventory_i - \inventory_i\estimateinventoryij]}
    \left(
    \pen\left(
    \frac{\inventory_i - \ell}{\inventory_i}
    \right)
    -
    \pen\left(
    \frac{\inventory_i - \ell + 1}{\inventory_i}
    \right)
    \right)
    -
    \pen\left(\estimateinventoryij\right)
    =
    -\pen(1)
    =
    0.
\end{align*}
Here, inequality (a) follows from an argument analogous to the baseline deterministic setting, leveraging the property that the projected capacity $a_{i,t_{j'}\rightarrow \tau'}$ is monotonically increasing in $\tau'$, coupled with the convexity of the penalty function $\pen(\cdot)$.\\
Inequality (b) is established by tracing the temporal trajectory of the projected capacity $a_{i,t\rightarrow \tau}$ at a fixed target time $\tau$ as the system time $t$ progresses. This trajectory evolves via two discrete, unit-sized mechanisms:
\begin{itemize}
    \item \textbf{Admission Decrements:} The admission of any online job $j$ whose maximum potential duration spans across $\tau$ (i.e., $t_j + \overestimateduration_{ij} > \tau$) decrements the projected available capacity by exactly $1/\inventory_i$.
    \item \textbf{Termination Increments:} Conversely, whenever an overlapping job terminates early prior to $\tau$, the server's capacity is incremented by $1/\inventory_i$.
\end{itemize}
Because all capacity updates occur strictly in discrete steps of $1/\inventory_i$ without arbitrary discontinuities, the trajectory satisfies a discrete intermediate value property. Consequently, for every integer capacity level $\ell \leq \inventory_i - \inventory_i\estimateinventoryij$, there must exist a unique pivotal job $j_\ell$ whose acceptance drives the precise step transition from an instantaneous pre-arrival state of 
\begin{align*}
    a_{i,t_{j_\ell}^-\rightarrow \tau} = \frac{\inventory_i - \ell + 1}{\inventory_i}
\end{align*}
to an instantaneous post-arrival state of 
\begin{align*}
    a_{i,t_{j_\ell}^+\rightarrow \tau} = \frac{\inventory_i - \ell}{\inventory_i},
\end{align*}
which validates the summation bounds in inequality (b). Note that while this inequality is structurally looser than its baseline counterpart due to potential upward capacity adjustments from early job completions, it remains sufficiently tight to guarantee approximate dual feasibility.
\end{proofof}
Finally, in \Cref{fig:noisyNumerics}, we evaluate the performance ratio of \FLB\ under varying noise levels $\estimatedurationerr, \estimaterewarderr \in \{1, 2,5\}$, using the synthetic problem instance detailed in \Cref{sec:RandomInstance} across capacities $\inventory_i \in \{5, 10, 20, 50\}$. As expected, the instance with $\estimaterewarderr = 2, \estimatedurationerr = 1$ exhibits the highest performance ratio across all capacity regimes over all noisy instances (i.e., either $\estimaterewarderr > 1$ or $\estimatedurationerr > 1$), followed by the instance with $\estimaterewarderr = 1, \estimatedurationerr = 2$. Interestingly, the performance trajectory for $\estimaterewarderr = \estimatedurationerr = 2$ closely tracks that of $\estimaterewarderr = 5, \estimatedurationerr = 1$. The performance ratio degrades further for the instance with $\estimaterewarderr = 1, \estimatedurationerr = 5$ case, ultimately reaching its minimum under the instance with $\estimaterewarderr = 5, \estimatedurationerr = 5$, which represents the highest level of joint uncertainty.}

\begin{figure}[ht]
    \centering
    \includegraphics[width=0.5\linewidth]{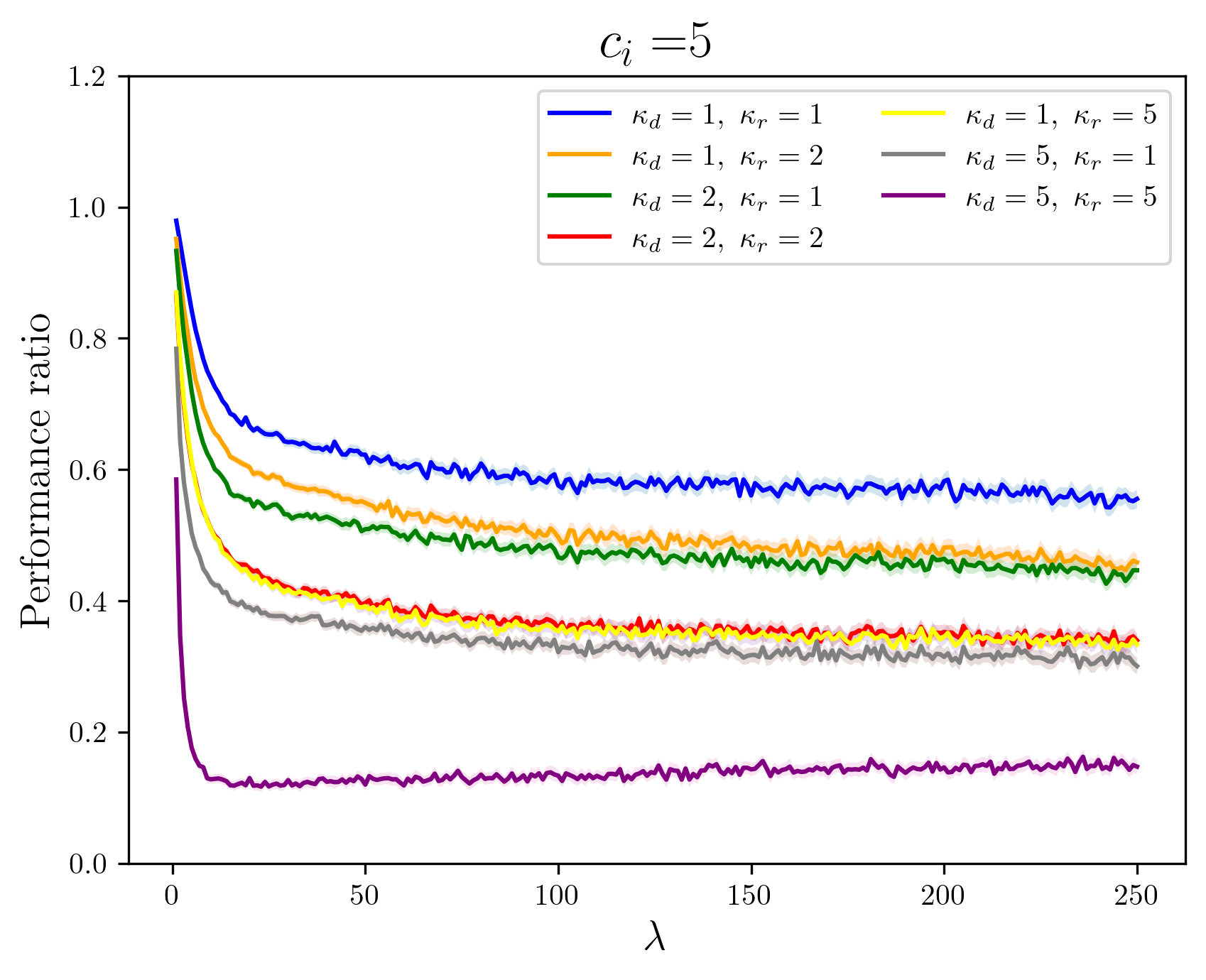}\includegraphics[width=0.5\linewidth]{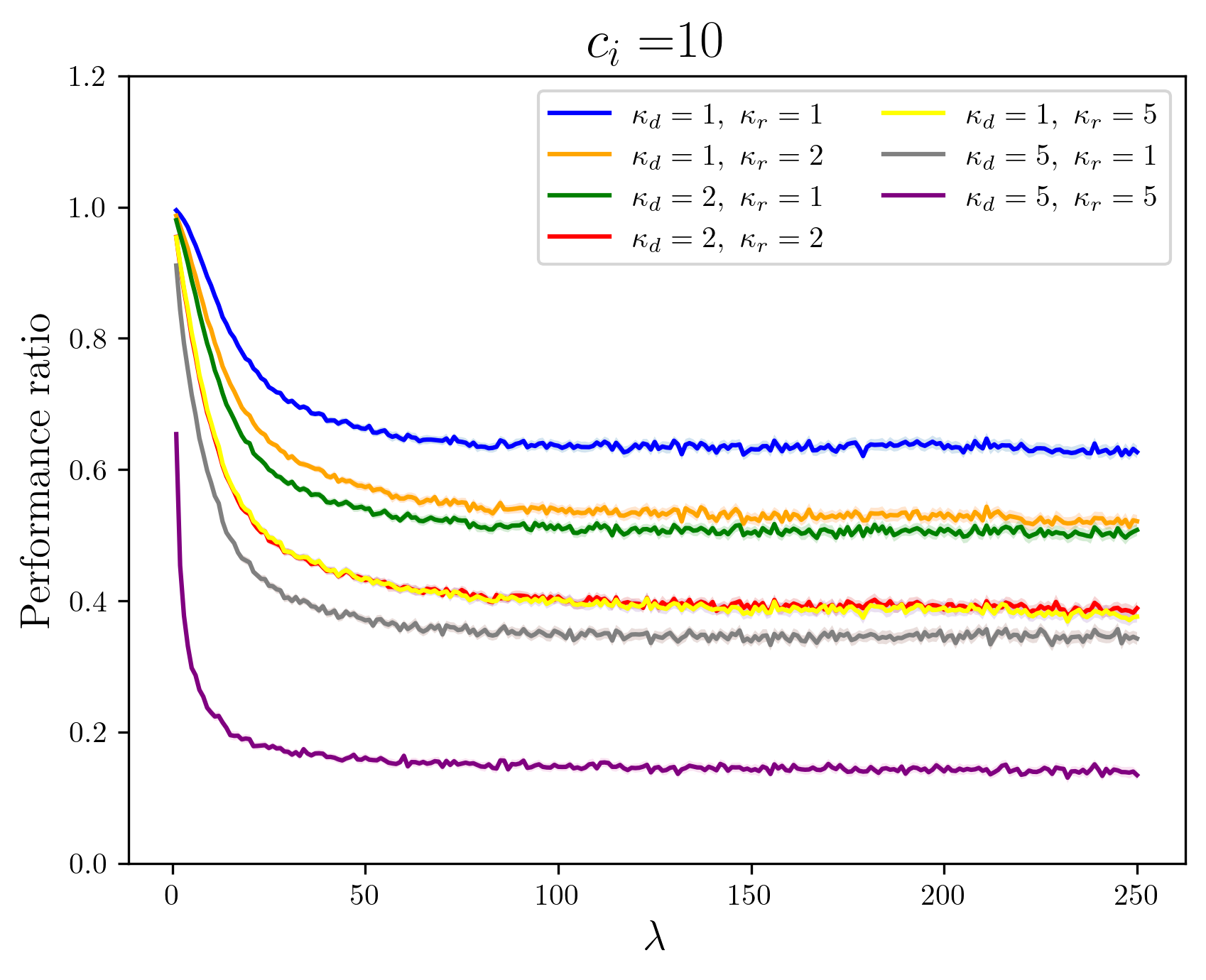}
    \\ \includegraphics[width=0.5\linewidth]{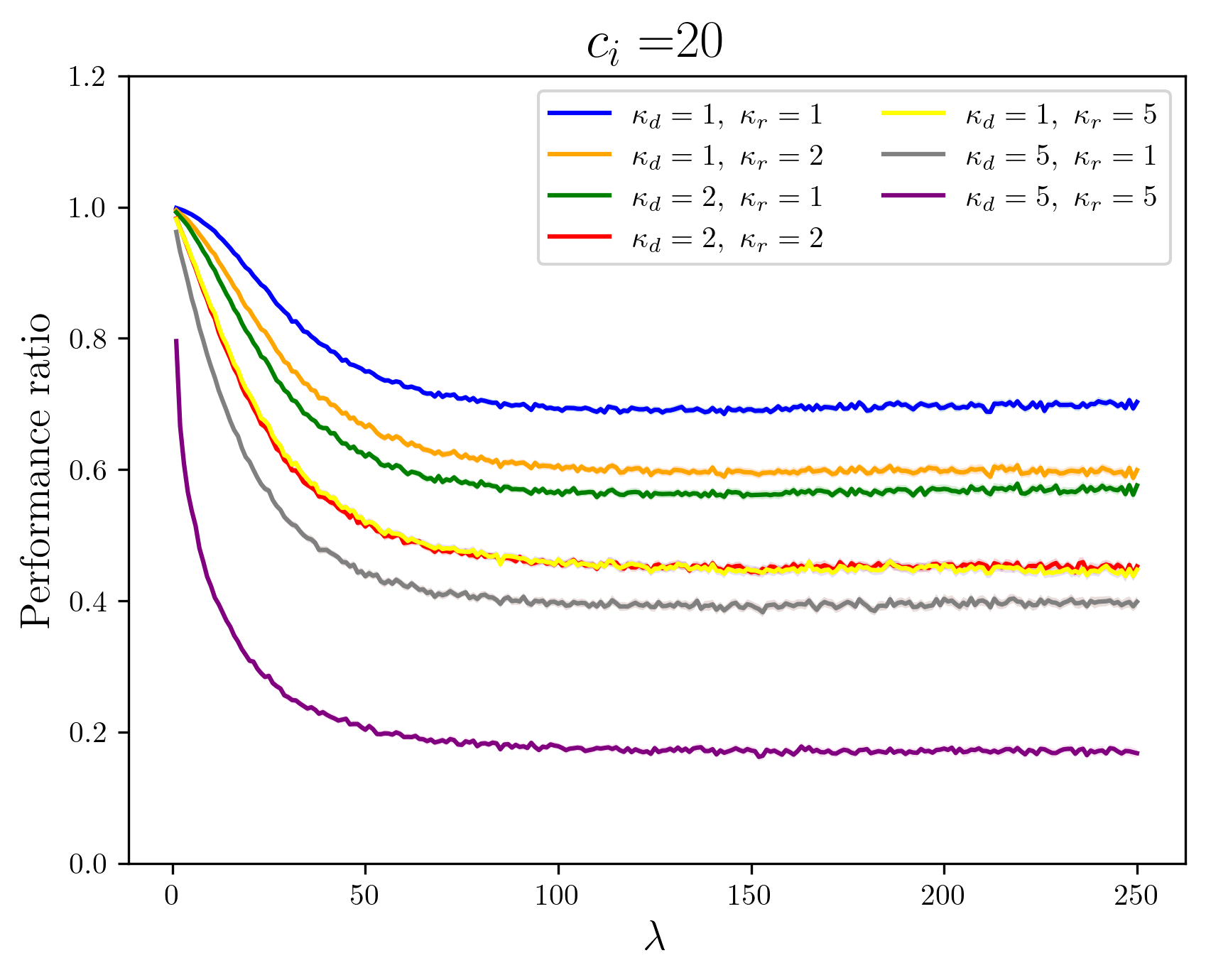}\includegraphics[width=0.5\linewidth]{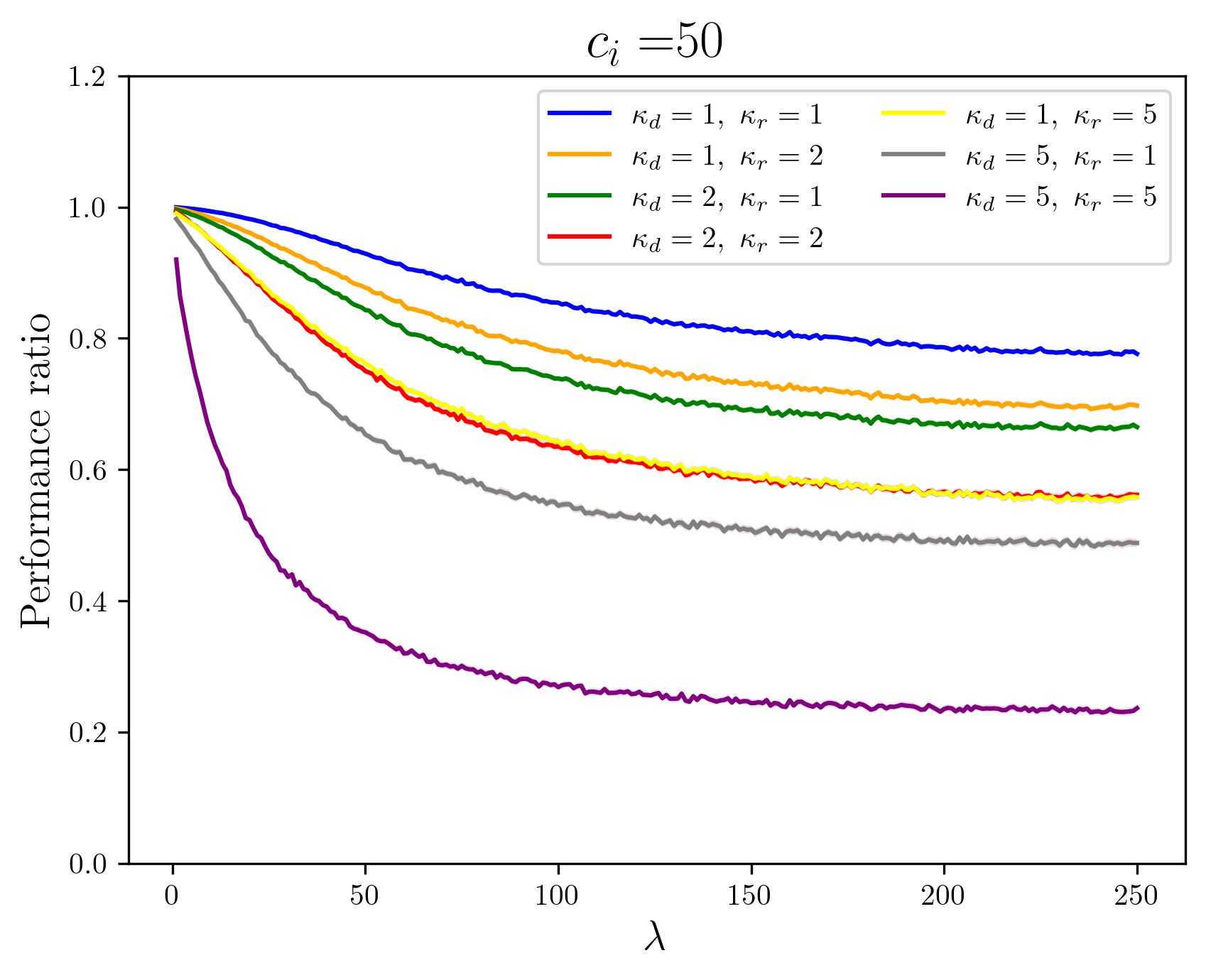}
    \caption{Performance ratio comparisons and empirical confidence intervals for \FLB\ under heterogeneous noise profiles. Evaluated on the random instance from Subsection~\ref{sec:RandomInstance} across different server capacities $\inventory_i \in \{5, 10, 20, 50\}$. Fixed parameters: $n = 3$, $\maxreward = \maxduration = 10$, $m = 500$, and $\inventory_1 = \inventory_2 = \inventory_3$.}
    \label{fig:noisyNumerics}
\end{figure}

\section{Competitive Ratio Analysis for Arbitrary Initial Capacity}

\label{apx:finiteCmin}
\yfedit{In this section, we illustrate the idea of choosing parameters $(\penscalar,\penscalarTwo)$ in {\FLB} for arbitrary initial capacity $\mininventory$ and its induced competitive ratio guarantee. For simplicity, we consider the integer-valued environments (similar to \Cref{sec:Heterogeneous weights}). Extension to real-valued durations can be done using an approach similar to \Cref{apx:FLB analysis real duration}.}

\begin{figure}[h]
    \centering
\includegraphics[width=0.5\linewidth]{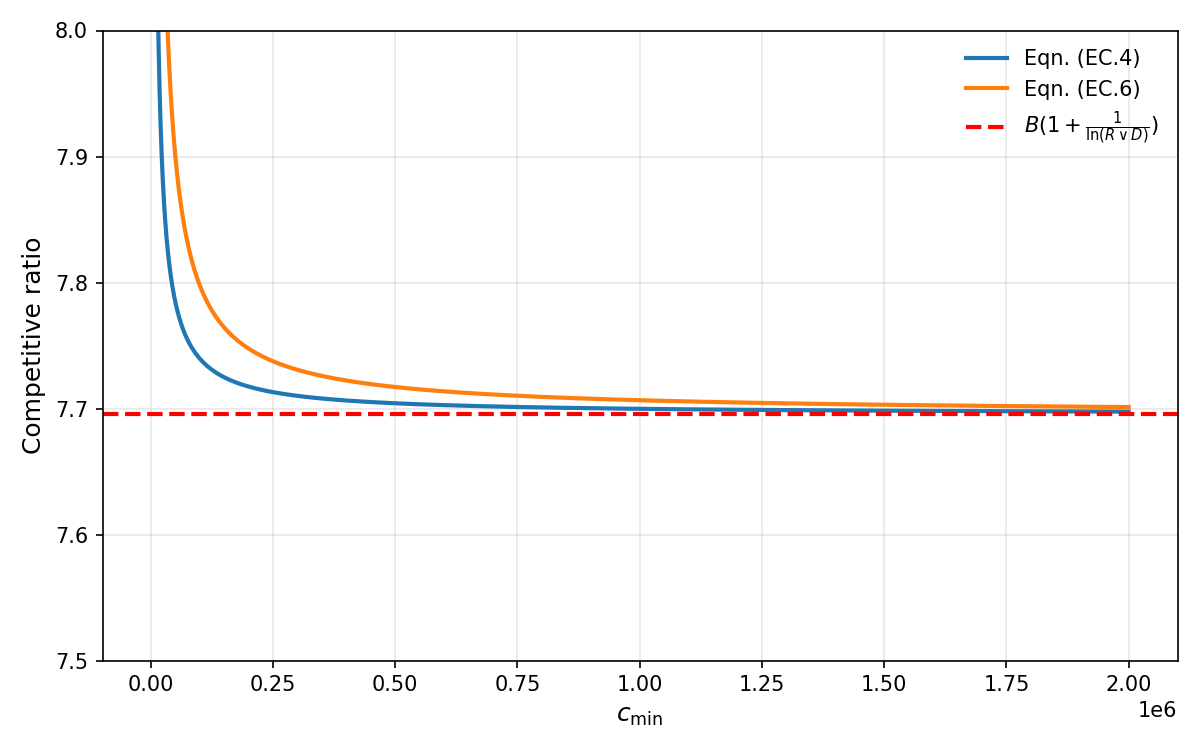}
    \caption{\revcolor{The competitive ratio of the instance with $R=D=10$ as a function of $c_{\min}$. The asymptotic bound is shown in red. The competitive ratio obtained by setting $\beta$ according to Eqn.~\ref{eq:tightCosntraint} and $\eta = 1/\log(R \vee D)$ is shown in blue. The upper bound from Eqn.~\ref{eq:competitiveFiniteCmin} is shown in orange. The algorithm is always feasible for $c_{\min} > 1898$.}}
    \label{fig:finiteCmin10}
\end{figure}

Recall the optimization \ref{eq:FLB parameter integer duration}: 
\begin{align*}
    &\begin{array}{lll}
    \min\limits_{\penscalarTwo,\penscalar}~ &
    \ln(\penscalar)\cdot \left(1 + \penscalarTwo\left(1 + \penscalar\left(\penscalar^{\frac{1}{\mininventory}}-1\right)\right)\right)
    &
    \text{s.t.}
    \vspace{5pt}
    \\
    &
    \ln(\penscalar) \geq - \ln\left(\prod\nolimits_{k \in[\maxduration]}\left(1-\frac{\maxreward}{k(\maxreward+\penscalarTwo
    )}\right)  
    - 
    \frac{(\maxreward+\penscalarTwo)\ln(\penscalar)}{\maxreward\mininventory}
    \right)
    &
    \vspace{5pt}
    \\
    &
    \penscalarTwo > 0,\penscalar\geq e.
    &
    \end{array}
\end{align*}
First notice that the constraint can be written as:
\begin{align*}
    -\ln\left(\frac{1}{\beta} + \frac{(\maxreward+\penscalarTwo)\ln(\penscalar)}{\maxreward\mininventory}\right) \geq -\ln \left(\prod\nolimits_{k \in[\maxduration]}\left(1-\frac{\maxreward}{k(\maxreward+\penscalarTwo
    )}\right) \right)
\end{align*}
We set $\penscalarTwo = \frac{1}{\ln(\maxreward\vee\maxduration)}$. \revcolor{Furthermore, as the objective value is increasing in $\beta$, we choose it to make the first constraint tight:}
\begin{align*}
    -\ln\left(\frac{1}{\beta} + \frac{(\maxreward\ln(\maxreward\vee\maxduration)+1)\ln(\penscalar)}{\maxreward\ln(\maxreward\vee\maxduration)\mininventory}\right) =& - \ln \left(\prod\nolimits_{k \in[\maxduration]}\left(1-\frac{\maxreward\ln(\maxreward\vee\maxduration)}{k(\maxreward\ln(\maxreward\vee\maxduration)+1
        )}\right)  
        \right)
\end{align*}
Looking at this equation as 
\begin{align*}
    -\ln\left(\frac{1}{\beta} + A\ln(\penscalar)\right) = B,
\end{align*}
\revcolor{Note that $B = \Theta(\ln (\maxduration))$}. We can solve it \revcolor{for $\beta$} and get
\begin{align}\label{eq:tightCosntraint}
    \frac{1}{\beta} = -AW_{-1}\left(-\frac{1}{Ae^{\frac{e^{-B}}{A}}}\right).
\end{align}
We chose the non-principal branch because a smaller $\beta$ leads to a lower objective value. For the equation to have a solution, the input to the Lambert function must be greater than $-1/e$, which is equivalent to:
\begin{align*}
    \frac{1}{A} \geq -e^BW_{-1}\left(-e^{-B-1}\right) \qquad \text{or} \qquad \frac{1}{A} \leq -e^BW_{0}\left(-e^{-B-1}\right).
\end{align*}
Therefore, if  
\begin{align}\label{eq:largeEnoughC}
    \mininventory  \geq \frac{\maxreward\ln(\maxreward\vee\maxduration)+1}{\maxreward\ln(\maxreward\vee\maxduration)}e^B2(B+1) \revcolor{= \Theta(\maxduration\log(\maxduration))},
\end{align}
using the fact that $2 \ln(-z) \leq W_{-1}(z) \leq \ln(-z)$ we will have:
\begin{align*}
    \frac{1}{A} = \frac{\maxreward\ln(\maxreward\vee\maxduration)\mininventory}{\maxreward\ln(\maxreward\vee\maxduration)+1} \geq -e^BW_{-1}\left(-e^{-B-1}\right)
\end{align*}
Therefore, for a large enough $\mininventory$ ($\revcolor{\mathcal{O}(\maxduration\log(\maxduration))}$ as in Eqn.~\eqref{eq:largeEnoughC}) we can set $\beta$ as in Eqn.~\eqref{eq:tightCosntraint} and get:
\begin{align*}
    \beta = \frac{1}{-AW_{-1}\left(-\frac{1}{Ae^{\frac{e^{-B}}{A}}}\right)} \leq \frac{1}{e^{-B} + A\ln(A)} = \frac{1}{e^{-B} + \frac{\maxreward\ln(\maxreward\vee\maxduration)+1}{\maxreward\ln(\maxreward\vee\maxduration)\mininventory}\ln(\frac{\maxreward\ln(\maxreward\vee\maxduration)+1}{\maxreward\ln(\maxreward\vee\maxduration)\mininventory})} \leq \frac{e^B}{1-2e^{B}\frac{\ln(\mininventory)}{\mininventory}}.
\end{align*}
Plugging this upper bound for $\beta$, $\eta = \frac{1}{\ln(\maxreward\vee\maxduration)}$, and using the inequality $\ln(1-x) \geq -\frac{x}{1-x}$, we get a competitive ratio better than:
\begin{align}\label{eq:competitiveFiniteCmin}
    \left(B-1+\frac{1}{1-2e^{B}\frac{\ln(\mininventory)}{\mininventory}}\right)\left(1+\frac{1}{\ln(\maxreward\vee\maxduration)}\left(1+ \frac{e^B}{1-2e^{B}\frac{\ln(\mininventory)}{\mininventory}}\left(\left(\frac{e^B}{1-2e^{B}\frac{\ln(\mininventory)}{\mininventory}}\right)^\frac{1}{\mininventory} - 1\right)\right)\right)
\end{align}
Notice that when $\mininventory \rightarrow \infty$ the above term simplifies to 
\begin{align*}
    B\left(1+\frac{1}{\ln(\maxreward\vee\maxduration)}\right),
\end{align*}
Here, both the parameter assignment and the competitive ratio converge to the ones in \Cref{subsec:CR optimization} for the large capacity regime (i.e., $\mininventory \rightarrow \infty$). \revcolor{In \Cref{fig:finiteCmin10} we draw the competitive ratio achieved by {\FLB} for finite values of $\mininventory$.}

\end{document}